\documentclass[11pt]{article}
\usepackage{jheppub}
\usepackage{amsmath,amssymb,amsthm,mathtools,bm,mathrsfs}
\usepackage{bold-extra}
\usepackage[dvipsnames]{xcolor}
\usepackage{comment}
\usepackage{hyperref}
\usepackage[utf8]{inputenc}
\usepackage[titletoc]{appendix}
\usepackage{tikz}
\usetikzlibrary{shapes,arrows,shadows}
\usepackage[colorinlistoftodos]{todonotes}
\usepackage{xspace}
\usepackage{etoolbox}
\usepackage{listofitems}
\usepackage{tikz-cd}
\usepackage{physics}
\usepackage[font={small,sf}]{caption, subcaption}

\makeatletter
\def\@fpheader{\relax}
\makeatother

\newcommand\be{\begin{equation}}
\newcommand\ee{\end{equation}}
\newcommand\bea{\begin{eqnarray}}
\newcommand\eea{\end{eqnarray}}
\newcommand\ba{\begin{array}}
\newcommand\ea{\end{array}}
\newcommand\nn{\nonumber}
\newcommand\eref[1]{(\ref{#1})}
\newcommand\bc{\begin{center}}
\newcommand\ec{\end{center}}

\renewcommand\comment[1]{}
\usepackage{blindtext}
\usepackage{enumitem}

\renewcommand\tilde{\widetilde}

\renewcommand{\dd}{\text{d}}

\renewcommand{\Tr}{\text{Tr}}

\newcounter{descriptcount}
\newlist{enumdescript}{description}{1}
\setlist[enumdescript,1]{%
  before={\setcounter{descriptcount}{0}
          \renewcommand*\thedescriptcount{\arabic{descriptcount}}},
        font={\bfseries\stepcounter{descriptcount} \thedescriptcount.~}
}

\newtheoremstyle{indented}{3pt}{3pt}{\addtolength{\leftskip}{2.5em}}{}{\bfseries}{.}{.5em}{}

\theoremstyle{indented}

 \allowdisplaybreaks
 \numberwithin{equation}{section}

\title{\begin{center} Machine Learning for Calabi--Yau Geometry\\ and Heterotic Compactification \end{center}}

\author[*,\dagger]{Vishnu Jejjala}
\author[\ddagger]{and Dami\'an Mayorga Pe\~na}

\affiliation[\,*]{Mandelstam Institute for Theoretical Physics, School of Physics, University of the Witwatersrand, Johannesburg, South Africa}
\affiliation[\,\dagger]{National Institute for Theoretical and Computational Sciences (NITheCS), Gauteng, South Africa}
\affiliation[\,\ddagger]{CAMSGD, Department of Mathematics, Instituto Superior T\'ecnico, Universidade de Lisboa, 1049-001 Lisboa, Portugal}

\emailAdd{v.jejjala@wits.ac.za}
\emailAdd{damian.mayorga.pena@tecnico.ulisboa.pt}

\abstract{
Quantitative predictions from string compactification depend on geometric data that are difficult to compute analytically.
This review surveys machine learning approaches to predicting topological properties of Calabi--Yau manifolds, discusses the search for new Calabi--Yau fourfolds, as well as numerical methods for approximating Ricci-flat metrics.
In the context of heterotic compactifications, we explain how bundle metrics and harmonic matter forms enter calculations of canonically normalized Yukawa couplings and axion–matter interactions.
We also outline applications to flux compactifications, warped and corrected backgrounds, and geometric structures beyond the Calabi--Yau setting.
Finally, we review available datasets and software, highlight connections with analytic methods, and discuss geometric constraints, numerical validation, and recent progress in rigorous error control.
}

\begin{document}

\setcounter{tocdepth}{2}
\maketitle

\section{Introduction and motivation}
\label{sec:intro}

The Standard Model of particle physics is famously the most precisely tested theory in the history of science.
As one example of the agreement between computation and experiment, the QED calculation of the electron magnetic moment includes $12,\!672$ tenth-order Feynman diagrams~\cite{Aoyama:2012wj}.
Combining the 2023 experimental measurement of this magnetic moment with the Standard Model calculation then available yielded the reciprocal fine structure constant~\cite{Fan:2022eto}
\be
\alpha^{-1} = 137.035999166(15) \,.
\ee
By comparison, the 2020 determination from Rb atom interferometry gave~\cite{Morel:2020dww}
\be
\alpha^{-1} = 137.035999206(11) \,.
\ee

The Standard Model describes the electromagnetic, weak, and strong interactions.
Its gauge group is $G = SU(3)\times SU(2)_L\times U(1)_Y$, with quarks and leptons transforming in fundamental, antifundamental, and singlet representations of its factors.
The gauge forces are mediated by vector bosons in adjoint representations.
A scalar Higgs doublet acquires a vacuum expectation value that breaks the electroweak symmetry to the $U(1)$ of electromagnetism.
Experiments establish three generations of quarks and leptons.
Neutrino masses and mixing already require an extension of the minimal Standard Model, but no new elementary particles, including sterile neutrinos, have been confirmed beyond its particle spectrum~\cite{ParticleDataGroup:2024cfk}.
In particular, despite its immense theoretical appeal, there is no experimental evidence for supersymmetry in the vicinity of the electroweak scale.

Gravity stands apart from the Standard Model.
Its gauge symmetry is diffeomorphism invariance.
General relativity has been tested from submillimeter distances, where experiments probe Newton's inverse square law, to cosmological scales.
General relativity is a classical theory, however.
When we quantize the theory, we introduce the coupling $\kappa = \sqrt{32\pi G_N}$.
This coupling, which sets the strength of the gravitational interaction, has negative mass dimension ($[\kappa]=M^{-1}$ in four dimensions).
The one-graviton exchange between two massive sources, proportional to $G_N$ because the coupling $\kappa$ appears twice, gives the Newtonian potential plus relativistic corrections, depending on kinematics.
As a tree diagram with a dimensionful coupling, it is ultraviolet (UV) finite for fixed external momenta.
Loop corrections to this scattering process generically contain UV divergences from integrating over internal momenta.
The superficial degree of divergence of an $L$-loop diagram in four dimensions grows like $D \sim 2 + 2L$.
Already at one-loop, we generically see UV divergences.
In the language of effective field theory, the divergences are absorbed into local higher derivative counterterms like $R^2$, $R_{\mu\nu}R^{\mu\nu}$, $R_{\mu\nu\rho\sigma}R^{\mu\nu\rho\sigma}$, $\dots$.
The non-analytic pieces that determine the long distance potential are finite and scheme independent; the local UV divergences are removed by renormalizing the counterterms.
At higher loop orders, this power counting permits counterterms with more derivatives.
A perturbative treatment at arbitrarily high energy therefore requires an unbounded set of higher curvature couplings.
In pure gravity there are additional on-shell cancellations: with no matter, the one-loop divergences in general relativity can be removed by field redefinitions/on-shell conditions~\cite{tHooft:1974toh}, but genuine new divergences appear at two loops~\cite{Goroff:1985sz}.
With matter present (as in the example with the two massive sources), we already need higher derivative counterterms at one-loop order.

As an effective field theory, gravity gives finite, predictive low energy observables once the counterterms relevant at the desired order have been fixed.
The difficulty is extending this description to arbitrarily high energies without an unlimited number of independent inputs.
One potential resolution is a non-trivial UV fixed point with finitely many relevant directions, as envisaged in the asymptotic safety proposal~\cite{Weinberg:1980gg}.
Evidence for such fixed points exists in truncations of the renormalization group flow~\cite{Lauscher:2001ya}, but whether this defines a consistent four-dimensional quantum theory of gravity remains unsettled.

The non-renormalizability of a theory with a four fermion interaction $G_F\, \overline\psi \psi \overline\psi \psi$ due to the negative mass dimension of the coupling $[G_F] = M^{-2}$ is accounted for in electroweak theory by the exchange of a $W$ boson so that $G_F \sim g^2/m_W^2$~\cite{Fermi:1934hr,Fermi:1934sk,Feynman:1958ty,Sudarshan:1958vf}.
The coupling to the intermediate vector boson smears out the interaction, and we regard the infinities from the four Fermi theory as arising from pushing an effective field theory beyond its regime of validity.
String theory offers a similar resolution to realize a well-behaved theory of quantum gravity.

Superstring theory is a quantum mechanical framework in which the fundamental degrees of freedom are one-dimensional extended objects (strings) rather than point particles.
We focus on the critical, spacetime supersymmetric theories in ten dimensions.
In addition, there are non-perturbative objects such as the D-branes on which open strings end.
Different choices of worldsheet boundary conditions and gauge groups give rise to five perturbatively distinct, but non-perturbatively dual, critical string theories (the imaginatively named type I, type IIA, and type IIB theories and the heterotic $SO(32)$ and $E_8\times E_8$ theories). These dualities also relate the perturbative theories to eleven-dimensional M-theory and to F-theory, in which an auxiliary elliptic fiber geometrizes the type IIB axiodilaton~\cite{Vafa:1996xn}.
(See, for example,~\cite{Polchinski:1998rq,Polchinski:1998rr} for a textbook review.)
At low energies, each theory reduces to a particular ten-dimensional supergravity coupled to additional fields.
The string spectrum contains a massless spin-two mode identified with the graviton and an infinite tower of massive excitations, thereby providing a candidate UV-complete theory of quantum gravity. 
Compactification of the extra dimensions on suitably chosen internal manifolds (such as Calabi--Yau threefolds) can, in principle, lead to four-dimensional effective theories with gauge interactions and chiral matter resembling the Standard Model.
This was the initial promise of the heterotic string~\cite{Gross:1985fr,Candelas:1985en}.
Beginning with the constructions of~\cite{Braun:2005bw,Bouchard:2005ag}, there have been numerous heterotic compactifications with Standard Model-like spectra.
A representative set of constructions and enumerations are in~\cite{Anderson:2011ns,Constantin:2018xkj,Deen:2020dlf}.
While algebraic methods determine spectra and holomorphic coupling data in many examples, canonically normalized couplings also require metric dependent matter kinetic terms.
Numerical methods applied to the internal geometry provide a route to these physical couplings.

The logic of compactification follows the Kaluza--Klein paradigm~\cite{Kaluza:1921tu,Klein:1926tv}.
One seeks to unify gravity with other interactions by enlarging spacetime to include extra compact dimensions and interpreting components of the higher dimensional metric as gauge fields and scalars in four dimensions.
Concretely, one starts with pure gravity (or gravity plus simple matter) in a $(4+n)$-dimensional spacetime $M_4 \times K_n$, where the internal manifold $K_n$ is compact and small, and assumes that fields are either independent of the extra coordinates or decompose into harmonics on $K_n$.
Upon dimensional reduction, the metric gives rise to ordinary four-dimensional gravity and scalar fields describing deformations of the internal geometry.
Its off-diagonal components yield massless gauge fields when $K_n$ has continuous isometries; the resulting gauge group is non-Abelian only when the isometry algebra is non-Abelian.
A smooth compact Calabi--Yau threefold with full $SU(3)$ holonomy has no continuous isometries.
In heterotic compactifications on such spaces, the non-Abelian gauge fields instead descend from the ten-dimensional gauge sector~\cite{Duff:1994tn}.
The spectrum of fields organizes into a tower of massive Kaluza--Klein modes whose masses are set by the inverse size of the extra dimensions, so that at low energies one sees only the massless sector (gravity plus gauge fields and moduli), while, in principle, the massive modes become relevant as probes of the higher dimensional geometry at large energies, corresponding to the inverse of the length scale associated to the compact directions.
The geometric origin of internal symmetries is compatible with the Coleman--Mandula theorem~\cite{Coleman:1967ad}: in the product background, the four-dimensional Poincar\'e symmetries commute with the internal isometries.
As $U(1)$ is the connected component of the isometry group of the circle, it is no accident that the original five-dimensional Kaluza--Klein model, which considered the Einstein--Hilbert action on $\mathbb{R}^{1,3}\times S^1$, yielded electromagnetism in four dimensions upon compactification.

The worldsheet $\Sigma$ of a string is $(1+1)$-dimensional, and the string is embedded in an ambient spacetime $\mathcal{M}$.
The bosonic sector of the worldsheet theory can be described in terms of fields $X^\mu$ that map points on the worldsheet to the target space:
\be
\begin{array}{cccc}
X^\mu: & \Sigma & \longrightarrow & \mathcal{M} \cr
& (\tau, \sigma) & \mapsto & X^\mu(\tau,\sigma) \,,
\end{array}
\ee
with $\mu=0,1,\ldots,d$ labeling coordinates in the ambient geometry.
The consistency of the worldsheet theory as a conformal field theory (CFT) is why $d=25$ in the bosonic theory and $d=9$ in the superstring theory.
In Lorentzian worldsheet signature, the bosonic part of the string action is
\be
\begin{split}
S_\sigma = -\frac{1}{4\pi\alpha'}\int \dd^2\sigma\,
\Big[&\sqrt{-\gamma}\,\gamma^{ab}g_{\mu\nu}(X)
\partial_aX^\mu\partial_bX^\nu \\
&+\varepsilon^{ab}B_{\mu\nu}(X)\partial_aX^\mu\partial_bX^\nu
+\alpha'\sqrt{-\gamma}\,R^{(2)}\Phi(X)\Big] \,,
\end{split}
\ee
where the spacetime metric $g_{\mu\nu}$, the antisymmetric tensor $B_{\mu\nu}$, and the dilaton $\Phi$ appear as couplings.
Here $R^{(2)}$ is the worldsheet Ricci scalar and $\varepsilon^{ab}$ is the antisymmetric tensor density with $\varepsilon^{01}=+1$.
The Weyl anomaly coefficient for the metric, conventionally referred to as its beta function, is~\cite{Polchinski:1998rq}
\be
\beta_{\mu\nu}^{(g)} = \alpha' \Big( R_{\mu\nu} + 2\nabla_\mu\nabla_\nu \Phi - \frac14 H_{\mu\lambda\rho} H_\nu^{\ \lambda\rho} \Big) + \mathcal{O}(\alpha^{\prime\,2}) \,,
\ee
where $H = \dd B$ at this order and the string coupling is $g_s=e^{\Phi_0}$ for a constant dilaton background $\Phi_0$.
The simultaneous vanishing of all Weyl anomaly coefficients ensures worldsheet conformal invariance and gives the spacetime equations for the metric and the other background fields.
In the critical dimension, with constant dilaton and $H=0$, these equations require a Ricci-flat spacetime at leading order in $\alpha'$.

In our everyday experience, we empirically observe three large spatial dimensions and time.\footnote{
The Brandenberger--Vafa mechanism~\cite{Brandenberger:1988aj} proposes an explanation for this observation.
Suppose that in the early Universe, all the spatial dimensions are compact.
Fundamental string worldsheets can generically intersect in at most three spatial dimensions, allowing winding modes to annihilate and those dimensions to expand while the others stay small.
Thus, string winding and their interaction probabilities may dynamically single out the existence of three large dimensions.
}
Because general relativity teaches us that the force of gravity is a consequence of the curvature of the geometry, the form of the gravitational interaction is sensitive to all the dimensions of spacetime.
The graviton propagates in the extra dimensions.
Precision torsion balance measurements agree with Newton's inverse square law at separations down to $52\,\mu\text{m}$, and constrain the range of a gravitational strength Yukawa correction to be below $38.6\,\mu\text{m}$ at 95\% confidence~\cite{Lee:2020zjt}.
Missing energy searches at the LHC also constrain models of extra dimensions, with bounds that depend on the compactification and on how the Standard Model fields propagate~\cite{ParticleDataGroup:2024cfk}.
In the compactifications considered here, the extra six dimensions are sufficiently small to evade these experimental constraints.
Where we use leading order supergravity, we assume weak string coupling and sufficiently small local curvature in string units.
String scale regions require separate control of higher derivative and quantum corrections.

Taking the spacetime to be $\mathcal{M} = \mathbb{R}^{1,3} \times K_6$, we aim to construct \textit{bona fide} solutions to superstring theory.
At leading order in $\alpha'$, with constant dilaton and vanishing flux, this leads us to compact spaces $K_6$ on which $R_{ab}=0$, with $a,b=1,\ldots,6$ denoting coordinates on $K_6$.
A rich class of such spaces are Calabi--Yau threefolds.
Knowledge of the Ricci-flat metric, together with the gauge bundle geometry and matter wavefunctions, allows metric dependent couplings in the low energy effective action to be calculated.

In order to study the effective field theory of particles in four dimensions upon compactification on a Calabi--Yau manifold $X$, we focus much of our attention on the heterotic string~\cite{Gross:1984dd,Gross:1985fr,Gross:1985rr}.
In the bosonic construction of the heterotic string, sixteen additional left-moving worldsheet bosons describe the gauge degrees of freedom.
Their momenta lie in an even self-dual sixteen-dimensional lattice, as required by modular invariance.
Choosing the $E_8\times E_8$ root lattice gives the corresponding ten-dimensional heterotic gauge symmetry.
Various grand unification theories (GUTs) have gauge groups that are subgroups of $E_8$.
A virtue of compactifications with full $SU(3)$ holonomy and a supersymmetric gauge bundle is that the low energy effective theory enjoys $\mathcal{N}=1$ supersymmetry.
There are four supercharges.
When supersymmetry is unbroken, each supermultiplet has equal numbers of bosonic and fermionic on-shell degrees of freedom, with equal masses.
Since we do not see a scalar particle with the same mass as the electron in Nature, this symmetry is broken.
We are agnostic about how this happens.
A motivation for these $\mathcal{N}=1$ constructions is that softly broken supersymmetry can protect the electroweak scale from large radiative corrections, provided the supersymmetry breaking scale is sufficiently low.

We use the term Calabi--Yau manifold for a compact connected K\"ahler manifold with holomorphically trivial canonical bundle, a condition that implies vanishing first Chern class.\footnote{
Recall that the total Chern class is $c(F) = \det(\mathbf{1}+\frac{\text{i}}{2\pi}F)$, where $F$ is the curvature two-form associated to a complex vector bundle $\pi:V\to X$.
Expanding this out, $c(F)=c_0+c_1+c_2+\ldots = 1+\frac{\text{i}}{2\pi}\Tr{F} + \frac{1}{8\pi^2}(\Tr{(F\wedge F)} - \Tr{F}\wedge \Tr{F})+\ldots$.
For a K\"ahler metric, the Ricci form represents $2\pi c_1(TX)$.
Thus $c_1(TX)=0$ makes this form cohomologically trivial; it does not imply pointwise vanishing of the Ricci form or scalar curvature for an arbitrary metric.
Yau's theorem supplies a Ricci-flat representative in each K\"ahler class.
}
We discuss this convention more precisely below.
We can ask whether we can make a more general choice that is merely compact and Ricci-flat without these extra conditions.
In a flux-free heterotic background of the form $\mathcal{M}=\mathbb{R}^{1,3}\times K_6$, the leading order supersymmetry equations require the internal space $K_6$ to admit a covariantly constant spinor:
\be
\nabla_a \eta = 0 \,.
\ee
The holonomy of the Levi--Civita connection must therefore preserve that spinor.
Ricci-flatness alone does not impose this reduction of holonomy; a metric with full $SO(6)$ holonomy has no invariant spinor.
Holonomy contained in $SU(3)$ preserves a spinor, while full $SU(3)$ holonomy gives four-dimensional $\mathcal{N}=1$ supersymmetry in the heterotic theory when the gauge background also satisfies its supersymmetry equations.
Smaller holonomy can give additional supercharges.
For comparison, an unprojected type II compactification on a Calabi--Yau threefold of full $SU(3)$ holonomy has $\mathcal{N}=2$ supersymmetry.
A compact six-dimensional Riemannian manifold with holonomy contained in $SU(3)$ has the complex, K\"ahler, and trivial canonical bundle structures used in our definition~\cite{Joyce:2001xt}.
Thus, the complex and K\"ahler structures are not extra aesthetic assumptions; they are what the supersymmetry equations enforce.\footnote{
We can as well consider M-theory compactifications on $G_2$ or $Spin(7)$ holonomy manifolds, giving four- or three-dimensional effective theories, respectively.
Invoking fluxes in heterotic theory, we can have complex but non-K\"ahler six manifolds with an $SU(3)$-structure (the Strominger system)~\cite{Strominger:1986uh}.
In type II string theory, the equivalent notion leads us to generalized complex geometry.
Moving from special holonomy to $G$-structures and generalized geometry, the supersymmetry conditions become differential constraints relating the intrinsic torsion to the fluxes.
The price we pay by including fluxes and torsion is that we lose most of the algebraic-geometric and worldsheet control that makes Calabi--Yau sigma models so attractive.
}
On top of this, Calabi--Yau geometry gives additional worldsheet structure.
In type II compactifications, and in the internal sector of heterotic compactifications with the standard embedding, one obtains $\mathcal{N}=(2,2)$ worldsheet supersymmetry, with spectral flow and a clean notion of chiral primaries.
A general heterotic compactification with a nonstandard gauge bundle instead has $\mathcal{N}=(0,2)$ worldsheet supersymmetry~\cite{Jardine:2018sft}.
In both settings, powerful tools from algebraic geometry (Hodge theory, sheaf cohomology, etc.) help compute spectra and couplings.
For heterotic strings, the requirement of a slope polystable holomorphic bundle admitting a Hermitian--Yang--Mills connection relies on the complex and K\"ahler structure.

To obtain the real world from string theory, we should also pay heed to cosmology.
On large scales, the Universe is homogeneous and isotropic.
This allows us to write a Friedmann--Robertson--Walker metric where the scale factor $a(t)$ is a function of time.
The cosmological history is described by solutions to the Friedmann equations for the scale factor.
In the current epoch, we have
\be
\Omega_r + \Omega_m + \Omega_\Lambda + \Omega_\kappa = 1 \,,
\ee
with radiation contributing $\Omega_r$ of order $10^{-4}$, the matter density parameter (including dark matter) $\Omega_m\simeq0.3$, and the dark energy density parameter $\Omega_\Lambda\simeq0.7$ in the standard cosmological model~\cite{ParticleDataGroup:2024cfk}.
The data are consistent with flat spatial sections.
If dark energy is a positive cosmological constant, the expanding Universe approaches de Sitter space as matter and radiation dilute.
Constructing controlled de Sitter solutions in compactifications of type II superstring theory remains a challenge.

While our motivation for studying string compactification on Calabi--Yau manifolds is rooted in the desire to derive realistic vacua from top down constructions in a controlled framework, we should be clear that this is not the only route available.
Other phenomenological approaches have attracted attention over the years and could be equally viable~\cite{Ibanez:2012zz}.

\subsection{Calabi--Yau manifolds}
Calabi conjectured the existence of K\"ahler metrics with prescribed Ricci form, including Ricci-flat metrics when the real first Chern class vanishes~\cite{Calabi:1954}.
Yau established the validity of the conjecture~\cite{Yau:1977CalabiSConjecture,Yau:1978cfy}.
We start by reviewing this statement.

\paragraph{Conjecture [Calabi (1954)]:}
Let $(X,\omega)$ be a compact K\"ahler manifold of complex dimension $n$, and let $\Omega$ be a smooth positive volume form on $X$ such that
\be
\int_X \Omega \;=\; \int_X \omega^n\,.
\ee
Then there exists a unique K\"ahler metric $\tilde\omega$ in the K\"ahler class $[\omega]$ such that
\be
\tilde\omega^n = \Omega\,.
\ee
Equivalently, fixing a real closed $(1,1)$-form $\hat\rho$ with $[\hat\rho]=2\pi c_1(X)$, there is a unique K\"ahler metric $\tilde\omega\in[\omega]$ whose Ricci form $\rho(\tilde\omega)$ equals $\hat\rho$.
In particular, if $c_1(X)=0$ in $H^2(X,\mathbb{R})$, there exists a unique Ricci-flat K\"ahler metric in any given K\"ahler class~\cite{Calabi:1954,Yau:1978cfy}.

\paragraph{Theorem [Yau (1977)]:}
Starting from Calabi's reformulation as a complex Monge--Amp\`ere equation on a compact K\"ahler manifold $(X,\omega)$, Yau's crucial contribution is the uniform \textit{a priori} estimates needed to solve it~\cite{Yau:1978cfy}.
Given a real closed $(1,1)$-form $\hat\rho$ in the same cohomology class as the Ricci form $\rho(\omega)$, or equivalently a smooth volume form $\Omega$ with $\int_X \Omega = \int_X \omega^n$, one seeks a K\"ahler potential $\varphi$ (normalized by $\int_X \omega^n\,\varphi=0$) such that the new K\"ahler form $\omega_\varphi := \omega + \text{i}\partial\bar\partial\varphi$ satisfies
\be
\omega_\varphi^n = e^{f}\,\omega^n \,, \quad \text{with}\ f\ \text{fixed and}\ \int_X \omega^n\,e^{f} = \int_X \omega^n \,.
\ee
This is a fully nonlinear elliptic partial differential equation for $\varphi$.
Yau applies the continuity method: he considers the family of equations
\be
\omega_{\varphi_t}^n=e^{tf+c_t}\,\omega^n \,,\qquad
c_t=\log\frac{\int_X\omega^n}{\int_X\omega^n\,e^{tf}} \,,\qquad t\in[0,1] \,,
\ee
where $c_t$ enforces the volume constraint and satisfies $c_0=c_1=0$.
He shows that the set of $t$ for which a smooth solution $\varphi_t$ exists is nonempty (trivial at $t=0$), open (by the implicit function theorem once linearized), and closed (using \textit{a priori} estimates)~\cite{Joyce:2001xt}.

The heart of the proof is the derivation of uniform \textit{a priori} estimates for $\varphi_t$ independent of $t$.
Yau first obtains a $C^0$ estimate on $\varphi_t$ via the maximum principle and Moser iteration, using the boundedness of $f$ and properties of the Green's function on $X$.
He then establishes second-order estimates by deriving a differential inequality for $\text{tr}_{\omega}(\omega_{\varphi_t})$, which yields control of the eigenvalues of the metric.
A Calabi-type third-order estimate and elliptic regularity give the higher derivative bounds.
Modern presentations can instead use Evans--Krylov estimates followed by Schauder estimates for the higher regularity step.
These estimates imply closedness, so a solution exists for $t=1$, giving the desired K\"ahler metric with prescribed Ricci form (and in particular a Ricci-flat metric when $c_1(X)=0$), thus confirming that the conjecture is true.
Though we know due to Yau that the Ricci-flat metric exists and due to Calabi that it is unique in each K\"ahler class, the theorem does not generally provide an explicit closed form expression.
Flat complex tori are an elementary exception; for compact spaces with nontrivial holonomy, finding the metric remains a difficult task.

For our purposes, a compact \emph{Calabi--Yau manifold} $X$ is a compact connected complex $n$-dimensional K\"ahler manifold\footnote{
A \emph{complex manifold} has an atlas of charts to open subsets of $\mathbb{C}^n$ with holomorphic transition functions.
For a Riemannian metric, we can choose normal coordinates centered at any point $p$ so that $g_{ab}(p)=\delta_{ab}$ and $\partial_cg_{ab}(p)=0$.
Near $p=0$, the expansions are $g_{ab}(x)=\delta_{ab}-\frac13R_{acbd}(0)x^cx^d+\mathcal{O}(|x|^3)$ and $\Gamma^a_{\ bc}=-\frac23R^a_{\ (bc)d}(0)x^d+\mathcal{O}(|x|^2)$.
For a Hermitian metric on a complex manifold, analogous normal coordinates cannot always be chosen holomorphically.
A Hermitian metric is \emph{K\"ahler} precisely when such holomorphic normal coordinates exist at every point; equivalently, its associated real $(1,1)$-form is closed.
Henceforth, unless otherwise stated, we take \emph{dimension} to mean complex dimension.
} whose canonical bundle is holomorphically trivial:
\be
K_X:=\bigwedge^n(T^{1,0}X)^*=\Omega_X^n\cong\mathcal{O}_X \,.
\ee
Equivalently, $X$ admits a nowhere vanishing holomorphic $(n,0)$-form, unique up to multiplication by a nonzero complex scalar.
This implies $c_1(TX)=0$, but it is stronger than vanishing of the real class $c_1(TX)\in H^2(X,\mathbb{R})$.
For example, an Enriques surface has a nontrivial canonical bundle whose square is trivial: its real first Chern class vanishes, but it has no nowhere vanishing holomorphic two-form.
Yau's theorem still supplies a Ricci-flat K\"ahler metric in the weaker real-Chern-class setting, without necessarily reducing the full holonomy to $SU(n)$.

Under our trivial canonical bundle convention, Yau's theorem gives a unique Ricci-flat K\"ahler metric in each K\"ahler class for every choice of complex structure, and its full holonomy is contained in $SU(n)$~\cite{Joyce:2001xt}.
This definition includes complex tori, product spaces, and hyperk\"ahler manifolds.
For a simply connected manifold whose Ricci-flat metric is irreducible, the holonomy is $SU(n)$ or, in even complex dimension $n=2m$, $Sp(m)$.
The two possibilities coincide for $n=2$, since $Sp(1)=SU(2)$, but differ for $m>1$.
Full $SU(n)$ holonomy implies
\be
H^0(X,\Omega_X^p)=0 \,,\quad\text{for }0<p<n \,,\qquad
\dim_{\mathbb{C}}H^0(X,\Omega_X^n)=1 \,.
\ee
The vanishing of intermediate holomorphic forms is an additional restriction, not part of our general definition.
In particular, the threefold and fourfold Hodge diamonds discussed below assume full $SU(3)$ and $SU(4)$ holonomy, respectively.

The first and second Stiefel--Whitney classes encode obstructions to orientability and spin.
The real tangent bundle of any complex manifold possesses a natural complex structure.
This means that the first Stiefel--Whitney class $w_1(X) \in H^1(X,\mathbb{Z}_2)$ automatically vanishes and ensures the orientability of the space.
For any complex manifold, the second Stiefel--Whitney class is the modulo $2$ reduction of the first Chern class: $w_{2}(X)\equiv c_{1}(X)\!\! \mod \,2$.
Since $c_1(X)$ vanishes by definition on a Calabi--Yau manifold, $X$ consequently admits a spin structure.
This is essential for defining spacetime fermions.

In K\"ahler geometry, the local metric components form a Hermitian matrix that can be derived from a local K\"ahler potential $K(z,\bar{z})$:
\be
g_{a\bar{b}}=\partial_a \partial_{\bar{b}} K(z,\bar{z}) \,.
\ee
On a compact K\"ahler manifold, such potentials are defined locally and differ on overlaps by K\"ahler transformations,
\be
K(z^\prime,\bar{z}^\prime)=K(z,\bar{z})+f(z)+\bar{f}(\bar{z})\,,    
\ee
with $f$ and $\bar{f}$ being holomorphic and antiholomorphic functions, respectively.
The metric can be used to construct the K\"ahler form as
\be
J=\frac{\text{i}}{2} g_{a\bar{b}}\, \dd z^a \wedge \dd\bar{z}^{\bar{b}} \,.
\ee
This is a closed $(1,1)$-form.
The corresponding Ricci tensor is given by:
\be
R_{a\bar{b}}=-\partial_a\partial_{\bar{b}}\log\det(g_{c\bar{d}}) \,.
\ee
For further details, we direct the reader to the classic references~\cite{Candelas:1987is,Hubsch:2025qhd}.

\subsection{Ricci-flat metrics}

We have seen that the Ricci-flat metric, by virtue of solving the equations of motion, supplies a leading order classical solution in the unwarped approximation with constant dilaton and vanishing three-form flux~\cite{Candelas:1985en}.
This metric also provides the geometric input needed to extract quantitative particle physics from a Calabi--Yau compactification.

Holomorphic superpotential couplings alone do not determine physical masses and interactions: the matter fields must also be canonically normalized.
This normalization can change coupling ratios and flavor hierarchies, and requires the Ricci-flat metric, the Hermitian bundle metric, and harmonic matter wavefunctions, as explained in Section~\ref{sec:yukawa}.
Special geometry provides shortcuts for certain standard embedding couplings, but general bundles require all of these geometric inputs.

Moduli stabilization requires the moduli K\"ahler potential, its metric, the superpotential, and $D$-terms.
Together these determine the scalar potential and canonically normalized moduli masses.
A family of Ricci-flat metrics supplies input for geometric corrections, local flux energies, and backreaction that can affect stabilization.
Some leading terms follow from periods or intersection data, but these shortcuts do not determine all corrections.
The metric alone does not generate a stabilizing potential.

The metric is also the starting point for considering $\alpha'$ corrections.
Ricci flatness allows nonzero Riemann curvature, which enters higher derivative terms and sources corrections to the background.
Computing this curvature tests local control, for example through $\alpha'\|\text{Riem}(g)\|\ll1$.
Corrections to the K\"ahler potential can modify the scalar potential and lift flat directions~\cite{Anguelova:2010ed}.
The corrected background need not remain Ricci-flat.

For supersymmetry breaking, the moduli dependence of the matter kinetic matrix enters soft scalar masses and trilinear couplings.
Calculating its derivatives and curvature over moduli space connects the internal geometry to the soft spectrum and its flavor structure.
This requires a specified breaking sector and its auxiliary fields, as well as the geometric data.

In type II constructions based on Calabi--Yau compactifications, the Ricci-flat metric supplies the Hodge star, Laplacian, and curvature data needed to compute local flux densities, warping, and $\alpha'$ corrections.
Together with the remaining compactification data, it enables quantitative tests of the balance between fluxes, nonperturbative effects, and uplifting contributions, including their effects on canonically normalized moduli masses and the Kaluza--Klein scale.
These calculations sharpen assessments of metastability and scale separation, although the metric alone does not establish a controlled de~Sitter vacuum.

Finally, the massive Kaluza--Klein spectrum depends on metric dependent differential operators.
For a minimally coupled scalar, $-\Delta_gf_n=m_n^2f_n$ determines the internal contribution to the masses~\cite{Braun:2008jp}.
The resulting eigenfunctions determine interactions with the heavy modes, while the mass gap sets the scale below which a four-dimensional effective theory is valid.

Thus, the construction of good numerical approximations to the Ricci-flat Calabi--Yau metric is a crucial input into string model building.
A complementary review discusses the calculations that become feasible once a Ricci-flat metric is available~\cite{Berglund:2026hdh}.

As an aside, Ricci-flat K\"ahler metrics on noncompact Calabi--Yau spaces are understood through extensive existence and uniqueness results, although explicit formul\ae\ remain restricted to special geometries.
In the noncompact setting, asymptotic conditions are essential: for suitable resolutions and smoothings of Calabi--Yau cones, complete asymptotically conical metrics exist and satisfy uniqueness theorems once the complex structure, K\"ahler class, and appropriate asymptotic data are fixed~\cite{Conlon:2012de}.
For the threefold cones relevant to D$3$-branes, the transverse metric takes the form
\be
 \dd s_X^2=\dd r^2+r^2 \dd s_Y^2 \,,\qquad
 \operatorname{Ric}(g_Y)=4g_Y \,,
\ee
where the five-dimensional link $Y$ is Sasaki--Einstein; this metric is Ricci-flat and K\"ahler away from the generally singular apex.
Explicit examples include the conifold with link $T^{1,1}$ and the infinite $Y^{p,q}$ family~\cite{Klebanov:1998hh,Gauntlett:2004yd}.
More generally, existence of a Ricci-flat cone metric on an algebraically specified normalized Fano cone is characterized by an appropriate K-stability condition~\cite{Collins:2015qsb}.
A stack of $N$ ordinary D$3$-branes at the apex gives, in the infrared (IR), a generically four-dimensional $\mathcal N=1$ superconformal theory with an $\text{AdS}_5\times Y$ near-horizon dual~\cite{Klebanov:1998hh}.
For studying this worldvolume theory, an explicit metric is often unnecessary: in toric examples, the quiver and superpotential can be constructed from toric geometry and brane tilings~\cite{Franco:2005sm}, while volume minimization determines the Reeb vector encoding the superconformal R-symmetry and the volume without reconstructing the full metric~\cite{Martelli:2005tp}.

\subsection{Outline}
A major wave of modern machine learning applications to string theory began in 2017~\cite{He:2017set,Krefl:2017yox,Ruehle:2017mzq,Carifio:2017bov}.
Since then, machine learning methods have been used to study the topology and differential geometry of Calabi--Yau manifolds and to search for new geometric constructions.
This review focuses on these geometric methods and their applications to heterotic compactification, particularly bundle geometry and canonically normalized couplings.
We also discuss selected extensions to fluxes, corrected backgrounds, and geometric structures beyond the Calabi--Yau setting.
Searches over flux vacua and brane configurations are outside our main scope.
For complementary perspectives, we refer the reader to other reviews such as~\cite{He:2018jtw,Anderson:2023viv}.

The review proceeds from discrete geometric data to differential geometry and physical couplings.
\begin{itemize}
\item Section~\ref{sec:topology} introduces Calabi--Yau constructions and invariants, their prediction with machine learning, and the problem of recognizing equivalent geometric presentations.
\item Section~\ref{sec:cy-databases} surveys the principal catalogs and computational tools, including searches for new geometries and the limitations of the resulting datasets.
\item Section~\ref{sec:points} develops point sampling and numerical integration, which underlie the metric computations that follow.
\item Section~\ref{sec:don} reviews Donaldson's algorithm and related finite dimensional approximations.
\item Section~\ref{sec:pinn} explains how differential equations supply training objectives for neural networks.
\item Section~\ref{sec:ml-metrics} applies this framework to Ricci-flat metrics, including their dependence on moduli and the role of sampling.
\item Section~\ref{sec:bundle-computations} connects the geometry to heterotic matter fields and physical couplings, treating both standard and non-standard embeddings and extending the discussion to axion--matter interactions.
\item Section~\ref{sec:differential-geometry} considers fluxes, warping, $\alpha'$ corrections, structures with torsion, $G_2$ geometry, and special Lagrangian submanifolds, before discussing analytic constructions and the rigorous validation of numerical metrics.
\item Section~\ref{sec:calabi-yau-software} surveys software and common benchmarks, and Section~\ref{sec:conclusion} concludes with open problems and prospects for string phenomenology.
\item Appendix~\ref{app:Appendix} summarizes the machine learning architectures and search methods used throughout the review.
\end{itemize}

\section{Topology of Calabi--Yau manifolds}
\label{sec:topology}

We start by reviewing the topological properties of Calabi--Yau spaces.

\paragraph{Calabi--Yau onefold:}
The simplest example of a Calabi--Yau space is an elliptic curve.
It is a one-dimensional Calabi--Yau manifold.
The defining equation
\be
z_1^3 + z_2^3 +z_3^3 = 0 \subset \mathbb{P}^2
\ee
defines a smooth cubic model of an elliptic curve in the complex projective plane.\footnote{
Recall that $\mathbb{P}^n = (\mathbb{C}^{n+1}\setminus\{0\})/\mathbb{C}^*$, where $z\sim\lambda z$ for $\lambda\in\mathbb{C}^*$.
That is to say, $\mathbb{P}^n$ is the set of lines through the origin in $\mathbb{C}^{n+1}$ and has dimension $n$.
It is a K\"ahler manifold endowed with the Fubini--Study metric.
}
Over $\mathbb{C}$, an elliptic curve $E$ is a smooth projective curve of genus $1$ with a chosen basepoint $O$; analytically, $E(\mathbb{C})$ is a complex torus.
The Euler character of the Calabi--Yau onefold is $\chi=2-2g=0$.
To pass from $E$ to the fundamental cell of the torus in the complex plane, one proceeds as follows.
We choose a nonzero holomorphic $1$-form $\omega$ on $E$ (for instance, if $E$ is given in Weierstrass form $y^2 = 4x^3 - g_2 x - g_3$, one can take $\omega = \frac{\dd x}{y}$).
Integrating $\omega$ along a positively oriented basis $\{\gamma_1,\gamma_2\}$ of $H_1(E,\mathbb{Z})$ gives the periods
\be
\omega_1 = \int_{\gamma_1} \omega \,, \qquad
\omega_2 = \int_{\gamma_2} \omega \,,
\ee
which generate a lattice $\Lambda = \mathbb{Z}\,\omega_1 \oplus \mathbb{Z}\,\omega_2 \subset \mathbb{C}$ with $\operatorname{Im}(\omega_2/\omega_1) > 0$.
The Abel map with respect to the differential $\omega$,
\be
\Phi : E(\mathbb{C}) \longrightarrow \mathbb{C}/\Lambda \,, \qquad
\Phi(P) = \int_{O}^{P} \omega \ \bmod \Lambda \,,
\ee
is a biholomorphism, so $E(\mathbb{C}) \cong \mathbb{C}/\Lambda$.
The fundamental cell of the torus is then the subset
\be
\mathcal{F} = \{ s\,\omega_1 + t\,\omega_2 ~|~ 0 \leq s,t < 1 \} \subset \mathbb{C} \,,
\ee
whose opposite edges are identified modulo $\Lambda$.
Thus, starting from the elliptic curve $E$, the periods of a holomorphic $1$-form determine a lattice in $\mathbb{C}$, and the corresponding fundamental parallelogram in the complex plane is the fundamental cell of the associated complex torus.

After rescaling the periods, we write $E=\mathbb{C}/(\mathbb{Z}+\tau\mathbb{Z})$, with $\operatorname{Im}\tau>0$.
The complex structure is specified by the complex parameter $\tau$ modulo $SL(2,\mathbb{Z})$, while the K\"ahler modulus is the independent area $A>0$.
The coordinate area of the normalized period parallelogram is $\operatorname{Im}\tau$; the metric scale fixes its physical area.
The numbers of independent deformations of various types can be expressed in terms of topological data.
As the number of deformations is given by the dimensions of cohomology groups, this information is summarized in the Hodge diamond:\footnote{
On a compact K\"ahler manifold $X$ of complex dimension $n$, the Hodge numbers $h^{p,q}(X)$ are defined by $h^{p,q}(X) := \dim H^{q}(X, \Omega_X^p)$, where $\Omega_X^p$ denotes the sheaf of holomorphic $p$-forms on $X$ and $H^q(X,\Omega_X^p)$ is its sheaf cohomology.
Equivalently, by Hodge theory one has $h^{p,q}(X) = \dim\mathcal{H}^{p,q}(X)$, the dimension of the space $\mathcal{H}^{p,q}(X)$ of harmonic $(p,q)$-forms, so $h^{p,q}$ counts independent harmonic representatives of $(p,q)$ cohomology classes.
To make the notation more compact, we will often simply write $h^{p,q}$.
The Hodge diamond is symmetric under reflection across both its vertical and horizontal axes, expressing Hodge symmetry $h^{p,q} = h^{q,p}$ and Serre duality $h^{p,q} = h^{n-p,n-q}$.
The former is the statement that complex conjugation gives an isomorphism between harmonic $(p,q)$-forms and harmonic $(q,p)$-forms.
}
\be
\begin{array}{ccc}
 & h^{0,0} & \cr
h^{1,0} &   & h^{0,1} \cr
 & h^{1,1} &
\end{array}
\quad = \quad
\begin{array}{ccc}
 & 1 & \cr
1 &   & 1 \cr
 & 1 &
\end{array} \,.
\ee

For the torus, this is all rather trivial.
More generally, the Bogomolov--Tian--Todorov theorem states that local complex deformations of a compact K\"ahler manifold with holomorphically trivial canonical bundle are unobstructed~\cite{Bogomolov:1978Hamiltonian,Tian:1987Smoothness,Todorov:1989umc}.
The number of independent real K\"ahler moduli (which govern the ``size'' of the manifold) is the real dimension of the space of K\"ahler classes.
This dimension is given by the Hodge number $h^{1,1}$.
The K\"ahler form $\omega$ is a real $(1,1)$-form, and its cohomology class $[\omega]\in H^2(X,\mathbb{R})$ determines the size and volume of the manifold.
The space of real $(1,1)$ cohomology classes, $H^{1,1}(X) \cap H^2(X,\mathbb{R})$, is then a real vector space of dimension $h^{1,1}$.
The K\"ahler cone is an open convex cone within this space, so the number of independent parameters is $h^{1,1}$.
The number of independent complex structure moduli (which govern the ``shape'' of the manifold) is the complex dimension of the moduli space of complex structures.
This dimension is given by the Hodge number $h^{n-1,1}$.
Since a Calabi--Yau $n$-fold has a nowhere vanishing holomorphic $n$-form $\Omega$, the tangent space to the space of complex structure deformations is isomorphic to the cohomology group $H^1(X,T_X)$, where $T_X$ is the holomorphic tangent bundle.
By the triviality of the canonical bundle $K_X \cong \Omega_X^n \cong \mathcal{O}_X$ and the Dolbeault isomorphism, we have $\dim H^1(X,T_X)=\dim H^1(X,\Omega_X^{n-1})=h^{n-1,1}$.
Thus, to summarize
\bea
\dim_{\mathbb{R}}\text{(K\"ahler cone)} & = & h^{1,1}(X) \,, \\
\dim_{\mathbb{C}}\text{(complex structure moduli space)} & = & h^{n-1,1}(X) \,,
\eea
for a Calabi--Yau $n$-fold $X$.
In string compactifications with a $B$-field, its periods can combine with the K\"ahler parameters into $h^{1,1}$ complexified K\"ahler moduli.

In the normalized period coordinate, the flat metric on the torus with area $A$ is
\be
\dd s^2 = \frac{A}{\operatorname{Im}\tau}\,\dd z\,\dd\overline{z} \,.
\ee
Reversing the change of coordinates to return to the elliptic curve, we obtain an analytic expression for the Ricci-flat metric on $E$.
This was realized by Clifford in 1873~\cite{Volkert:2013SpaceFormsHistory}.
Flat complex tori admit explicit metrics in every dimension.
For compact Calabi--Yau manifolds with nontrivial holonomy, no comparable general explicit formula is known.

\paragraph{Calabi--Yau twofold:}
The simply connected Calabi--Yau twofold with holonomy $SU(2)$ is a K3 surface; complex two-tori also have trivial canonical bundle under the broader definition used above.
It is essentially unique because any two complex analytic K3 surfaces are diffeomorphic as smooth four real dimensional manifolds, as shown by Kodaira~\cite{Kodaira:1964Structure}.
Its Hodge diamond is
\be
\begin{array}{ccccc}
&& h^{0,0} \cr
& h^{1,0} && h^{0,1} \cr
h^{2,0} && h^{1,1} && h^{0,2} \cr
& h^{2,1} && h^{1,2} \cr
&& h^{2,2} \cr
\end{array}
\quad = \quad
\begin{array}{ccccc}
&& 1 \cr
& 0 && 0 \cr
1 && 20 && 1 \cr
& 0 && 0 \cr
&& 1 \cr
\end{array} \,. \label{eq:k3hodge}
\ee
This Hodge diamond encodes the ways to deform the K3 surface.
The general formul\ae\ give $20$ complex structure moduli and, at fixed complex structure, a K\"ahler cone of real dimension $20$.
Quartic hypersurfaces in $\mathbb{P}^3$ instead describe a polarized family: the hyperplane class must remain of type $(1,1)$, imposing one complex condition and leaving $19$ complex structure moduli~\cite{Gritsenko:2007K3}.
For example, consider the Fermat quartic in $\mathbb{P}^3$:
\be
z_1^4 + z_2^4 + z_3^4 + z_4^4 = 0 \subset \mathbb{P}^3 \,.
\ee
Homogeneous deformations of the defining equation involve tweaking the coefficients in front of each of the quartic monomials and introducing deformations $z_i z_j^3$, $z_i^2 z_j^2$, $z_i z_j z_k^2$, and $z_1 z_2 z_3 z_4$.
The three mixed monomial families have, respectively, $4\times3=12$, ${4\choose2}=6$, and $4{3\choose2}=12$ members, with distinct indices in each monomial; including the four $z_i^4$ terms and $z_1z_2z_3z_4$ gives $35$ coefficients before quotienting by redundancies.
We can perform a $GL(4,\mathbb{C})$ coordinate transformation on the $z_i$s to remove $16$ of these parameters as redundant.
The remaining $19$ deformations account for the complex structure moduli of this polarized quartic family.
If we also restrict the K\"ahler class to a positive multiple of the hyperplane class, only its overall scale remains as a K\"ahler parameter; this restriction does not describe the full K\"ahler cone of K3.
The Euler character of K3 is $\chi=24$.
The $1$ on the left and right sides of the Hodge diamond is the statement that there is a nowhere vanishing top holomorphic form and its antiholomorphic conjugate, each unique up to scale.
Similarly, the $1$ at the top and the bottom of the Hodge diagram means that there is exactly one independent constant function and exactly one independent top degree $(n,n)$ cohomology class (the volume or fundamental class).

\paragraph{Calabi--Yau threefolds:}
The Hodge diamond of a Calabi--Yau threefold with holonomy $SU(3)$ is
\be
\begin{array}{ccccccc}
&&& h^{0,0} \cr
&& h^{1,0} && h^{0,1} \cr
& h^{2,0} && h^{1,1} && h^{0,2} \cr
h^{3,0} && h^{2,1} && h^{1,2} && h^{0,3} \cr
& h^{3,1} && h^{2,2} && h^{1,3} \cr
&& h^{3,2} && h^{2,3} \cr
&&& h^{3,3} \cr  
\end{array}
\quad = \quad
\begin{array}{ccccccc}
&&& 1 \cr
&& 0 && 0 \cr
& 0 && h^{1,1} && 0 \cr
1 && h^{2,1} && h^{2,1} && 1 \cr
& 0 && h^{1,1} && 0 \cr
&& 0 && 0 \cr
&&& 1 \cr  
\end{array} \,. \label{eq:hd}
\ee
The Hodge numbers $h^{1,1}$ and $h^{2,1}$ count, respectively, the number of K\"ahler and complex structure deformations of the manifold.
The ranks of the independent two-, three-, and four-cycle classes are $b_2=b_4=h^{1,1}$ and $b_3=2h^{2,1}+2$.
The equality $b_2=b_4$ follows from Poincar\'e duality.
The Euler character of a Calabi--Yau threefold is
\be
\chi = 2(h^{1,1} - h^{2,1}) \in 2\mathbb{Z} \,.
\ee

Two important classes of Calabi--Yau threefolds are complete intersections in products of projective spaces~\cite{Candelas:1987kf} and anticanonical hypersurfaces in toric varieties~\cite{Batyrev:1993oya,Kreuzer:2000xy}.
We describe these constructions below, their use in machine learning in Section~\ref{sec:mltop}, and the associated databases in Section~\ref{sec:cy-databases}.

\begin{figure}[tbp]
\centering
\begin{minipage}[b]{.49\textwidth}
\centering
\includegraphics[height=.78\linewidth]{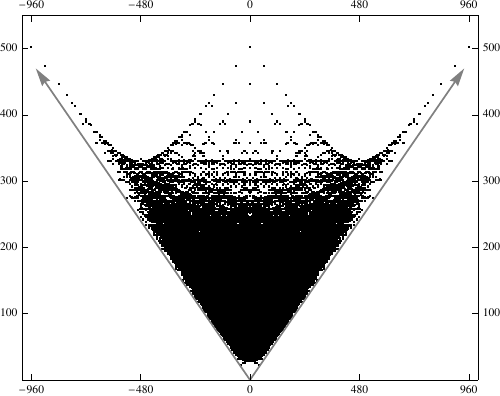}\\[3pt]
\textnormal{(a)}
\end{minipage}\hfill
\begin{minipage}[b]{.49\textwidth}
\centering
\includegraphics[height=.78\linewidth]{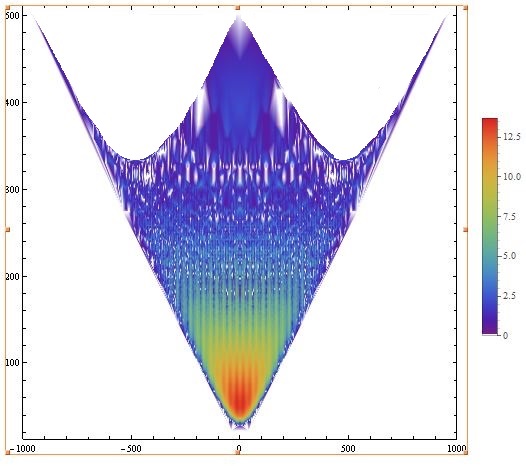}\\[3pt]
\textnormal{(b)}
\end{minipage}
\caption{\small The shield of Hodge numbers for Calabi--Yau threefold hypersurfaces associated with the Kreuzer--Skarke database of four-dimensional reflexive polytopes.
In both panels the horizontal axis is the Euler character $\chi=2(h^{1,1}-h^{2,1})$, and the vertical axis is $h^{1,1}+h^{2,1}$.
Panel (a) shows the $30108$ distinct Hodge pairs, while panel (b) colors each pair by the natural logarithm of its multiplicity in the polytope database.
The multiplicity counts reflexive polytopes yielding the pair; it does not count distinct Calabi--Yau geometries.
Figures reproduced from~\cite{He:2015fif}.}
\label{fig:hodge-shield}
\end{figure}

At present there is no proven universal bound on the Hodge numbers of all smooth compact Calabi--Yau threefolds: we do not even know whether the set of deformation types of Calabi--Yau threefolds is finite, nor whether their Hodge numbers are globally bounded, outside special subclasses such as elliptically fibered threefolds~\cite{Borisov:2015CompleteIntersectionsTrivial,He:2018jtw}.
The Kreuzer--Skarke classification of four-dimensional reflexive polytopes produces Calabi--Yau threefold hypersurfaces with $30108$ distinct Hodge pairs $(h^{1,1},h^{2,1})$; in this dataset one finds
\be
1 \le h^{1,1},~h^{2,1} \le 491 \,,\qquad |\chi|\le 960 \,,
\ee
and no example with larger $|\chi|$ is known~\cite{Kreuzer:2000xy,He:2018jtw,Esser:2022CalabiYauVarietiesLarge}.
The characteristic shield of Hodge pairs and their multiplicities are shown in Figure~\ref{fig:hodge-shield}.
The Hodge pairs of generic elliptically fibered threefolds over the bases studied by Taylor form a subset of the Kreuzer--Skarke Hodge data and reproduce much of its ``shield'' pattern~\cite{Taylor:2012dr}.
This empirical pattern does not establish a universal bound for all Calabi--Yau threefolds.
Patterns in the Hodge numbers are studied in~\cite{He:2015fif}.

For Calabi--Yau threefolds that are elliptically fibered (\textit{i.e.}, admitting a global section) there are now rigorous bounds: using F-theory and the classification of allowed bases, Taylor proved $h^{2,1}\le 491$ for any such threefold~\cite{Taylor:2012dr}.
More recently, Kim, Vafa, and Xu used six-dimensional $\mathcal N=(1,0)$ supergravity plus F-theory to show that for elliptically fibered Calabi--Yau threefolds one has
\be
h^{1,1}(\text{CY}_3)\le 491 \,, \qquad h^{1,1}(\text{base})\le 194 \,,
\ee
and these bounds are sharp~\cite{Kim:2024eoa}.
In parallel, Borisov and Li proved birational boundedness results for complete intersections of base point free divisors with trivial canonical class in the Fano settings they consider~\cite{Batyrev:1994pg}. 
In particular, the Batyrev--Borisov construction produces only a bounded set of Hodge numbers in each fixed dimension, even when the codimension varies~\cite{Borisov:2015CompleteIntersectionsTrivial}.
These results motivate the conjecture that Calabi--Yau threefold Hodge numbers are globally bounded, but do not prove a universal Kreuzer--Skarke shield bound~\cite{He:2018jtw}.
More specifically, Johnson and Taylor enumerate elliptically fibered threefolds with section in the regime $h^{2,1}\ge350$ and conjecture that their construction exhausts Calabi--Yau threefolds in that regime, including those not assumed to be elliptically fibered~\cite{Johnson:2014xpa}.
Global boundedness beyond the established subclasses remains open.

For a mirror pair of Calabi--Yau threefolds, reflecting the Hodge diamond in~\eref{eq:hd} interchanges $h^{1,1}$ and $h^{2,1}$.
The symmetry of Figure~\ref{fig:hodge-shield} illustrates this exchange at the level of Hodge numbers.
The geometric construction of mirror pairs and the associated physical duality contain additional information beyond this numerical symmetry.
Historically, mirror symmetry emerged in the late 1980s as an observation in $\mathcal{N}=(2,2)$ superconformal field theory, with explicit geometric mirror pairs constructed via orbifolds by Greene and Plesser and their collaborators, and systematically explored in higher dimensions in~\cite{Greene:1993vm,Greene:1991iv}.
Candelas, de la Ossa, Green, and Parkes' computation of instanton corrected Yukawa couplings on the quintic~\cite{Candelas:1990rm} sparked major mathematical interest, while later developments provided deeper conceptual frameworks: Strominger--Yau--Zaslow proposed a geometric explanation in terms of dual special Lagrangian $T^3$-fibrations~\cite{Strominger:1996it}, and Kontsevich reformulated mirror symmetry as an equivalence of categories (homological mirror symmetry) between the derived category of coherent sheaves on one manifold and the Fukaya category of its mirror~\cite{Kontsevich:1995joz}.

Mirror symmetry for Calabi--Yau threefolds is a duality between pairs of Calabi--Yau manifolds $(X,Y)$ such that type IIA string theory compactified on $X$ is physically equivalent to type IIB compactified on $Y$.
At the level of geometry, this predicts an exchange of Hodge numbers
\be
h^{1,1}(X) = h^{2,1}(Y) \,, \qquad h^{2,1}(X) = h^{1,1}(Y) \,,
\ee
and relates the complex structure moduli space of $X$ to the complexified, quantum corrected K\"ahler moduli space of $Y$, so that quantities like Yukawa couplings and Gromov--Witten invariants on one side can be computed from classical period integrals on the mirror~\cite{Candelas:1990rm}.
For the quintic threefold, this led to striking predictions for the numbers of rational curves of given degree, providing a powerful new tool in enumerative geometry and giving some of the first concrete evidence for the mirror conjecture.

\paragraph{Calabi--Yau fourfolds:}
The Hodge diamond of a Calabi--Yau fourfold with holonomy $SU(4)$ is
\be
\begin{array}{ccccccccc}
&&&& h^{0,0} \cr
&&& h^{1,0} && h^{0,1} \cr
&& h^{2,0} && h^{1,1} && h^{0,2} \cr
& h^{3,0} && h^{2,1} && h^{1,2} && h^{0,3} \cr
h^{4,0} && h^{3,1} && h^{2,2} && h^{1,3} && h^{0,4} \cr
& h^{4,1} && h^{3,2} && h^{2,3} && h^{1,4} \cr
&& h^{4,2} && h^{3,3} && h^{2,4} \cr
&&& h^{4,3} && h^{3,4} \cr
&&&& h^{4,4}
\end{array}
\quad = \quad
\begin{array}{ccccccccc}
&&&& 1 \cr
&&& 0 && 0 \cr
&& 0 && h^{1,1} && 0 \cr
& 0 && h^{2,1} && h^{2,1} && 0 \cr
1 && h^{3,1} && h^{2,2} && h^{3,1} && 1 \cr
& 0 && h^{2,1} && h^{2,1} && 0 \cr
&& 0 && h^{1,1} && 0 \cr
&&& 0 && 0 \cr
&&&& 1
\end{array} \,.
\ee
From the general formul\ae, such a Calabi--Yau fourfold has $h^{1,1}$ real K\"ahler parameters and $h^{3,1}$ complex structure moduli; when a $B$-field is included, its periods complexify the K\"ahler parameters.
But we now have two additional topological invariants.
The Hodge number $h^{2,1}$ measures the dimension of the space of harmonic $(2,1)$-forms.
Since
\be
H^{3}(X,\mathbb{C}) \cong H^{2,1}(X)\oplus H^{1,2}(X) \,,
\quad \text{and} \quad
h^{3,0}(X)=h^{0,3}(X)=0 \,,
\ee
one has $b_3(X)=2h^{2,1}$: $h^{2,1}$ counts the complex $(2,1)$ component, while there are $2h^{2,1}$ independent real three-form cohomology classes.
For example, expanding the M-theory three-form potential along these internal harmonic three-forms gives scalar zero modes in the lower dimensional theory, subject to the projections and constraints of the compactification.
They do not correspond to deformations of the complex structure or K\"ahler class.
Similarly, $h^{2,2}$ counts harmonic $(2,2)$-forms, \textit{i.e.}, a large part of the middle cohomology
\be
H^{4}(X,\mathbb{C}) \cong H^{4,0}\oplus H^{3,1}\oplus H^{2,2}\oplus H^{1,3}\oplus H^{0,4} \,,
\ee
with $h^{4,0}=h^{0,4}=1$ and $h^{1,3}=h^{3,1}$.
Geometrically, these classes enter intersection theory on $X$ and the description of four-form fluxes in string/F-theory compactifications; allowed fluxes are further constrained by quantization, Hodge type, primitivity, and other consistency conditions such as tadpole cancelation.
These cohomology classes, including the ``vertical'' subspace generated by wedge products of $(1,1)$-classes, are not additional independent metric moduli.
In fact, we have the following relation:
\be
h^{2,2} = 44 + 4h^{1,1} - 2h^{2,1} + 4h^{3,1} \,.
\ee
The Euler character is
\be
\chi = 48 + 6h^{1,1} - 6h^{2,1} + 6h^{3,1} \,.
\ee
Mirror symmetry interchanges $h^{1,1}$ and $h^{3,1}$ while leaving $h^{2,1}$ and $h^{2,2}$ unaffected.
Thus, under mirror symmetry, the Euler character is invariant.

\paragraph{Complete Intersection Calabi--Yau (CICY) manifolds:}
A \emph{complete intersection Calabi--Yau (CICY)} manifold $X$ of complex dimension $n$ is defined as the common locus of $K$ transverse homogeneous polynomials $\{p_1, \dots, p_K\}$ within a product of complex projective spaces:
\be
X \subset \mathbb{P}^{n_1} \times \mathbb{P}^{n_2} \times \dots \times \mathbb{P}^{n_m} \,,
\ee
where the total dimension of the ambient space is $\sum n_r = n + K$.
Together with transversality, this ensures that the common zero locus is a smooth $n$-fold of codimension $K$.
For $X$ to be a Calabi--Yau manifold, it must satisfy the condition $c_1(TX) = 0$.
In this setting, the condition is satisfied if the sum of the degrees of the defining polynomials in each projective factor equals the dimension of that factor plus one:
\be
\sum_{j=1}^{K} q^r_j = n_r + 1 \,, \quad \forall\ r \in \{1, \dots, m\} \,,
\ee
where $q^r_j$ is the degree of the $j$-th polynomial in the $r$-th projective space $\mathbb{P}^{n_r}$.
These data are conveniently expressed as a configuration matrix
\be
X = \left( \begin{array}{l|rcc}
\mathbb{P}^{n_1} \quad & q^1_1 & \ldots & q^1_K \cr
\vdots & \vdots & \ddots & \vdots \cr
\mathbb{P}^{n_m} & q^m_1 & \ldots & q^m_K 
\end{array} \right) \,.
\ee
For Calabi--Yau threefolds, the simplest such configuration matrices correspond to the zero locus of polynomials in a single projective space:
\be
(\mathbb{P}^4 | 5) \,,\quad (\mathbb{P}^5 | 3, 3) \,,\quad (\mathbb{P}^5 | 4, 2) \,,\quad (\mathbb{P}^6 | 3, 2, 2) \,,\quad (\mathbb{P}^7 | 2, 2, 2, 2) \,. \label{eq:cicy}
\ee
A slightly more complicated threefold is the Tian--Yau manifold
\be
X = \left( \begin{array}{l|ccc} \mathbb{P}^3 \  & 3 & 0 & 1 \cr \mathbb{P}^3 \  & 0 & 3 & 1 \end{array} \right) \qquad \Longrightarrow \qquad \left\{ \begin{array}{ccc} a^{\alpha\beta\gamma} w_\alpha w_\beta w_\gamma &=& 0 \,, \cr b^{\alpha \beta \gamma} z_\alpha z_\beta z_\gamma &=& 0 \,, \cr c^{\alpha\beta} w_\alpha z_\beta &=& 0 \,, \end{array} \right.
\ee
where we have written the homogeneous polynomials encoded by the configuration matrix explicitly in terms of coordinates $w$ and $z$ on each of the two $\mathbb{P}^3$s; these are indexed by $\alpha=1,\ldots,4$.
For further examples the reader can refer to the CICY surveys and corresponding catalogs discussed in Section~\ref{sec:cicy-catalogs}.

\paragraph{Toric Calabi--Yau manifolds:}\label{pg:toric}
Let $M \cong \mathbb{Z}^d$ be a lattice and set $M_{\mathbb{R}} := M\otimes_{\mathbb{Z}}\mathbb{R}\cong \mathbb{R}^d$.
A \emph{lattice polytope} is the convex hull $\Delta=\text{Conv}(v_1,\dots,v_k)\subset M_{\mathbb{R}}$ of finitely many points $v_i\in M$.
A lattice polytope $\Delta\subset M_{\mathbb{R}}$ is called \emph{reflexive} if
\begin{enumerate}
\item the origin lies in its strict interior, $0\in \text{int}(\Delta)$, and
\item its \emph{polar (dual) polytope}
\be
\Delta^\circ := \bigl\{\, y\in N_{\mathbb{R}} \;\big|\; \langle y,x\rangle \ge -1 \text{ for all } x\in \Delta \,\bigr\} \,,
\qquad N:=\text{Hom}(M,\mathbb{Z}) \,,
\ee
is also a lattice polytope (\textit{i.e.}, all vertices of $\Delta^\circ$ lie in the dual lattice $N$).
\end{enumerate}
The angle brackets denote the usual dot product on $\mathbb{R}^d$.
Equivalently, $\Delta$ is reflexive iff $0\in\text{int}(\Delta)$ and every facet of $\Delta$ is at unit lattice distance from the origin: for each facet $F$ there exists a primitive inward normal $u_F\in N$ such that
\be
\Delta = \{ x\in M_{\mathbb{R}} \mid \langle u_F,x\rangle \ge -1 \text{ for all facets } F \} \,,
\ee
and $F=\{x\in\Delta\mid \langle u_F,x\rangle=-1\}$.
In this case $(\Delta^\circ)^\circ=\Delta$, and $\Delta^\circ$ is reflexive as well.

One associates to $\Delta$ a projective Gorenstein toric Fano variety $X_\Delta$ by taking the fan $\Sigma_\Delta$ in $N_{\mathbb R}$ whose cones are the cones over faces of $\Delta^\circ$; its rays are generated by the vertices of $\Delta^\circ$.
Reflexivity is precisely the condition ensuring that the anticanonical linear system $|-K_{X_\Delta}|$ corresponds to $\Delta$ and hence contains Calabi--Yau hypersurfaces~\cite{Batyrev:1993oya}.

The Calabi--Yau hypersurface $Z$ is constructed as the zero locus of a generic section of the anticanonical bundle $-K_{X_\Delta}$ on the associated toric variety $X_\Delta$.
Explicitly, $Z$ is given by
\be
Z\,:\quad \sum_{m\in \Delta\cap M} a_m \prod_i x_i^{\langle m, v_i^\circ\rangle + 1}=0\,,
\ee
where the $v_i^\circ$ are the vertices of the dual (polar) polytope $\Delta^\circ$, the $x_i$ are homogeneous coordinates on $X_\Delta$, and the coefficients $a_m$ describe polynomial complex structure deformations of $Z$, with redundancies from coordinate transformations and overall scaling.
Nonpolynomial complex structure deformations can also occur.
Exchanging $\Delta$ and $\Delta^\circ$ yields an analogous construction of the mirror Calabi--Yau hypersurface $W$; in the threefold case $d=4$, $h^{1,1}$ and $h^{2,1}$ are interchanged:
\be
0 = \sum_{n\in \Delta^\circ\cap N} b_n \prod_i y_i^{\langle n, v_i\rangle + 1}\,,
\ee
where now the $v_i$ are the vertices of $\Delta$, the $y_i$ are homogeneous coordinates on the mirror toric variety $X_{\Delta^\circ}$, and the coefficients $b_n$ similarly describe polynomial complex structure deformations of $W$.
After refining either fan, the corresponding homogeneous coordinate product runs over all ray generators of the refined fan, including any additional boundary lattice points introduced by the subdivision.

To be slightly more precise, we choose a $\Delta$-regular (non-degenerate) Laurent polynomial
\be
f=\sum_{m\in \Delta\cap M} a_m \chi^m \,.
\ee
Its closure $Z_f\subset X_\Delta$ is an anticanonical hypersurface $Z_f\in |-K_{X_\Delta}|$ with (at worst) canonical singularities.
To desingularize the Calabi--Yau hypersurface, Batyrev~\cite{Batyrev:1993oya} constructs a maximal projective crepant partial desingularization (MPCP) using a fine, regular, star triangulation (FRST) $T$ of $\Delta^\circ$:
\begin{itemize}
\item A \emph{triangulation} of a $d$-dimensional polytope is a collection of $d$-simplices and their faces such that the intersection of any two simplices is a face of each and their union is the polytope.
\item A triangulation is \emph{fine} when it uses all points in the point configuration.
Using all boundary lattice points supplies the rays needed for an MPCP desingularization; fineness alone does not imply that the ambient toric variety is smooth.
\item A triangulation is \emph{regular} when it is induced by projecting the lower faces of the convex hull of points lifted by a height function to one higher dimension.
This ensures projectivity, and hence K\"ahlerity, of the resulting toric construction.
\item A triangulation is \emph{star} when the origin is a vertex of each full dimensional simplex.
Coning the boundary simplices from the origin then gives a fan refining the original face fan.
\end{itemize}
The resulting simplicial refinement $\Sigma_T$ of $\Sigma_\Delta$ defines a projective birational toric morphism
\be
\varphi_T:\ \widehat X_\Delta:=X_{\Sigma_T}\longrightarrow X_\Delta
\ee
that is crepant and yields an MPCP-desingularization of $X_\Delta$.
Define the proper transform (or total transform) of the hypersurface by $\widehat Z_f:=\varphi_T^{-1}(Z_f)$.
For $\Delta$-regular $Z_f$, the restriction $\varphi_T:\widehat Z_f\to Z_f$ is again an MPCP-desingularization (a simultaneous toric crepant partial desingularization of the hypersurface family).
In particular, in the cases relevant for Calabi--Yau threefolds (\textit{i.e.}, $d=4$ so $\dim \widehat Z_f=3$), Batyrev shows that a toroidal MPCP-desingularization is \emph{smooth}, and $\widehat Z_f$ is a smooth Calabi--Yau manifold.
In higher dimensions this construction may leave terminal singularities; for a Calabi--Yau fourfold these can be isolated points.
Different FRSTs specify different crepant toric phases, but some changes affect only the ambient variety and leave the Calabi--Yau geometry unchanged.

In dimensions two and three, reflexive polytopes yield elliptic curves and K3 surfaces, respectively.
The higher dimensional catalogs are discussed in Section~\ref{sec:toric-catalogs}, and searches for new five-dimensional polytopes as progenitors of Calabi--Yau fourfolds in Section~\ref{sec:newcy}.

\paragraph{Elliptically fibered Calabi--Yau manifolds:}
A Calabi--Yau manifold $X$ of complex dimension $n$ is \emph{genus-one fibered} if there exists a surjective holomorphic map $\pi:X\to B$ with connected fibers onto a base variety $B$ of dimension $n-1$ such that the generic fiber $F_b=\pi^{-1}(b)$ is a smooth curve of genus one.
We reserve the term \emph{elliptically fibered} for a genus-one fibration equipped with a global section $s:B\to X$, which supplies the origin on each smooth fiber and permits a Weierstrass model.
Likewise, $X$ is \emph{K$3$-fibered} if there is a surjective holomorphic map $\pi:X\to B$ with $\dim B=n-2$ whose generic fiber is a smooth K$3$ surface; in this case the fibration may have singular fibers over a discriminant locus in $B$, but on a dense open set the fibers are all deformation equivalent to K$3$.
In both cases ``fibered'' means that, at least birationally, $X$ is assembled from a family of lower dimensional Calabi--Yau manifolds (elliptic curves or K$3$ surfaces) varying over the base, providing extra structure that is central in duality constructions (\textit{e.g.}, F-theory for elliptic fibrations) and in organizing moduli and topological data.

Fibration structures also provide a useful statistical organization of known Calabi--Yau threefolds; we discuss the associated catalogs in Section~\ref{sec:derived-datasets}.

\subsection{Machine learning topological properties}\label{sec:mltop}
The $7890$ configuration matrices of CICY threefolds are severely constrained:
\be
K \le N_1 + N_a + 3 \,, \qquad N_1 \le 9 \,, \qquad N_a \le 6 \,,
\ee
where $N_1$ denotes the number of $\mathbb{P}^1$ factors and $N_a$ the number of remaining projective space factors in the ambient space.
The matrices range from $1\times1$ (the quintic) to $12\times15$ and have entries $q_r^j\in\{0,1,2,3,4,5\}$.
Excluding the $22$ direct product configurations leaves $7868$ configurations.\footnote{
The original database assigns the pair $(0,0)$ to these direct products as a placeholder; these entries do not give their actual Hodge numbers.
}
The remaining configurations realize $70$ distinct Euler characteristics $\chi\in[-200,0]$ and $265$ distinct Hodge pairs $(h^{1,1},h^{2,1})$, with
\be
1 \le h^{1,1} \le 19 \,, \qquad 15 \le h^{2,1} \le 101 \,. \label{eq:ch11}
\ee
This dataset is not closed under mirror symmetry: $\chi$ is non-positive, so members with negative Euler characteristic would require mirrors with positive Euler characteristic.

For CICY threefolds, the Euler characteristic can be computed explicitly from the configuration matrix using Chern classes and intersection theory.
Although the coefficients of $c_3(TX)$ are cubic in the multidegrees, the intersection numbers introduce additional dependence on the configuration data.
For example, for the manifolds in a single projective space that we have listed in~\eref{eq:cicy}, the Euler characters are:
\bea
(\mathbb{P}^4 | 5) : & \qquad \chi = & \frac13 \cdot 5 \cdot ((4+1)-5^3) = -200 \,, \nn \\
(\mathbb{P}^5 | 3, 3) : & \qquad \chi = & \frac13 \cdot 3^2 \cdot ((5+1) - 2\times 3^3 ) = -144 \,, \nn \\
(\mathbb{P}^5 | 4, 2) : & \qquad \chi = & \frac13 \cdot (4\times 2) \cdot ((5+1) - 4^3 - 2^3) = -176 \,, \\
(\mathbb{P}^6 | 3, 2, 2) : & \qquad \chi = & \frac13 \cdot (3\times 2^2) \cdot ((6+1) - 3^3 - 2\times 2^3 ) = -144 \,, \nn \\
(\mathbb{P}^7 | 2, 2, 2, 2) : & \qquad \chi = & \frac13 \cdot 2^4 \cdot ((7+1) - 4\times 2^3) = -128 \,. \nn
\eea
Entering the configuration matrix as a flattened vector into a fully connected feedforward neural network, $\chi$ can be machine learned with essentially $100\%$ accuracy.
Since the Euler character $\chi = 2 (h^{1,1}-h^{2,1})$, if we know one of the Hodge numbers, we know the other.
From~\eref{eq:ch11}, $h^{1,1}$ is a natural initial learning target because it takes values in a much smaller range than $h^{2,1}$.

The initial work in this direction~\cite{He:2017aed} used a fully connected multilayer perceptron to predict $h^{1,1}$ from padded CICY configuration matrices.
Subsequently,~\cite{Bull:2018uow} showed that neural networks and support vector machines can predict $h^{1,1}$ (and, by relation to $\chi$, indirectly constrain $h^{2,1}$) from raw configuration matrices with strong performance, motivating the view that nontrivial topological information is statistically encoded in the combinatorics of the defining data.
Subsequent work refines both architectures and data handling, emphasizing that apparent gains often come from choosing a representation that respects symmetries (\textit{e.g.}, row/column permutations) and from tuning baselines before deploying deeper models~\cite{Erbin:2020tks,Erbin:2023ncy}.
A consistent observation is that $h^{1,1}$ is significantly easier to learn than $h^{2,1}$ at fixed training budgets, and that the latter benefits disproportionately from careful preprocessing and validation~\cite{Erbin:2020tks}.
The $88\%$ accuracies for predicting $h^{1,1}$ on the complementary test set based on training with $70\%$ of the dataset reported in~\cite{Bull:2018uow} become $99\%$ accuracies using more sophisticated architectures such as the inception neural network~\cite{Erbin:2020srm}.

A key scientific question is whether models trained on ``easy'' parts of the dataset generalize to harder regions, \textit{e.g.}, training at small $h^{1,1}$ and validating at larger $h^{1,1}$.
The CICY threefolds provide a controlled laboratory for extrapolation: learned predictors can capture trends but may require a small amount of representative data from the target regime to recover strong accuracy~\cite{Bull:2019cij}.
Erbin and Finotello consolidate and extend these tests (including the reverse direction, high-to-low) and frame them as a diagnostic for when the machine learning model is genuinely learning structure rather than interpolating within a narrow domain~\cite{Erbin:2023ncy}.

Beyond recognizing Hodge numbers, the same input data support classification tasks relevant to string model building.
In particular, one can train binary classifiers to recognize whether a CICY is \emph{favorably embedded} (roughly, whether $H^{1,1}$ descends from the ambient product of projective spaces) and whether it admits a quotient by a freely acting discrete symmetry~\cite{Bull:2018uow}.
These queries illustrate a pragmatic use case: when an explicit computation is expensive or incomplete, machine learning can quickly shortlist candidates with desired properties for subsequent exact verification~\cite{Bull:2018uow}.
The symmetry problem is typically highly imbalanced (most CICYs do not admit a given symmetry), so the literature emphasizes metrics and training strategies appropriate to imbalance rather than raw accuracy~\cite{Bull:2018uow,Erbin:2023ncy}.

A recurring message is that ``feeding the brute matrix to a network'' is not a complete machine learning recipe.
Augmenting the data with permutations or providing additional features such as the cubes of entries in the configuration matrices improves performance~\cite{Bull:2019cij}.
Moreover, a robust workflow starts with exploratory data analysis, principled train/validation splits, and baseline models, and only then compares more expressive learners~\cite{Erbin:2020tks}.
In this setting, feature engineering (or architectures that implicitly learn invariances) can make linear or shallow models unexpectedly competitive in certain regimes, while deeper models can dominate when the representation is closer to the underlying geometric structure~\cite{Erbin:2020tks,Erbin:2023ncy}.
Overall, the machine learning outcome is not only predictions, but also a set of best practices for treating integer valued algebraic geometry data as a string dataset~\cite{Erbin:2023ncy}.

He and Lukas initiated systematic supervised learning on the dataset of $921,497$ CICY fourfold configuration matrices.
They found that $h^{1,1}$ is readily learnable from a padded or enhanced configuration matrix using fully connected classifiers/regressors, whereas $h^{3,1}$ is substantially more challenging and requires more structured architectures (\textit{e.g.}, two-branch networks together with feature enhancement), achieving high precision at least on controlled subsets of the data~\cite{He:2020lbz}.
Interpreted broadly, these results suggest that the success/failure modes of different networks can act as a probe of hidden structure in the Hodge number map, potentially hinting at simplified (and perhaps unknown) formul\ae\ in favorable regimes.
Multi-task architectures can be trained to predict them jointly, sharing an internal representation across outputs~\cite{Erbin:2021hmx}.
This improves sample efficiency and stabilizes learning compared with separate single-target models, especially for the harder invariants~\cite{Erbin:2021hmx,Erbin:2023ncy}.
Moreover, incorporating known exact constraints (\textit{e.g.}, linear relations among Hodge numbers and the Euler characteristic when available) can dramatically sharpen predictions: the network effectively learns a compatible subset of degrees of freedom while consistency fixes the rest~\cite{Erbin:2021hmx}.
Dao~\cite{Dao:2024zab} applies hybrid convolutional neural networks and recurrent neural networks (GRU/LSTM) and pure LSTM architectures to nearly one million CICY fourfold configuration matrices to predict the four Hodge numbers of each manifold.\footnote{
A convolutional neural network (CNN) first reads the CICY configuration matrix like a small ``image,'' learning local patterns (\textit{e.g.}, recurring blocks/row--column motifs) and compressing it into a feature vector.
A recurrent neural network (RNN) then processes that feature vector as a sequence to capture longer-range dependencies.
The RNN cell used is either an LSTM (Long Short-Term Memory), which has gating (input/forget/output) designed to keep important information over many steps and reduce vanishing gradients, or a GRU (Gated Recurrent Unit), which is a lighter gated variant (fewer gates/parameters than LSTM) that often trains faster while still handling long-range structure.
Pure LSTM means one skips the CNN front-end and feeds a sequence representation of the configuration matrix directly into an LSTM-only network, relying on the LSTM's internal memory/gates to learn both local and global structure without explicit convolutional feature extraction.
}
At an $80\%$ training ratio, the best individual recurrent model attains exact prediction accuracies of $99.85\%$, $98.66\%$, $96.26\%$, and $84.77\%$ for $(h^{1,1},h^{2,1},h^{3,1},h^{2,2})$, respectively.
The best ensemble improves these scores to $99.90\%$, $99.03\%$, $97.97\%$, and $87.34\%$.
These results demonstrate competitive performance for recurrent architectures while retaining a substantial gap between the easiest and hardest targets.

Calabi--Yau manifolds of higher dimension are studied using machine learning methods in~\cite{Alawadhi:2023gxa,Hirst:2023kdl}.
Likewise,~\cite{Aggarwal:2023swe} studies topological invariants of contact Calabi--Yau links carrying $G_2$ structures.
We discuss learning their differential geometry in Section~\ref{sec:g2-structure-metrics}.

Comparatively, the Kreuzer--Skarke database of reflexive polytopes is less well studied than the CICYs using machine learning.
The paper~\cite{Berglund:2021ztg} considers four-dimensional reflexive polytopes in the Kreuzer--Skarke list and asks how much of the topological data can be predicted from minimal combinatorial features of a polytope/dual pair $(\Delta,\Delta^\circ)$.
Using only the four integers $\big(l(\Delta),\,l(\Delta^\circ),\,l(2\Delta),\,l(2\Delta^\circ)\big)$ (numbers of lattice points in the polytope, its dual, and their $k=2$ dilations), the authors train a fully connected neural network and find that the Euler characteristic $\chi$ is predicted with high accuracy (\textit{e.g.}, $\sim 97\%$ accuracy with small mean absolute error on large test sets), while $h^{1,1}$ and $h^{2,1}$ are predicted less accurately but with typically small absolute error.
Motivated by the machine learning result, the authors perform linear regression on a random sample of $200,\!000$ polytopes and mirrors and identify approximate linear expressions for $h^{1,1}$ and $h^{2,1}$ in terms of these four lattice point counts, and obtain a new exact integer coefficient relation for the Euler character:
\be
\frac{1}{2}\chi = 9\big(l(\Delta^\circ)-l(\Delta)\bigr) + \bigl(l(2\Delta)-l(2\Delta^\circ)\big) \,, \label{eq:polytopechi}
\ee
which holds for every four-dimensional reflexive polytope.
This relation is proven analytically using Ehrhart theoretic topological input and emphasizes that the comparatively poorer learnability of $h^{1,1}$, $h^{2,1}$, and $h^{1,1}+h^{2,1}$ from the same minimal features suggests there is no similarly simple universal formula for those quantities at this level of coarse polytope data.

The paper~\cite{Ewert:2024ucr} studies supervised machine learning models that predict Hodge numbers of toric Calabi--Yau threefolds from the vertex matrix encoding of a four-dimensional reflexive polytope.
A key point is that these targets are invariant under natural group actions on the input, in particular, reordering vertices and swapping coordinate axes (row/column permutations).
Using a medium sized subset of the Kreuzer--Skarke data ($\sim 78{,}000$ polytopes with $26$ vertices), they find that enforcing symmetry awareness improves predictive performance, with their best group invariant approach achieving close to $97\%$ accuracy, supporting the use of symmetry respecting machine learning as a fast screening tool on large Calabi--Yau datasets.

Artificial intelligence methods are effective at detecting nontrivial correlations among features in large datasets.
This applies to topology.
In favorable situations, neural network based regression and symbolic post-processing of learned patterns can even guide the discovery of new analytic formul\ae.
This has occurred, for example, in studies of line bundle cohomology on surfaces and Calabi--Yau threefolds~\cite{Klaewer:2018sfl,Brodie:2019dfx,Brodie:2019pnz,Larfors:2020ugo}, as well as in the machine assisted analysis of knot invariants~\cite{Hughes:2020NeuralNetworkApproach,Jejjala:2019kio,Gukov:2020qaj,Craven:2020bdz,Davies:2021AdvancingMathematicsGuiding,Craven:2021ckk,Craven:2022cxe,Gukov:2024buj}.

The Hodge numbers are calculated analytically by sequence chasing~\cite{Hubsch:2025qhd}.
Typically, one uses short exact sequences of sheaves to produce long exact sequences in cohomology and then calculates the dimensions of the resulting kernels and cokernels.
That is to say, from
\be
0\to \mathcal{F}' \to \mathcal{F} \to \mathcal{F}'' \to 0 \,,
\ee
one gets
\be
\cdots \to H^i(\mathcal{F}') \to H^i(\mathcal{F}) \to H^i(\mathcal{F}'') \to H^{i+1}(\mathcal{F}')
\to \cdots \,,
\ee
so if two of the three graded cohomology groups are known (using ambient cohomology calculations and vanishing theorems), together with the relevant map ranks or vanishing results, one can deduce the third by exactness and dimension counts.
For a smooth hypersurface (or complete intersection) $X\subset A$, the standard inputs are the Koszul resolution (\textit{e.g.}, $0\to \mathcal{O}_A(-X)\to \mathcal{O}_A\to \mathcal{O}_X\to 0$ in the hypersurface case) and the conormal sequence
\be
0\to N^*_{X/A}\to \Omega_A^1|_X \to \Omega_X^1 \to 0 \,,
\ee
where $N^*_{X/A}$ is the conormal bundle, equal to $\mathcal{O}_X(-X)$ for a hypersurface.
The induced exterior power sequences relate $H^q(X,\Omega_X^p)$ to computable ambient groups.
Applying vanishing theorems and Serre duality can collapse many terms; the remaining calculations require the dimensions of the relevant kernels and cokernels to determine the Hodge numbers $h^{p,q}(X)=\dim H^q(X,\Omega_X^p)$.
This can be algorithmized and implemented in software such as \href{https://www.macaulay2.com/}{\texttt{Macaulay2}}~\cite{Grayson:ndMacaulay2}.
Exact cohomology calculations can be computationally demanding.
Once trained, a neural network provides rapid approximate predictions on the tested dataset~\cite{Bull:2019cij}, although this does not establish an asymptotic complexity bound or replace exact verification.
These experiments may motivate searches for unknown simple formul\ae\ for Hodge numbers, as in~\eref{eq:polytopechi} and the analytic results in line bundle cohomology.

\subsection{Machine learning enumerative invariants}
\label{sec:ml-enumerative-invariants}

Machine learning has already revealed structure in enumerative geometry: Coates, Kasprzyk, and Veneziale~\cite{Coates:2023wlw} inferred the dimension of Fano varieties from their regularized quantum periods, whose coefficients encode genus zero Gromov--Witten invariants, and used the learned patterns to establish asymptotic results.
We can envision extending this approach to predicting curve counting invariants of Calabi--Yau threefolds.
An initial target could be the integer Gopakumar--Vafa invariant $n^0_\beta$ associated with a nonzero effective curve class $\beta$.
Training data are available from the complete intersection database of Carta \textit{et al.}~\cite{Carta:2021sms} and the computational mirror symmetry algorithms of Demirtas \textit{et al.}~\cite{Demirtas:2023als} for compact toric hypersurfaces.
A model could take the defining geometric data, the chosen K\"ahler chamber, and the curve class as inputs, with equivalent geometric presentations identified consistently.
The corresponding rational Gromov--Witten invariants would then follow from the multicover relation~\cite{Gopakumar:1998jq}
\be
 N^{\mathrm{GW}}_{0,\beta}
 =\sum_{k\mid\beta}\frac{n^0_{\beta/k}}{k^3}
 \,.
 \label{eq:ml-gw-gv}
\ee
Here, $k\mid\beta$ means that $\beta=k\beta'$ for an integral effective curve class $\beta'$.
This relation should be imposed exactly, rather than learned independently.

The principal challenge is to recover exact integers, including their signs and vanishing, across the large dynamic range of these invariants.
Tests should withhold entire geometries and degrees beyond the training range, and measure exact agreement with independently computed invariants.
Uncertainty estimates could guide additional mirror symmetry calculations, while symbolic regression could suggest recurrences or generating functions for subsequent proof.
Integer-valued output alone would not establish correctness.
For noncompact toric Calabi--Yau threefolds, the topological vertex supplies a complementary source of exact data at higher genus~\cite{Aganagic:2003db}.
Learning the higher genus Gopakumar--Vafa invariants would also provide a route to Donaldson--Thomas theory through the Gromov--Witten/Donaldson--Thomas correspondence, in settings where it is established~\cite{Maulik:2003rzb}.
This comparison requires the appropriate change of variables and removal of the degree zero contribution from the partition functions; genus zero data alone do not determine the full Donaldson--Thomas series.
The objective is therefore to accelerate verified computations and discover exact structure in curve counting, with learned predictions serving as candidates for mathematical validation.

\subsection{Uniqueness of Calabi--Yau geometries}\label{sec:wall}

In low complex dimensions, the familiar examples are essentially unique at the level of oriented smooth manifolds: every elliptic curve has the underlying smooth torus $T^2$, and all complex K3 surfaces are orientation preservingly diffeomorphic~\cite{Kodaira:1964Structure,Kodaira:1963CompactAnalyticSurfaces,Joyce:2001xt}.
K3 is the simply connected case among Calabi--Yau surfaces; complex two-tori also occur under our broader trivial canonical bundle convention.
This uniqueness does not extend to complex structures or Ricci-flat metrics, which have continuous moduli, nor does the K3 statement identify every exotic smooth structure on the underlying topological four-manifold with a complex K3 surface.

What does it mean for two Calabi--Yau manifolds to be the same?
Different equations, configuration matrices, polytopes, or triangulations can describe the same geometry, but the relevant equivalence relation must first be specified.
For counting smooth compactification spaces, we shall call two Calabi--Yau manifolds the same \emph{as oriented smooth manifolds} when there is an orientation preserving diffeomorphism between them.
This is distinct from being the same \emph{complex manifold}, which requires a biholomorphism, or the same \emph{polarized variety}, which additionally requires preservation of the polarization.
Fibers in a connected smooth proper deformation family are diffeomorphic, but a diffeomorphism alone does not identify their complex structures.
Birational equivalence is another relation: a flop connects birational Calabi--Yau models, but does not by itself certify that they are diffeomorphic.
A chosen fibration is additional structure, so equivalence of fibered spaces must also specify how their projections and bases are identified.
Finally, equivalence of Ricci-flat metrics means isometry of the corresponding Riemannian manifolds.
Yau's uniqueness theorem~\cite{Yau:1978cfy} concerns a fixed complex manifold and a fixed K\"ahler class; it does not say that the topology determines a unique Calabi--Yau metric.
Thus, counting manifolds, complex structures, fibrations, and metrics are different problems, even before bundle or flux data are supplied for a compactification.

Wall's classification applies to closed, oriented, simply connected, smooth, spin, six real dimensional manifolds with torsion-free integral cohomology~\cite{Wall:1966rcd}.
Their oriented diffeomorphism classes are determined by $H^2(X,\mathbb Z)$, the rank $b_3$, the integral cubic cup product form, and the first Pontryagin class functional, up to a common integral lattice isomorphism.
For a Calabi--Yau threefold, $c_1=0$ implies that the manifold is spin and
\be
 p_1(TX_{\mathbb R})=c_1(TX)^2-2c_2(TX)=-2c_2(TX) \,.
\ee
For the full $SU(3)$ holonomy threefolds considered here, $b_2=h^{1,1}$ and $b_3=2h^{2,1}+2$.
Consequently, under the stated simple connectedness and torsion assumptions, the familiar Wall data are
\be\label{eq:wall-data}
 \left(h^{1,1},h^{2,1},\kappa_{ijk},c_{2,i}\right)\,,
 \qquad
 \kappa_{ijk}=\int_X D_i\wedge D_j\wedge D_k\,,
 \qquad
 c_{2,i}=\int_X c_2(X)\wedge D_i \,,
\ee
where the $D_i$ form an integral basis of $H^2(X,\mathbb Z)$.
The pairing on $H^3$ is unimodular and alternating, so its rank supplies the remaining information in this torsion-free setting.

In practice, two such threefolds have the same oriented smooth type if and only if their Hodge numbers agree and there exists a single matrix $A\in GL(h^{1,1},\mathbb Z)$ such that
\be\label{eq:wall-matching}
 \kappa'_{ijk}=A_i{}^a A_j{}^b A_k{}^c\kappa_{abc}\,,
 \qquad
 c'_{2,i}=A_i{}^a c_{2,a} \,.
\ee
Here, $\det A=\pm1$: orientation of the six-manifold does not select an orientation of its $H^2$ lattice.
The same integral change of basis must match both tensors.
Agreement of Hodge numbers alone or a rational change of basis is not a complete equivalence test.
Agreement of the readily computed basis independent invariants used in these searches is also insufficient in general~\cite{Hubsch:2025qhd}.
Conversely, an explicit matrix satisfying~\eqref{eq:wall-matching} is a guarantee of diffeomorphism equivalence under Wall's hypotheses.
For non-simply connected examples or cohomology with torsion, this short list is not a complete classification criterion.

The difficulty of finding integral transformations at large $h^{1,1}$ motivates both invariant based methods and machine learning~\cite{Jejjala:2022lxh}.
He, Yao, and Yau use an Inception convolutional neural network to predict divisibility invariants constructed from CICY triple intersection numbers, reporting prediction accuracies of approximately $90\%$ under fivefold cross-validation~\cite{He:2024coy}.
This extends learning beyond Hodge numbers to finer topological data, while the predicted invariants remain incomplete and do not replace the exact matching test in~\eqref{eq:wall-matching}.

A mismatch of genuine diffeomorphism invariants proves inequivalence, whereas matching these invariants only identifies candidates for the exact test.
Likewise, failure of a bounded search for $A$ does not prove that no such matrix exists, and a learned prediction is not itself a certificate.
Gendler \textit{et al.}\ obtain respectively $4$, $27$, $183$, $1\,184$, and $8\,036$ diffeomorphism classes of simply connected toric hypersurface threefolds in the Kreuzer--Skarke construction at $h^{1,1}=1,\ldots,5$, including favorable and non-favorable cases~\cite{Gendler:2023ujl}.
Chandra \textit{et al.}\ develop complementary lower bounds from invariants and upper bounds from demonstrated equivalences, extending their analysis through $h^{1,1}=6$~\cite{Chandra:2023afu}.
These results concern overlapping ensembles and should not be added together.
Gopakumar--Vafa invariants can help investigate phases and find candidate identifications, but they contain symplectic or deformation information; differences in such invariants alone are not a general obstruction to a smooth diffeomorphism.

The distinction between construction data and smooth manifolds is particularly important for global estimates.
For threefold hypersurfaces obtained from fine, regular, star triangulations of four-dimensional reflexive polytopes, the bound of Demirtas \textit{et al.}\ has recently been strengthened by MacFadden, Orevkov, and Stepniczka~\cite{Demirtas:2020dbm,MacFadden:2026xvp}:
\be\label{eq:ks-topology-bounds}
 N_{\text{CY}}^{\text{KS}}<1.65\times10^{428}
 \quad\longrightarrow\quad
 N_{\text{CY}}^{\text{KS}}<1.211\times10^{296}
 \,.
\ee
These are upper bounds for this construction, not for all Calabi--Yau threefolds.
The lower bound of $10^{276}$ counts equivalence classes of triangulations defined by their restrictions to two-faces; it is not a lower bound on distinct diffeomorphism classes~\cite{MacFadden:2026xvp}.
The earlier extrapolation of order $10^{400}$ manifolds in~\cite{Chandra:2023afu} was heuristic and is excluded for this hypersurface ensemble by the newer rigorous upper bound.
Similarly, the fibration catalogs discussed in Section~\ref{sec:derived-datasets} count additional structures on the manifolds: a single manifold may admit several fibrations, and quotienting by polytope automorphisms need not identify every geometrically equivalent presentation~\cite{Abbasi:2025lvn}.

There is no comparably practical criterion for Calabi--Yau fourfold diffeomorphism expressed solely in the divisor intersection and Chern data used here.
A first guess extends the intersection and characteristic class data.
See, for example,~\cite{Anderson:2018pui}.
Assume that a Calabi--Yau fourfold $Y$ has full $SU(4)$ holonomy, is simply connected, and has torsion-free integral cohomology.
In addition to its Betti numbers, natural quantities to compute are the full oriented integral cohomology ring and characteristic classes.
By Poincar\'e duality, the ring can be encoded by the free groups $H^2$, $H^3$, and $H^4$, their duality pairings, and the maps
\begin{align}\label{eq:fourfold-ring}
 Q(a,b)&=\int_Y a\wedge b \,, & a,b \in H^4(Y,\mathbb Z)\,,\nonumber\\
 m(x,y)&=x\smile y \,,  & m:\operatorname{Sym}^2 H^2(Y,\mathbb Z)\longrightarrow H^4(Y,\mathbb Z) \,,\\
 T(x,u,v)&=\int_Y x\wedge u\wedge v \,, & x\in H^2(Y,\mathbb Z) \,,\quad u,v\in H^3(Y,\mathbb Z) \,.\nonumber
\end{align}
Here, $Q$ is a unimodular symmetric lattice form and $T$ is alternating in the two odd-degree arguments.
Compatible integral isomorphisms must preserve all these maps.
Choose an integral divisor basis $D_i$, $i=1,\ldots,r$, where $r=b_2(Y)=h^{1,1}(Y)$.
The quadruple intersection data are the components of a symmetric integral rank-four tensor,
\be\label{eq:fourfold-quartic}
 d_{ijkl}=\int_Y D_i\smile D_j\smile D_k\smile D_l\,,
 \qquad
 \mathcal I_Y(t)=\int_Y(t^iD_i)^4=d_{ijkl}t^it^jt^kt^l \,.
\ee
Repeated indices are summed over the full range, so the quartic polynomial includes the usual multinomial multiplicities.
Geometrically, the tensor gives the oriented intersection number of four divisor classes; it is not a single number attached to $Y$.
For a K\"ahler class $J=t^iD_i$, the volume is $\mathcal I_Y(t)/4!$.
In terms of~\eqref{eq:fourfold-ring},
\be
 d(x_1,x_2,x_3,x_4)
 =Q\bigl(m(x_1,x_2),m(x_3,x_4)\bigr) \,.
\ee
These intersections probe only the sublattice generated by products of degree-two classes, and cannot be assumed to determine the full middle lattice, its embedding data, or the mixed product $T$.
Likewise, the numbers $c_{2,ij}:=\int_Y c_2\smile D_i\smile D_j$ need not determine $c_2$ as a class in the whole of $H^4(Y,\mathbb Z)$.
The direct analog of the threefold change of basis is a single $A\in GL(r,\mathbb Z)$ satisfying
\begin{align}\label{eq:fourfold-divisor-matching}
 d'_{ijkl}&=A_i{}^aA_j{}^bA_k{}^cA_l{}^d\,d_{abcd}\,,\nonumber\\*
 c'_{2,ij}&=A_i{}^aA_j{}^b\,c_{2,ab} \,,
\end{align}
where $A$ identifies $D'_i$ with $A_i{}^aD_a$, so $\mathcal I_{Y'}(t)=\mathcal I_Y(A^Tt)$.
This ``$GL$ rotation'' is an integral basis change with determinant $\pm1$; it need not be orthogonal.
Matching the quartic and $c_2$ pairings with different matrices does not suffice, and a solution of~\eqref{eq:fourfold-divisor-matching} only certifies agreement of the divisor data.

For the underlying real tangent bundle of a Calabi--Yau fourfold,
\be\label{eq:fourfold-pontryagin}
 p_1=-2c_2\,,\qquad p_2=c_2^2+2c_4\,,\qquad c_4=e(TY_{\mathbb R}) \,.
\ee
Thus, once the full oriented ring and $c_2$ are specified, $p_2$ supplies no further independent top degree number: $\int_Y c_4=\chi(Y)$ is already determined by the Betti numbers.
One should also distinguish real and complex characteristic data.
An arbitrary oriented diffeomorphism need not preserve the integral $c_3$ of a chosen complex tangent bundle: conjugating a complex structure in complex dimension four preserves the real orientation but reverses the sign of $c_3$.
Requiring preservation of all complex Chern classes would therefore address additional structure, rather than just the underlying oriented smooth manifold.
Necessary consistency checks on the proposed data include
\be\label{eq:fourfold-consistency}
 Q(a,a)\equiv Q(a,c_2)\pmod 2 \,,
 \qquad
 \int_Y(3c_2^2-c_4)=1440 \,.
\ee
The first relation holds for every integral degree-four class $a$.
The second assumes full $SU(4)$ holonomy, for which $\chi(\mathcal O_Y)=2$.
These follow from the Wu formula and Hirzebruch--Riemann--Roch, respectively; they are constraints, not sufficient conditions for diffeomorphism.
The same index identities give useful numerical checks~\cite{Klemm:1996ts}:
\be\label{eq:fourfold-index-checks}
 \int_Yc_2^2=480+\frac{\chi(Y)}{3}\,,
 \qquad \operatorname{sign}(Q)=32+\frac{\chi(Y)}{3} \,.
\ee
For every integral divisor class $D$, the line bundle index further requires
\be\label{eq:fourfold-line-index}
 \chi(\mathcal O_Y(D))=2+\frac{\int_Y(D^4+c_2D^2)}{24}\in\mathbb Z \,.
\ee
These formulas retain the assumption $\chi(\mathcal O_Y)=2$; they must not be imposed unchanged on products with reduced holonomy.

To go beyond the divisor test, choose integral bases $u_s$ of $H^3(Y,\mathbb Z)$ and $E_\alpha$ of $H^4(Y,\mathbb Z)$, and write
\be
 m(D_i,D_j)=m_{ij}{}^\alpha E_\alpha \,,
 \quad Q_{\alpha\beta}=\int_YE_\alpha E_\beta \,,
 \quad T_{ist}=\int_YD_i u_su_t\,,
 \quad c_2=v^\alpha E_\alpha \,.
\ee
A prospective oriented ring isomorphism $F:H^*(Y',\mathbb Z)\to H^*(Y,\mathbb Z)$ must identify the degree-two, degree-four, and degree-three bases through one compatible triple
\be
 F(D'_i)=A_i{}^aD_a \,,
 \quad F(E'_\alpha)=B_\alpha{}^\beta E_\beta \,,
 \quad F(u'_s)=C_s{}^pu_p \,,
\ee
where $A\in GL(b_2,\mathbb Z)$, $B\in GL(b_4,\mathbb Z)$, and $C\in GL(b_3,\mathbb Z)$.
Preserving the products, the orientation, and $c_2$ gives the explicit matching equations
\begin{align}\label{eq:fourfold-full-matching}
 Q'_{\alpha\beta}&=B_\alpha{}^\gamma B_\beta{}^\delta Q_{\gamma\delta}\,,\nonumber\\*
 m'_{ij}{}^\alpha B_\alpha{}^\gamma&=A_i{}^a A_j{}^b m_{ab}{}^\gamma\,,\nonumber\\*
 T'_{ist}&=A_i{}^a C_s{}^p C_t{}^q T_{apq}\,,\nonumber\\*
 v^\beta&=v'^\alpha B_\alpha{}^\beta \,.
\end{align}
The duality pairings determine the maps in degrees five and six.
Thus, this is an explicit test for the oriented integral cohomology ring together with $c_2$; the quartic transformation follows from its first two equations.
If $h^{2,1}=0$, the $H^3$ group and its matching equation are absent.
For a refinement preserving $c_3$, its divisor covector $c_{3,i}=\int_Yc_3D_i$ must additionally obey $c'_{3,i}=A_i{}^a c_{3,a}$.
This is an extra necessary condition for preserving that characteristic class, not a classification of almost complex or complex structures.

There is a useful distinction between the abstract lattice and the way these classes sit inside it.
For a fourfold with full $SU(4)$ holonomy, $Q$ is indefinite precisely when $h^{1,1}+2h^{3,1}>1$.
In this case, its rank, signature, and parity determine its abstract integral isometry class~\cite{Milnor:1973SymmetricBilinearForms}.
Even then, one must match the product map $m$, the embedding of its image, the class $c_2$, and, when present, $T$.
The preceding equations do not by themselves establish a fourfold analog of Wall's diffeomorphism criterion.

There is now a precise eight-dimensional framework beyond the expectation expressed in~\cite{Anderson:2018pui}.
Building on modified surgery~\cite{Kreck:1999Surgery}, Nagy's theorem applies to simply connected closed oriented eight-manifolds with normal $3$-type $B$ satisfying that $H_4(B,\mathbb Z)$ is free~\cite{Nagy:2026ExtendedSurgery}.
The normal $3$-type records low degree homotopy together with stable normal bundle information.
Within this class, oriented diffeomorphism is characterized by a common normal $3$-type and the existence of bordant normal $3$-smoothings $f:Y\to B$ with isomorphic extended intersection data
\be\label{eq:fourfold-surgery}
 \left(H_4(Y,\mathbb Z),\lambda_Y,
 f_*:H_4(Y,\mathbb Z)\longrightarrow H_4(B,\mathbb Z)\right)\,,
\ee
where $\lambda_Y$ is the middle-dimensional intersection pairing.
The freeness assumption concerns $H_4(B)$, not merely $H_4(Y)$, and the normal bordism condition is essential.

An effective Calabi--Yau counterpart of~\eqref{eq:wall-matching} would therefore require more than guessing a longer list of Chern numbers.
\emph{A plausible formulation to investigate} is that, within a specified class of simply connected, torsion-free Calabi--Yau fourfolds, the full integral ring~\eqref{eq:fourfold-ring}, the class $c_2$, and finitely many computable discrete refinements determine the normal $3$-type, normal bordism class, and extended intersection data required by surgery theory.
Whether the Calabi--Yau conditions force those refinements, so that the ring and $c_2$ alone suffice in any broad class, remains to be established.
This is the speculative step; the conditional surgery theorem is established.
Even a successful smooth classification would still leave the separate questions of complex structures, K\"ahler chambers, and Ricci-flat metrics on each manifold.

\section{Databases and computational tools for Calabi--Yau geometry}
\label{sec:cy-databases}

The databases underlying the constructions discussed above constitute systematic searches over large classes of Calabi--Yau geometries possible.
Their entries count presentations, phases, or additional structures according to the conventions of each catalog; identifying the underlying manifolds requires the equivalence tests discussed in Section~\ref{sec:wall}.
We first assess the principal catalogs and geometric software, then explain how generative methods extend their reach and how derived datasets organize additional structures relevant to physics.

\subsection{Complete intersection catalogs}\label{sec:cicy-catalogs}

The complete intersection Calabi--Yau threefold catalog was the first large database in string theory and the first used for machine learning; it remains an especially useful starting point.
The list was first generated by Candelas, Dale, L\"utken, and Schimmrigk~\cite{Candelas:1987kf}, with Hodge numbers subsequently computed by Green, H\"ubsch, and L\"utken~\cite{Green:1987cr}.
The \href{https://www-thphys.physics.ox.ac.uk/projects/CalabiYau/cicylist/index.html}{Oxford archive} distributes the resulting $7890$ configuration matrices describing complete intersections in products of ordinary projective spaces.
The archive supplies Hodge numbers and, in its \href{https://www.wolfram.com/mathematica/}{Mathematica} distribution, the freely acting symmetry data of Braun~\cite{Braun:2010vc}.
Configurations provide concrete input for intersection theory, bundle constructions, and sheaf cohomology calculations.
The catalog's completeness is restricted to this construction, and the list retains equivalent presentations.
It is therefore a tractable laboratory for heterotic model building and computational benchmarking, rather than a representative sample of all Calabi--Yau threefolds.

Subsequent work has substantially increased the usefulness of the CICY collection.
Anderson, Gao, Gray, and Lee supplied favorable presentations for most entries, together with intersection and fibration data~\cite{Anderson:2017aux}.
Here, ``favorable'' means that the divisor classes needed on the Calabi--Yau manifold descend from the ambient space; the stronger property that the ambient construction also determines the full K\"ahler cone must be checked separately.
Catalogs of free quotients provide examples with nontrivial fundamental group relevant to Wilson line breaking of heterotic gauge groups, but symmetry searches depend on the chosen presentation.
Failure to find an action induced by ambient automorphisms does not establish the absence of a free action on the manifold~\cite{Braun:2010vc,Gray:2021kax}.
Generalized CICY constructions extend the configuration matrix approach further, allowing negative multidegrees when the corresponding sections exist on an intermediate subvariety, although the resulting collections do not have the same classification status as the ordinary CICY list~\cite{Anderson:2015iia}.

The analogous \href{https://www-thphys.physics.ox.ac.uk/projects/CalabiYau/Cicy4folds/index.html}{CICY fourfold database}, constructed by Gray, Haupt, and Lukas, contains $921,497$ configurations~\cite{Gray:2013mja}.
This total includes $15,813$ product configurations, leaving $905,684$ indecomposable entries.
The associated studies provide Hodge data, Chern classes, intersection information for divisors inherited from the ambient space, and extensive fibration data.
In particular, they record more than fifty million obvious genus-one fibration presentations, with such structures found on 99.95\% of the configurations~\cite{Gray:2014fla}.
This makes the collection valuable for investigating the prevalence of fibrations and identifying candidate F-theory geometries.
However, the fibration need not possess a section, and the recorded fibrations are not all inequivalent. Their existence alone does not provide resolutions of singular fibers in the various codimensions (data needed to determine the gauge and matter content) nor the flux information required to specify a complete F-theory compactification.

\subsection{Toric and weighted hypersurfaces}\label{sec:toric-catalogs}

For toric hypersurface constructions, the central resource is the \href{https://hep.itp.tuwien.ac.at/~kreuzer/CY/}{Kreuzer--Skarke archive}.
There are $16$ reflexive polytopes in two dimensions and $4319$ in three dimensions.
In four dimensions, the archive gives the complete classification, up to lattice equivalence, of $473,800,776$ reflexive polytopes~\cite{Kreuzer:2000xy}, whose hypersurfaces produce the Hodge pairs shown in Figure~\ref{fig:hodge-shield}.
This is an exhaustive classification of the relevant polytopes, not a classification of all Calabi--Yau threefolds or a fully resolved catalog of all geometries arising from those polytopes.
Its breadth and combinatorially accessible invariants are major strengths, but obtaining detailed geometric information requires further choices and computations.

Resolving the hypersurfaces requires the FRST choices described on page~\pageref{pg:toric}.
Even for a fixed polytope, the number of FRSTs can be enormous.
The upper bound across the Kreuzer--Skarke list is approximately $10^{928}$~\cite{Demirtas:2020dbm}.
The current upper bound on topologically inequivalent hypersurfaces is given in~\eqref{eq:ks-topology-bounds}.
Generating and sampling triangulations is consequently a central computational problem, and an exhaustive catalog across the list is impractical with present methods.

The established \href{https://hep.itp.tuwien.ac.at/~kreuzer/CY/CYpalp.html}{\texttt{PALP}} software, developed by Kreuzer and Skarke, supplies lattice polytope calculations, normal forms, Hodge data, and related toric constructions~\cite{Kreuzer:2002uu}.
Earlier PALP-based triangulation workflows were limited to dual polytopes with at most three points interior to their facets, excluding even some examples with $h^{1,1}=5$~\cite{Altman:2014bfa}.
The \href{https://www.rossealtman.com/toriccy}{ToricCY database}, originating in the work of Altman, Gray, He, Jejjala, and Nelson, extended systematic triangulation through $h^{1,1}=6$, with three cases at $h^{1,1}=6$ unresolved in the original dataset and partial results at $h^{1,1}=7$~\cite{Altman:2014bfa}.
It supplies divisor bases, intersection tensors, Chern classes, Stanley--Reisner ideals, and Mori and K\"ahler cone data, while grouping appropriate phases.
These quantities directly support effective field theory calculations.
For example, for $J=t^aD_a$, the triple intersections determine the classical volume:
\be\label{eq:database-volume}
 \mathcal V=\frac{1}{6}\kappa_{abc}t^at^bt^c\,,
 \qquad
 \kappa_{abc}=\int_X D_a\wedge D_b\wedge D_c\,.
\ee
A motivation for this work was to identify geometries supporting the large volume scenario~\cite{Balasubramanian:2005zx} by cataloging threefolds with parametrically large cycles~\cite{Gray:2012jy,Altman:2017vzk}.
The current interface also exposes orientifold and ``Swiss cheese'' volume structure data.
Its strength is the depth of the stored information; coverage and completeness must be assessed for the particular computation.

The complementary role of \href{https://cy.tools/}{\texttt{CYTools}} is to compute geometric data on demand.
Developed by Demirtas, Rios-Tascon, and McAllister, it provides a \href{https://www.python.org/}{Python} framework for manipulating polytopes, constructing triangulations, and calculating Calabi--Yau invariants~\cite{Demirtas:2022hqf}.
Its algorithms construct representative triangulations even for the largest polytopes in the Kreuzer--Skarke collection, substantially extending the range accessible to systematic computation.
Related sampling and counting methods underlie the bounds quoted above~\cite{Demirtas:2020dbm}; neither these bounds nor the ability to handle individual large examples constitutes an exhaustive triangulation catalog.
The documentation describes cone constructions inferred from toric geometry, volume and K\"ahler moduli space metric calculations, and interfaces for Gromov--Witten and Gopakumar--Vafa invariants~\cite{CYTools:2026Documentation}.
A K\"ahler cone inferred from the ambient construction should not automatically be identified with the full cone of the hypersurface, especially when nontoric information is missing.
Likewise, a metric on K\"ahler moduli space is different from a pointwise Ricci-flat metric on the compactification manifold.
These distinctions determine which physical questions the software can answer directly and which require additional calculations.

Other large catalogs deserve inclusion alongside CICYs and toric hypersurfaces.
The classical lists of weighted projective hypersurfaces contain $7555$ transverse weight systems for Calabi--Yau threefolds, classified by Klemm and Schimmrigk and by Kreuzer and Skarke, and $1,100,055$ for fourfolds, classified by Lynker, Schimmrigk, and Wi\ss kirchen~\cite{Klemm:1992bx,Kreuzer:1992np,Lynker:1998pb}.
Their compact encoding makes them useful for studying Hodge number distributions, Landau--Ginzburg descriptions, and machine learning methods~\cite{Hirst:2023kdl}.
The Kreuzer--Skarke site distributes the weight data, while the \href{https://www.grdb.co.uk/}{Graded Ring Database} supplies searchable algebraic information, including singularity data.
Brown and Kasprzyk give a mathematical classification of the fourfold hypersurfaces and their quotient singularities~\cite{Brown:2015orbifold}.
Quasi-smoothness allows for quotient singularities inherited from the ambient weighted projective space, so these entries require a distinction between an orbifold presentation and a smooth resolved manifold, particularly in complex dimension four.

A much larger fourfold resource is the Sch\"oller--Skarke classification of $322,383,760,930$ \href{https://huggingface.co/datasets/calabi-yau-data/ws-5d}{weight systems} with six components and the relevant interior point property~\cite{Scholler:2018apc}.
Of these, $185,269,499,015$ directly determine reflexive polytopes, producing $532,600,483$ distinct Hodge number sets.
This classifies weight systems used in the construction, not all five-dimensional reflexive polytopes; further combinatorial steps would be required for the latter.
Extrapolating trends, Sch\"oller and Skarke estimate that there are $1.15\times10^{18}$ five-dimensional reflexive polytopes, so the available catalogs remain manifestly incomplete~\cite{Scholler:2018apc}.
Moreover, reflexivity in five dimensions does not guarantee that the associated Calabi--Yau fourfold admits a smooth crepant resolution.
These data consequently provide a large source of fourfold constructions whose smoothness and physical suitability require further investigation.

\subsection{Machine learning new Calabi--Yau manifolds}\label{sec:newcy}

Generative methods address the difficulty of exhaustive enumeration by searching for polytopes or triangulations with desired properties.
The \href{https://github.com/elliheyes/Polytope-Generation}{\texttt{Polytope-Generation}} repository, accompanying work by Berglund, He, Heyes, Hirst, Jejjala, and Lukas, implements a genetic algorithm (GA) for finding reflexive lattice polytopes~\cite{Berglund:2023ztk}.
It provides C and Python implementations and generated five-dimensional data.
The vertices are encoded as binary strings with a fixed maximum number of bits, constraining the polytope to a hypercube around the origin; an upper bound on the number of vertices is also set \textit{a priori}.
The resulting environment contains
\be
\#\ \text{states} = 2^{n_\text{max}} \,, \qquad n_\text{max} = \text{dimension} \times \text{maximum vertices} \times \text{maximum bits} \,,
\ee
possible binary encodings, although distinct encodings can represent the same polytope.
A fitness function measures how close the convex hull of a candidate collection of vertices is to being reflexive:
\be
f(\Delta) = w_1 (\text{IP}(\Delta) - 1) - \frac{w_2}{k} \sum_{i=1}^k |d_i(\Delta)-1| \,.
\ee
Here, $\text{IP}(\Delta)$ equals $1$ if the origin is the unique interior lattice point and $0$ otherwise, while $d_i(\Delta)$ is the lattice distance of the $i$th of the $k$ facets from the origin.
For positive $w_1$ and $w_2$, the fitness vanishes precisely when the origin is the unique interior lattice point and all facets have unit lattice distance from it; otherwise it is negative.
The binary encoding of a candidate's vertices is its genotype.
Ranking a population by fitness and applying selection, crossover, and mutation provides an evolutionary search for reflexive polytopes.
Additional fitness terms can target geometric properties, including Hodge numbers and divisibility conditions on the Euler number $\chi$ motivated by compactification.

As a proof of principle, the genetic algorithm recovers complete classifications in low dimensions: all $16$ two-dimensional reflexive polytopes are found essentially immediately, and all three-dimensional examples are recovered after sufficiently many evolutions, with approximately $10^4$ evolutions reported in~\cite{Berglund:2023ztk}.
These searches visit fractions of approximately $10^{-6}$ of $10^{11}$ encoded states in two dimensions and $10^{-40}$ of $10^{51}$ states in three dimensions.
In four dimensions, the study restricts to subsets with small numbers of lattice points or vertices and recovers the corresponding known cases after removing redundancies using normal forms.
Targeted searches in five dimensions recover the known examples within the small lattice point domains studied and discover lattice inequivalent polytopes absent from the comparison datasets, including examples with previously unrecorded Hodge data.
One such polytope is
\be \Delta =
\left(\begin{array}{cccccc}
0 & 4 & -1 & 2 & -1 & -2 \cr
3 & -1 & -4 & 9 & 0 & -3 \cr
2 & 2 & -2 & 2 & -2 & 2 \cr
1 & -3 & 0 & 3 & 0 & -1 \cr
3 & -5 & 1 & -1 & 1 & -1
\end{array}\right) \,.
\ee
Its associated Calabi--Yau fourfold construction has Hodge data $h^{1,1}=1$, $h^{1,2}=0$, $h^{1,3}=111$, $h^{2,2}=492$, and Euler number $\chi=720$.
The authors conjecture that the $15$ cases found with $h^{1,1}=1$ exhaust this class~\cite{Berglund:2023ztk}.
Empirical saturation of a stochastic search does not prove completeness, however, and lattice inequivalent polytopes need not define inequivalent Calabi--Yau topologies.
The resulting fourfolds may also retain terminal singularities after MPCP desingularization.
The method should therefore be assessed by validated discoveries and coverage within a stated search domain.

Machine learning also supplies tools for exploring triangulations of known polytopes.
Three approaches have been proposed: genetic algorithms~\cite{MacFadden:2024him}, reinforcement learning~\cite{Berglund:2024reu}, and transformers~\cite{Yip:2025hon}.
The earlier study by MacFadden, Schachner, and Sheridan introduced genetic optimization of triangulations with prescribed physical properties~\cite{MacFadden:2024him}, preceding the reinforcement learning study~\cite{Berglund:2024reu}.
Each chromosome records triangulations of the two-dimensional faces, using the extension algorithm of~\cite{MacFadden:2023cyf} to test whether these choices admit a globally regular triangulation.
This removes the redundancy from FRSTs sharing the same two-face restrictions, although further topological equivalences can remain.
Selection, crossover, and mutation then explore this reduced space, with mutations and crossover permitting changes across several faces simultaneously.
The authors benchmarked volume maximization at $h^{1,1}=23$, evaluating the volume at the tip of each stretched K\"ahler cone, against random sampling, Markov chain Monte Carlo, and simulated annealing.
This demonstrated the effectiveness of genetic search and Bayesian hyperparameter optimization.
They also solved an inverse problem at $h^{1,1}=60$, searching for prescribed axion decay constants, and optimized axion-photon couplings at $h^{1,1}=491$.
The latter calculation fixed Delaunay triangulations on the three largest two-faces and used untuned hyperparameters, demonstrating scalability within a restricted search space rather than exhaustive exploration.

The \href{https://github.com/elliheyes/Triangulation-Generation}{\texttt{Triangulation-Generation}} repository implements the reinforcement learning approach of Berglund, Butbaia, He, Heyes, Hirst, and Jejjala~\cite{Berglund:2024reu}.
It formulates FRST construction as a sequential decision problem, with states and rewards that can incorporate geometric and physical constraints.
The study demonstrates triangulation generation, searches for reflexive subpolytopes and compatible triangulations exhibiting fibrations, including K3 fibrations, and searches for triangulation--bundle pairs in heterotic model building.
Targeting additional structures can avoid spending most computational effort on geometries irrelevant to a particular application.
The implemented reward must nevertheless be distinguished from the complete physical requirement.
For the line bundle sums studied, the search imposes anomaly cancellation and a mixed sign matrix test that is necessary for poly-stability, rather than solving the full simultaneous conditions for vanishing slopes inside the K\"ahler cone.
The resulting pairs remain candidates requiring these further checks.
Neither favorable benchmark performance nor recovery of known small collections establishes general completeness or unbiased sampling.

The \href{https://github.com/jhtyip/cytransformer}{\texttt{CYTransformer}} model of Yip, Arnal, Charton, and Shiu uses an encoder-decoder transformer to generate FRSTs from encodings of favorable four-dimensional reflexive polytopes~\cite{Yip:2025hon}.
Its emphasis is efficient sampling at scale, reducing sampling bias, and improving the generator by retraining on its own validated outputs, rather than optimizing one observable.
The associated site \href{https://aicy.physics.wisc.edu/}{AICY} (AI-enabled living Calabi--Yau repositories) distributes data, software, and pretrained checkpoints.
The released training data span $h^{1,1}=5,\ldots,10$, contain millions of polytope--triangulation pairs, and separate training, validation, and test sets by polytope~\cite{AICY:2026Resources}.
The distinction between polytopes and pairs matters because many records describe triangulations of the same polytope.
The site reports exhaustive triangulation data for the included configurations with fewer points and capped sampling for the larger ones; the separate catalog of model generated FRSTs is advertised as forthcoming.
AICY is therefore a reproducible generative research resource with substantial reference data, while its proposed continually expanding catalog remains under development.
The three approaches are potentially complementary: a transformer can supply candidate triangulations for subsequent genetic algorithm or reinforcement learning searches, whose validated results can in turn augment its training data.
Supervised fine tuning or reward shaping could also introduce more specialized objectives, but the sampling distribution and geometric validity of generated outputs require explicit assessment.

\subsection{Derived datasets and physical structures}\label{sec:derived-datasets}

Large derived databases reorganize known geometries around structures that matter directly in physics.
For F-theory, Morrison and Taylor classified $61,539$ base configurations arising from toric surfaces for elliptically fibered Calabi--Yau threefolds, including cases requiring additional non-toric blow-ups~\cite{Morrison:2012js}.
Complementarily, Huang and Taylor found that all but $29,223$ of the four-dimensional reflexive polytopes in the Kreuzer--Skarke list contain a two-dimensional reflexive subpolytope through the origin: approximately $99.994\%$ therefore exhibit the toric structure associated with an elliptic or genus-one fibration in a suitable birational phase~\cite{Huang:2019pne}.
Extending this analysis, Abbasi, Nally, and Taylor identify $2,264,992,252$ toric elliptic and genus-one fibrations across the list, or $2,250,744,657$ classes after accounting for automorphisms of the ambient polytope~\cite{Abbasi:2025lvn}.
The accompanying \href{https://doi.org/10.5281/zenodo.18500236}{data} record fiber and base structures.
These statistics count bases, polytopes, and fibration structures, respectively.
The Abbasi--Nally--Taylor catalog gives a detailed view of fibration multiplicity, but its completeness claim concerns the toric fibrations analyzed in that construction, realized in suitable phases.
It does not identify every possible fibration of every underlying Calabi--Yau manifold or assert that every resolution admits the specified fibration.
Conversely, the absence of a reflexive two-dimensional subpolytope does not exclude a non-toric fibration~\cite{Huang:2019pne}.
The distinction between elliptic fibrations and genus-one fibrations (without a section) remains important for interpreting the associated physics.

For type IIB compactifications, orientifold catalogs add involutions, fixed loci, parity decompositions, and tadpole information.
The \href{https://github.com/AndreasSchachner/CY_Orientifold_database}{Crin\`o--Quevedo--Schachner--Valandro database} studies reflection orientifolds on favorable toric hypersurfaces, with exhaustive triangulation coverage at low $h^{1,1}$ and sampling extending to $h^{1,1}=12$~\cite{Crino:2022zjk}.
Cao, Gao, and Gao subsequently reported 320,386,067 \href{https://github.com/GroupofXG/anewcydatabase/}{orientifold constructions} using divisor exchanges and multiple reflections, again with explicit sampling restrictions at larger Hodge number~\cite{Cao:2024oqx}.
These resources are closer to phenomenological input than bare lists of Hodge numbers because orientifold planes and brane charges directly constrain compactification choices.
Their totals count constructions equipped with involutions and should not be added to counts of underlying Calabi--Yau manifolds.
Nor does an orientifold construction establish a fully stabilized vacuum.

Recent projects for distributing data further reduce the gap between geometry and computation.
Schachner's \href{https://huggingface.co/datasets/aschachner/cy-database}{\texttt{cy-database}}, accessed through \href{https://github.com/StringJAX/stringforge}{StringForge}, packages precomputed toric and CICY information, with additional enumerative and conifold data where available; its current documentation reports more than $10^8$ FRST phase records~\cite{Schachner:2026CYDatabase}.
Coverage, equivalence conventions, and access mechanisms vary between components, making the documentation for each release essential.
At the level of physical models, the \href{https://www-thphys.physics.ox.ac.uk/projects/CalabiYau/linebundlemodels/}{Oxford heterotic line bundle archive} records $63,325$ $SU(5)$ GUT line bundle models on $68$ favorable CICYs~\cite{Anderson:2013xka}.
Here, the multiplicity arises from additional bundle choices over a comparatively small geometric collection.

\subsection{Uses and limitations}

These resources support different research strategies.
CICY catalogs offer explicit presentations and established tools for bundle calculations; the Kreuzer--Skarke archive supplies combinatorial breadth; ToricCY and CYTools provide detailed geometry through stored results and computation on demand; and generative methods offer targeted access to regions where enumeration is impractical.
For statistical applications, document the construction and sampling biases and the equivalence relation used to group entries.
For machine learning applications, equivalent presentations and multiple triangulations require careful handling when constructing independent test sets.

All of these catalogs can be data mined, \textit{e.g.}, for novel relationships between features.
The most valuable future extensions will combine increased coverage with stronger identification and validation.
Records should preserve the defining construction, the chosen phase, divisor basis conventions, known equivalences, and the status of smoothness and cone calculations.
Generated candidates should retain enough information in order be validated, while physical datasets should distinguish geometric information from choices of bundles, branes, involutions, and fluxes.
The scientific value of a Calabi--Yau database ultimately depends on how reliably an entry can be converted into a well-defined calculation, and on how precisely the limits of that calculation are documented.

\section{Point sampling and numerical integration}\label{sec:points}

Numerical integration enters both Donaldson's algorithm and the training and assessment of learned metrics, approaches which we explain below.
The starting point is a sample with a known probability distribution on the Calabi--Yau manifold.
Uniformity is always relative to a specified measure: the volume of an ambient Fubini--Study metric, the volume determined by the holomorphic top form, and Lebesgue measure in a chosen local coordinate chart (\textit{i.e.}, sampling in some coordinate chart on the manifold with constant density) generally give different distributions.
A convenient sampling measure need not coincide with the measure in the desired integral, provided the difference is included through integration weights.

\subsection{Random sections and Fubini--Study sampling}

The mathematical foundation is the random zero construction of Shiffman and Zelditch~\cite{Shiffman:1998drw}.
Let $X$ have complex dimension $n$, and choose a Hermitian inner product and a suitable 
basis $s_\alpha$ of the base point free space $H^0(X,L^k)$.
A random section is
\be
 s=\sum_{\alpha=1}^{N_k}c_\alpha s_\alpha \,,\qquad
 c_\alpha\sim\mathcal N_{\mathbb C}(0,1) \,,
 \quad\hbox{independently} \,.
 \label{eq:random-section}
\ee
Gaussian coefficients and a uniformly distributed unit coefficient vector give the same zero statistics, since multiplying a section by a nonzero scalar leaves its zeros unchanged.
For the projective map $\Phi_k=[s_1:\cdots:s_{N_k}]$, the expected zero currents satisfy
\be
 \mathbb E[Z_s]=\Phi_k^*\omega_{\text{FS}} \,,\qquad
 \mathbb E[Z_{s^{(1)},\ldots,s^{(\ell)}}]
   =(\Phi_k^*\omega_{\text{FS}})^\ell \,,
 \label{eq:expected-zero-current}
\ee
for independent sections and proper intersections.
Here, $Z$ denotes integration over the zero locus; when $\ell=n$, it counts isolated points with multiplicity.
Locally, the first identity follows by applying Poincar\'e--Lelong to the Gaussian average of $\log|s|^2$.

These identities at finite $k$ should be distinguished from Shiffman--Zelditch's equidistribution theorem.
For the $L^2$ inner products induced by a fixed positive Hermitian metric on $L$ and its curvature volume, with curvature form $\omega$, they prove almost sure weak convergence $k^{-1}Z_{s_k}\to\omega$ as $k\to\infty$.
The practical sampler below instead repeats intersections at fixed degree, whose expected measure is already known exactly.

Douglas, Karp, Lukic, and Reinbacher~\cite{Douglas:2006rr} apply this construction to a smooth hypersurface $X=\{Q=0\}\subset\mathbb P^{n+1}$ of degree $d$.
With $n=\dim_{\mathbb C}X$, the Calabi--Yau condition is $d=n+2$.
The algorithm is particularly simple:
\begin{enumerate}
\item Draw independent isotropic complex Gaussian vectors $u,v\in\mathbb C^{n+2}$.
Their span defines a projective line
\be
 L_{u,v}=\{[a u+b v]:[a:b]\in\mathbb P^1\} \,.
 \label{eq:pp}
 \ee
This line has the unitary invariant distribution.
Normalizing the vectors is optional; it gives uniform vectors on $S^{2n+3}$ without changing their span.
\item Solve the homogeneous polynomial equation $Q(a u+b v)=0$ on $\mathbb P^1$ and retain all $d$ intersections.
They are distinct for a generic line.
The Fermat cubic in $\mathbb P^2$, for example, supplies three points per line.
\item Repeat with independent lines until the desired sample size is reached.
\end{enumerate}
The Gaussian construction avoids rejection from a high dimensional cube.
Drawing Cartesian coordinates uniformly in a cube and merely normalizing the resulting vector would not produce the required invariant ensemble.

Write $\omega_{\text{FS},X}=\iota^*\omega_{\text{FS}}$, where $\iota:X\hookrightarrow\mathbb P^{n+1}$ and $[\omega_{\text{FS}}]=c_1(\mathcal O(1))$, as in Section~\ref{sec:don}.
A line is the intersection of $n$ hyperplanes, so~\eqref{eq:expected-zero-current} gives
\begin{align}
 \mathbb E\!\left[\sum_{p\in L\cap X}\delta_p\right]
   &=\omega_{\text{FS},X}^{n} \,,\nonumber\\*
 \dd\rho_{\text{FS}}
   &=\frac{\omega_{\text{FS},X}^{n}}{d}
    =\frac{\dd\text{Vol}_{\text{FS}}}{V_{\text{FS}}} \,,
 \qquad V_{\text{FS}}=\frac{d}{n!} \,.
 \label{eq:fs-probability}
\end{align}
Thus, the points sample the normalized restricted Fubini--Study volume, generally different from the Ricci flat volume.
Points from the same line are correlated.
This does not bias a sample average, but its statistical uncertainty should be estimated using independent lines or batches, rather than treating all roots as independent observations.

\subsection{Local coordinates and integration weights}

For $p=[Z_0:\cdots:Z_{n+1}]$, choose $m$ for which $|Z_m|$ is largest and set $x_a=Z_a/Z_m$ for $a\ne m$.
All these affine coordinates have modulus at most one.
Among them, choose a dependent coordinate $x_j$ maximizing $|\partial_j Q_m|$, where $Q_m$ is the affine defining polynomial.
On a smooth hypersurface this derivative can be chosen nonzero, and the implicit function theorem gives
\be
 \frac{\partial x_j}{\partial x_a}
   =-\frac{\partial_a Q_m}{\partial_j Q_m} \,,\qquad a\ne j \,.
 \label{eq:sampling-pullback}
\ee
These derivatives pull ambient tensors back to the tangent space of $X$.
The pair $(m,j)$ selects a convenient local representation, with ties resolved consistently; it does not change the sampling measure.
For a complete intersection one instead selects an invertible, well conditioned Jacobian block for the dependent coordinates.
This chart construction is used in numerical metric computations~\cite{Douglas:2006rr,Ashmore:2019wzb,Cui:2019uhy}.

\begin{figure}[tbp]
\centering
\includegraphics[width=.43\textwidth]{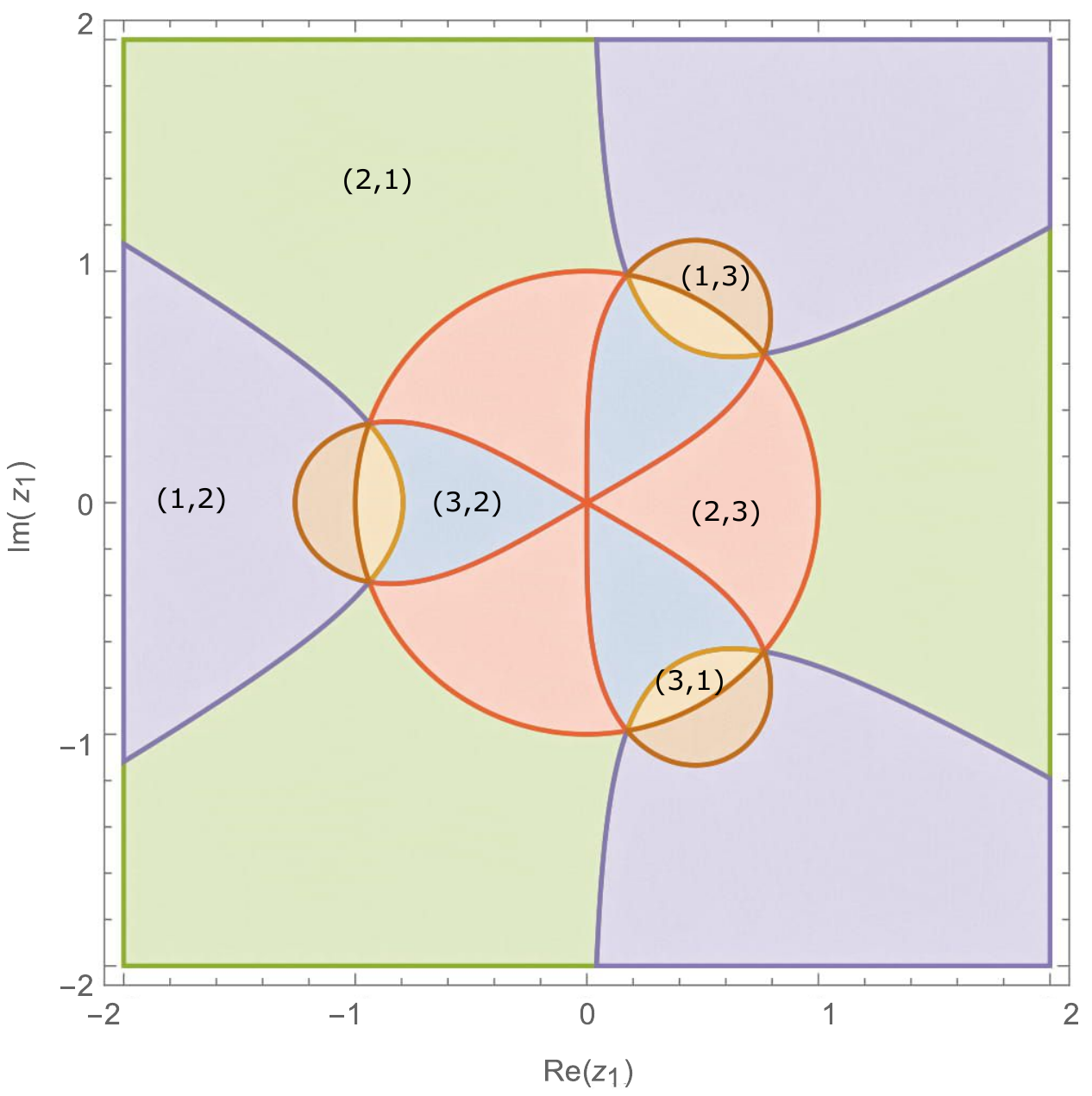}\hfill
\includegraphics[width=.49\textwidth]{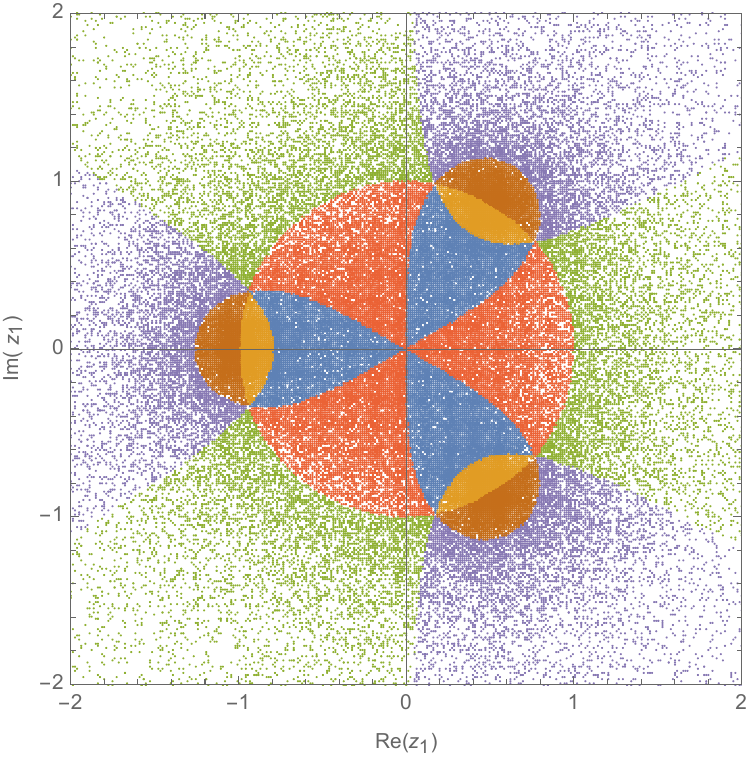}
\caption{\small Local representations and point sampling for a cubic in $\mathbb P^2$.
The six choices $(i,j)$, $i\ne j$, specify the projective coordinate fixed to one and the dependent coordinate, respectively; the diagrams use labels $1,2,3$.
Left: the three sheets in the representation $(3,2)$, with colors indicating overlaps with other representations.
Right: points generated by random line intersections.
Their density is uniform with respect to the restricted Fubini--Study volume, not with respect to Euclidean area in the displayed coordinate.}
\label{fig:T2patches}
\end{figure}

The holomorphic form is available algebraically, without knowing the Ricci flat metric.
For an anticanonical hypersurface, its local residue expression is, up to a constant and an orientation sign,
\be
 \Omega \wedge_m \partial_j Q_m \dd x^j =\wedge_{a}\dd x_a \,.
 \label{eq:sampling-residue}
\ee
Define the positive volume form and its normalized measure by
\be
 \dd\mu_\Omega=\text{i}^{n^2}\Omega\wedge\bar\Omega \,,\qquad
 V_\Omega=\int_X\dd\mu_\Omega \,,\qquad
 \dd\nu=\frac{\dd\mu_\Omega}{V_\Omega} \,.
 \label{eq:sampling-target}
\ee
This is the same $\dd\nu$ as in~\eqref{eq:donaldson-volume}.
At fixed K\"ahler class, the Ricci flat volume is proportional to $\dd\mu_\Omega$; knowing this volume does not determine all components of the metric in higher dimension.

For points sampled from~\eqref{eq:fs-probability} one has the following numerical integration formula: 
\begin{align}
 w(p)&=\frac{\dd\mu_\Omega}{\dd\text{Vol}_{\text{FS}}}(p) \,,\nonumber\\*
 \int_X \dd\mu_\Omega\,f
   &\simeq\frac{V_{\text{FS}}}{N}\sum_{i=1}^N f(p_i)w(p_i) \,,
 \qquad
 V_\Omega\simeq\frac{V_{\text{FS}}}{N}\sum_{i=1}^Nw(p_i) \,.
 \label{eq:sampling-integrals}
\end{align}
The factor $V_{\text{FS}}$ is necessary because $\dd\text{Vol}_{\text{FS}}$ is not a probability measure.
In local coordinates with $\Omega=h\,\dd y^1\wedge\cdots\wedge \dd y^n$ and $\omega_{\text{FS},X}=\text{i}g^{\text{FS}}_{a\bar b}\dd y^a\wedge \dd\bar y^{\bar b}$, the ratio is $w=|h|^2/\det g^{\text{FS}}$.
If only normalized averages are needed, the unknown overall normalization cancels:
\be
 \int_X \dd\nu\,f\simeq
 \frac{\sum_iw(p_i)f(p_i)}{\sum_iw(p_i)} \,.
 \label{eq:sampling-normalized}
\ee
This ratio estimator is consistent, although generally biased at finite $N$.
When the normalized density ratio $\dd\nu/\dd\rho$ is known, the ordinary average of $f\,\dd\nu/\dd\rho$ is unbiased.
The same distinction applies to training losses: an unweighted average over Fubini--Study points estimates a Fubini--Study average, while a loss defined using $\dd\nu$ requires the corresponding weights.

\section{Donaldson's algorithm}\label{sec:don}

An early direct approach to numerical Calabi--Yau metrics was developed by Headrick and Wiseman~\cite{Headrick:2005ch}.
They solved the complex Monge--Amp\`ere equation on a family of K3 surfaces obtained by resolving $T^4/\mathbb{Z}_2$, using finite differences in overlapping coordinate patches and a relaxation algorithm for the K\"ahler potential.
This demonstrated that one could compute compact Ricci-flat metrics away from analytically controlled limits, while exposing the practical cost of a coordinate lattice and the need to match potentials between patches.
Donaldson's construction~\cite{Donaldson:2005hvr} replaces this local representation by a finite dimensional family of projectively induced metrics, with numerical integration supplying the matrix updates.

Let $X$ be a smooth connected compact Calabi--Yau manifold of complex dimension $n$, with an ample line bundle $L\to X$.
Choose $k$ so that $L^{\otimes k}$ is very ample, and let $s=(s_1,\ldots,s_{N_k})^T$ be a basis of $H^0(X,L^{\otimes k})$.
A positive Hermitian matrix $G$ defines a Fubini--Study metric $FS(G)$ on this line bundle.
In a local trivialization, its pointwise norm and the induced K\"ahler potential are
\begin{align}
 |u|_{FS(G)}^2&=\frac{|u|^2}{s^\dagger G^{-1}s} \,,\nonumber\\*
 K_G^{(k)}&=\frac{1}{2\pi k}\log(s^\dagger G^{-1}s) \,,\qquad
 \omega_G^{(k)}=\text{i}\partial\bar\partial K_G^{(k)} \,.
 \label{eq:FSk}
\end{align}
Here, $u$ is a local representative of a section; changes of trivialization change the potential by a K\"ahler transformation.
With this convention $[\omega_G^{(k)}]=c_1(L)$.
Multiplying $G$ by a positive constant only adds a constant to the potential and leaves the K\"ahler metric unchanged.

For the Calabi--Yau problem we prescribe the normalized positive measure
\be
 \dd\nu=\frac{\text{i}^{n^2}\Omega\wedge\bar\Omega}
 {\int_X\text{i}^{n^2}\Omega\wedge\bar\Omega} \,,\qquad \int_X\dd\nu=1 \,,
 \label{eq:donaldson-volume}
\ee
where $\Omega$ is a nowhere vanishing holomorphic top form.
A Hermitian line bundle metric $h$ then defines the Hilbert norm
\be\label{eq:one}
 \|u\|_{\operatorname{Hilb}_\nu(h)}^2
 =N_k\int_X \dd\nu\,|u|_h^2 \,.
\ee
Composing the two constructions gives the fixed measure $T$-operator,
\be
 T_\nu(G)=\operatorname{Hilb}_\nu(FS(G))
 =N_k\int_X \dd\nu\,\frac{s\,s^\dagger}{s^\dagger G^{-1}s} \,.
 \label{eq:donaldson-T}
\ee
Its fixed points are called $\nu$-balanced metrics.
Donaldson's fixed measure existence and convergence theorem applies to a smooth positive measure on a compact, nondegenerately embedded $X$~\cite{Donaldson:2005hvr}.
The balanced form is unique up to overall scale, and iteration from a positive initial matrix converges after fixing this scale, for example by setting $\det G=1$ after each update.
This is a statement about iteration at fixed $k$.
A separate large-$k$ result shows that the induced balanced metrics converge smoothly to the solution of the prescribed volume equation in $c_1(L)$~\cite{Cao:2013AboutCalabiProblem}.
For \eqref{eq:donaldson-volume}, that solution is the unique Ricci-flat K\"ahler metric in the class.
At finite $k$, the balanced metric is generally only an approximation to it.

It is essential to distinguish this construction from the usual balanced metric problem, where the Hilbert map uses the volume of the evolving metric itself.
The latter is related to constant \emph{scalar} curvature: under the discrete automorphism hypothesis, existence of a constant scalar curvature K\"ahler metric implies the existence of convergent balanced approximations for sufficiently large $k$~\cite{Donaldson:2001ScalarCurvatureProjective,Donaldson:2005ScalarCurvatureProjective}.
In the fixed measure construction, choosing a different prescribed volume generally changes the limiting metric.
In particular, fixing a Fubini--Study volume does not generally target the Ricci-flat metric.
Sampling points from a Fubini--Study distribution is nevertheless compatible with \eqref{eq:donaldson-T}, provided the integrand is weighted to estimate the prescribed $\dd\nu$ integral~\cite{Douglas:2006rr}.
Section~\ref{sec:points} discusses this distinction between the sampling distribution and integration measure.

To assess an approximate metric, define
\be
 \dd\text{Vol}_\omega=\frac{\omega^n}{n!} \,,\qquad
 V_\omega=\int_X\dd\text{Vol}_\omega \,,\qquad
 \eta_\omega=\frac{\dd\text{Vol}_\omega}{V_\omega\,\dd\nu} \,.
 \label{eq:donaldson-eta}
\ee
Thus, $\int_X \dd\nu\,\eta_\omega=1$, and the normalized Monge--Amp\`ere equation is $\eta_\omega=1$.
Two useful diagnostics are the volume residual and an integrated scalar curvature residual~\cite{Douglas:2006rr,Braun:2007sn,Braun:2008jp}:
\be
 \sigma(\omega)=\int_X \dd\nu\,|1-\eta_\omega| \,,\qquad
 r(\omega)=V_\omega^{1/n-1}\int_X \dd\text{Vol}_\omega\,|R_\omega| \,.
 \label{eq:donaldson-diagnostics}
\ee
Both are invariant under an overall rescaling of $\omega$ and vanish for the target metric.
Computing $\sigma$ requires two derivatives of the potential, whereas $r$ requires four.
They therefore probe different aspects of numerical accuracy: a small volume residual alone does not bound curvature errors, which are sensitive to higher derivatives.

Residuals should be evaluated on an independent validation sample with sufficient resolution in regions of rapid variation. As the approximate metric converges to the Ricci-flat metric, both the  $\sigma$ and $r$ residuals approach zero. Only in this limit do the two residuals carry the same information; for Donaldson’s algorithm, this corresponds to large $k$.

The matrix size is governed by the Hilbert polynomial, rather than factorial growth.
For fixed $X$ and $L$, Riemann--Roch gives
\be
 N_k=\frac{k^n}{n!}\int_Xc_1(L)^n+O(k^{n-1}) \,.
\ee
For a smooth quintic hypersurface with $L=\mathcal{O}_X(1)$,
\be
 N_k=\begin{cases}\binom{k+4}{4}\, \,,\qquad (k<5)\,,\\\binom{k+4}{4}-\binom{k-1}{4}
     =\frac{5}{6}k^3+\frac{25}{6}k \,,\quad (k\geq5)\,, \end{cases}
\ee
An unrestricted Hermitian form has $N_k^2$ real parameters, before removing its irrelevant scale or exploiting symmetries, and dense matrix inversion costs $O(N_k^3)$.
The quintic calculations of Braun, Brelidze, Douglas, and Ovrut~\cite{Braun:2008jp} used the practical sampling prescription
\be
 N_p=10N_k^2+50,000 \,.
\ee
This gives $N_p=50,250$ at $k=1$ and $N_p=7,706,250$ at $k=10$, where $N_{10}=875$.
It is an empirical accuracy choice, not a universal lower bound on the number of integration points.
Convergence of the matrix iteration, Monte Carlo error, and the finite-$k$ approximation error must be checked separately.

At $k=1$ the Fermat quintic provides a useful symmetry check.
Independent fifth root phase symmetries force the balanced matrix in the coordinate basis to be diagonal, and permutations force all diagonal entries to agree.
Consequently, for either the normalized $\Omega$ measure or the fixed, symmetry invariant Fubini--Study measure, a determinant one representative is
\be
 G_*^\Omega=G_*^{FS}=I_5 \,.
\ee
Indeed, the off-diagonal integrals vanish by phase symmetry, while the five equal diagonal integrals in \eqref{eq:donaldson-T} sum to five.
This coincidence at the lowest level does not make the Fubini--Study metric Ricci-flat or imply that the two prescribed measures have the same large-$k$ limit.

\subsection{Variations on Donaldson's algorithm}
\label{sec:donaldson-variations}

Douglas, Karp, Lukic, and Reinbacher~\cite{Douglas:2006rr} developed the numerical implementation of balanced Calabi--Yau metrics, including Monte Carlo integration on projective hypersurfaces.
Braun, Brelidze, Douglas, and Ovrut~\cite{Braun:2007sn} extended these computations to complete intersections and free quotients, bringing geometries relevant to heterotic compactification within reach.
Their subsequent computation of scalar Laplace eigenvalues and eigenfunctions~\cite{Braun:2008jp} illustrates how a numerical metric becomes an input to spectral calculations.

A parallel development concerns metrics on holomorphic vector bundles.
Douglas, Karp, Lukic, and Reinbacher~\cite{Douglas:2006hz} adapted the balanced construction to approximate Hermitian Yang--Mills connections on the Fermat quintic.
Anderson, Braun, Karp, and Ovrut~\cite{Anderson:2010ke} developed numerical diagnostics of slope stability and studied stable, semistable, and unstable examples; Anderson, Braun, and Ovrut~\cite{Anderson:2011ed} subsequently followed the dependence on the K\"ahler class and its stability chambers.
These calculations separate the Ricci-flat metric on the base from the Hermitian metric on the bundle, while linking the two numerical problems.

A different variation optimizes the algebraic metric directly.
Headrick and Nassar~\cite{Headrick:2009jz} introduced energies built from the volume ratio \eqref{eq:donaldson-eta}, such as
\be
 E_F[\omega]=\int_X \dd\nu\,F(\eta_\omega) \,,\qquad
 E_2[\omega]=\int_X \dd\nu\,(\eta_\omega-1)^2 \,.
\ee
For smooth strictly convex $F$, the unrestricted variational problem on a fixed K\"ahler class has the Ricci-flat metric as its only critical point.
Restricting to \eqref{eq:FSk} gives the finite dimensional stationarity condition
\be\label{eq:galerkin}
 \frac{\partial E_2}{\partial q^A}
 =2\int_X \dd\nu\,(\eta_\omega-1)\frac{\partial\eta_\omega}{\partial q^A}=0 \,,
\ee
where $q^A$ are independent real parameters of the positive Hermitian form, with its scale fixed.
This enforces orthogonality of the residual to the available variations, rather than exact pointwise satisfaction of the equation.
The restricted optimization can have additional stationary points.
The authors observed exponential improvement with degree for optimized algebraic metrics in smooth examples, but polynomial decay in their singular conifold example.
These observations are not a universal convergence guarantee for a numerical optimizer.
Notice also that $E_2$ is a squared $L^2$ residual, whereas $\sigma$ is an $L^1$ residual.
Since $\nu$ is normalized, $\sigma\leq\sqrt{E_2}$, but the two objectives are distinct.

Cui and Gray~\cite{Cui:2019uhy} used metrics optimized by the Headrick--Nassar method to locate regions of large curvature relative to the overall compactification scale and assess the reliability of curvature expansions.
They checked curvature maxima and local Ricci-flatness, varying the section degree and number of sampled points.

The early machine learning study of Ashmore, He, and Ovrut~\cite{Ashmore:2019wzb} used Donaldson calculations as training data to accelerate metric approximation.
In particular, they combined extrapolation in the section degree with supervised regression to predict metric determinants and components from less expensive balanced metric data.
Their principal examples used gradient boosted decision trees; this was an acceleration of an existing numerical method using labeled data.
Anderson, Gerdes, Gray, Krippendorf, Raghuram, and Ruehle~\cite{Anderson:2020hux} subsequently explored moduli dependent metrics, including both supervised learning of algebraic metric coefficients and direct training with geometric losses.
Their treatment of $SU(3)$ structures with nonzero intrinsic torsion is discussed in Section~\ref{sec:su3-structure-metrics}.
This development connects the algebraic approximation scheme to the PINN logic described below: one may learn from an existing solver, or use geometric equations to train a metric representation directly.
The metric and potential networks discussed below pursue the latter approach.

\subsection{Ambient balanced metrics and the SYZ limit}
\label{sec:ambient-balanced}

Berglund \textit{et al.}~\cite{Berglund:2026hel} retain the full ambient monomial vector $S$, including sections related by the defining ideal, through a Moore--Penrose inverse.
Write $S|_X=Rs$, with $R$ an $N_{\text{amb}}\times N_k$ matrix of full column rank.
Their iteration is
\be
 T_X^+(G_{\text{amb}})
 =N_k\int_X \dd\nu\,\frac{S\,S^\dagger}{S^\dagger G_{\text{amb}}^+S} \,,
 \label{eq:ambient-T}
\ee
where $+$ denotes the pseudoinverse; the normalization counts independent sections on $X$.
When restriction is surjective, this map is conjugate to \eqref{eq:donaldson-T} under $G_{\text{amb}}=RGR^\dagger$.
For $H=G^{-1}$, the corresponding coefficient matrix is the minimum Frobenius norm lift
\be
 H_{\text{amb}}=(R^+)^\dagger H R^+ \,,\qquad
 R^\dagger H_{\text{amb}}R=H \,.
\ee
It is positive semidefinite of rank $N_k$ and independent of the quotient basis, with the ambient coordinates and inner product fixed.
After scale normalization, it yields the same balanced metric on $X$.

The retained monomial labels make symmetries and complex structure dependence accessible.
In the Strominger--Yau--Zaslow (SYZ) picture, suitably normalized metrics collapse along special Lagrangian torus fibers toward a real base.
Near large complex structure,~\cite{Berglund:2026hel} observes power laws in normalized coefficients and proposes a Legendre dual interpretation of their limiting exponents in terms of the SYZ base potential.
These asymptotic identifications are, for the moment, conjectural: the finite level conjugacy does not establish uniform convergence in the simultaneous large-$k$ and degeneration limits.

\section{Physics informed neural networks}\label{sec:pinn}

Before discussing numerical Calabi--Yau metrics, let us explain how a neural network can approximate the solution of a differential equation.
A physics informed neural network (PINN) represents the unknown field by a differentiable trial function $u_\theta(x)$, with trainable weights and biases $\theta$~\cite{Raissi:2017zsi}.
For an equation $\mathcal{D}[u]=f$, automatic differentiation evaluates the derivatives of this trial function and hence its residual,
\be
 r_\theta(x)=\mathcal{D}[u_\theta](x)-f(x) \,.
\ee
The activation functions must have the differentiability required by $\mathcal{D}$.
Sampling interior points $x_i$ turns residual minimization into a finite optimization problem:
\be\label{eq:pinnloss}
 \mathcal{L}(\theta)
 =\frac{\lambda_r}{N_r}\sum_{i=1}^{N_r}|r_\theta(x_i)|^2
 +\lambda_b\mathcal{L}_{\text{boundary}}
 +\lambda_0\mathcal{L}_{\text{initial}}
 +\lambda_d\mathcal{L}_{\text{data}} \,.
\ee
Here, the additional terms penalize violations of prescribed boundary conditions, initial conditions for evolution equations, and any available observations; irrelevant terms are omitted.
Each is typically a mean squared discrepancy on the corresponding sample points, and the nonnegative coefficients set their relative importance.
Training differentiates this loss with respect to $\theta$ and updates the network parameters.
For a forward problem with sufficient prescribed conditions, no labeled solutions in the interior are required: the differential equation supplies the training signal.
Observational data can nevertheless be included, particularly in inverse problems~\cite{Raissi:2017zsi}.

This construction separates representation of the solution from enforcement of its defining equations.
When possible, constraints can instead be built into the trial function, reducing the number of competing loss terms.
The sampling distribution also matters: the residual sum estimates an integral with respect to that distribution, so a different target measure requires appropriate integration weights.

A small training loss alone is not a guarantor of accuracy.
Finite sampling can miss regions with large residuals, and the relation between residual and solution error depends on stability of the underlying problem.
Moreover, the nonconvex optimization can be difficult even when the network can represent an accurate solution~\cite{Krishnapriyan:2021failure}.
We should therefore evaluate residuals and constraints on independent points, test sensitivity to sampling and loss weights, and compare with known solutions or independent numerical methods where available.

For a compact Calabi--Yau manifold there is no physical boundary; compatibility on overlapping coordinate charts takes the place of boundary matching.
The trial object may be a metric or a real scalar correction $\varphi_\theta$ to a reference K\"ahler form,
\be
 \omega_\theta=\omega_0+\text{i}\partial\bar\partial\varphi_\theta \,.
\ee
A globally defined smooth correction preserves closure and the K\"ahler class, while positivity must still be ensured.
Ricci-flatness, equivalently the normalized complex Monge--Amp\`ere equation within this class, then supplies the differential constraint.
This is the logic behind learning metrics from geometry rather than from a collection of known Ricci-flat metrics.
We now turn to neural representations of Calabi--Yau metrics.

\section{Machine learned metric approximations}
\label{sec:ml-metrics}

\subsection{Learning the metric directly}\label{sec:direct-metric}

The PINN framework of Section~\ref{sec:pinn} applies directly to the Calabi--Yau metric problem: a neural network represents the unknown geometry, and its defining equations supply the training signal.
Several representations are possible.
We illustrate direct metric learning with the architecture of~\cite{Jejjala:2020wcc}.
Contemporaneous and complementary approaches are~\cite{Anderson:2020hux} and~\cite{Douglas:2020hpv}.
The approaches share this training philosophy, but differ in which geometric constraints are built into the architecture and which must be imposed through the loss.

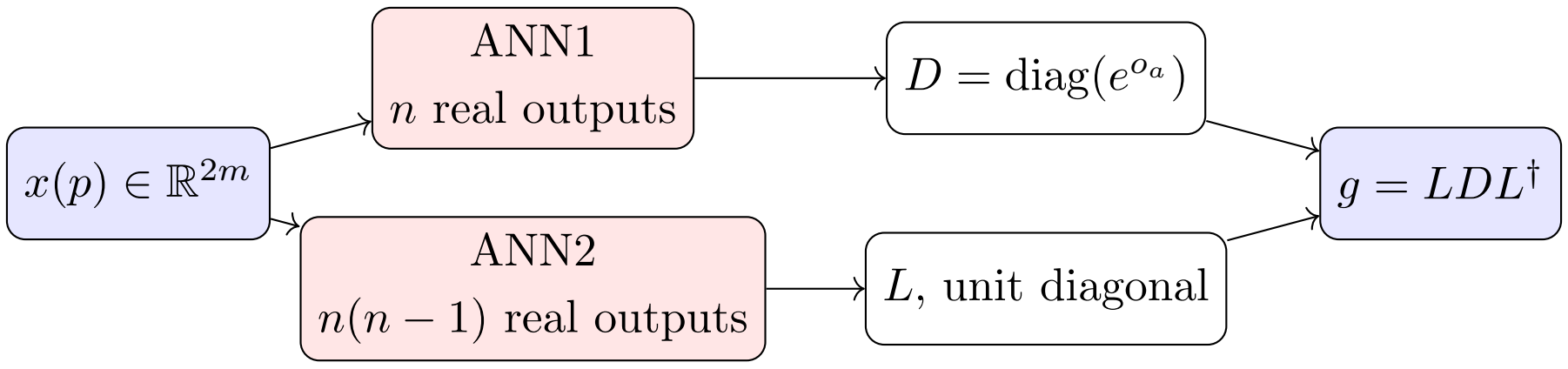
\begin{figure}[htbp]
\centering
\begin{tikzpicture}[every node/.style={font=\small},
 box/.style={draw,rounded corners,align=center,minimum height=0.85cm}]
 \node[box,fill=blue!10] (input) at (0,0) {$x(p)\in\mathbb R^{2m}$};
 \node[box,fill=red!10] (ann1) at (3,0.8) {ANN1\\$n$ real outputs};
 \node[box,fill=red!10] (ann2) at (3,-0.8) {ANN2\\$n(n-1)$ real outputs};
 \node[box] (d) at (6.9,0.8) {$D=\operatorname{diag}(e^{o_a})$};
 \node[box] (l) at (6.9,-0.8) {$L$, unit diagonal};
 \node[box,fill=blue!10] (g) at (9.9,0) {$g=LDL^\dagger$};
 \draw[->] (input) -- (ann1);
 \draw[->] (input) -- (ann2);
 \draw[->] (ann1) -- (d);
 \draw[->] (ann2) -- (l);
 \draw[->] (d) -- (g);
 \draw[->] (l) -- (g);
\end{tikzpicture}
\caption{A schematic version of the direct metric architecture of~\cite{Jejjala:2020wcc}.
Both networks receive coordinate data for the same point $p\in X$.
Their outputs specify a positive Hermitian $n\times n$ matrix, while training imposes the remaining geometric conditions.}
\label{fig:arch}
\end{figure}

On a complex $n$-dimensional manifold, a positive Hermitian metric has the decomposition
\be
 g=L D L^\dagger\,,\qquad
 D=\operatorname{diag}\!\left(e^{o_1},\ldots,e^{o_n}\right)\,.
\ee
Here, $L$ is lower triangular with unit diagonal, and its remaining entries contain $n(n-1)$ real parameters.
Following~\cite{Jejjala:2020wcc}, two networks supply these parameters and the $n$ outputs $o_a$, respectively.
Exponentiation makes $D$ strictly positive and hence ensures positive definiteness of $g$.
The diagonal entries of $D$ are not, in general, the eigenvalues of $g$.
For $X\subset\mathcal A$, with $\dim_{\mathbb C}\mathcal A=m$, the inputs can be the real and imaginary parts of the $m$ affine ambient coordinates, including the dependent coordinates, as in Figure~\ref{fig:arch}.

This factorization alone imposes neither the K\"ahler condition nor compatibility between charts.
Writing $\omega=\mathrm i g_{a\bar b}\,\dd z^a\wedge\dd\bar z^{\bar b}$, closure requires
\be
 \partial_a g_{b\bar c}-\partial_b g_{a\bar c}=0\,.
\ee
To describe the residuals, choose a reference metric $g_0$, set $V_0=\operatorname{Vol}_{g_0}$ and $\ell_0=V_0^{1/(2n)}$, and use the fixed probability measure
\be
 \dd\mu=\frac{\dd\operatorname{Vol}_\Omega}{\operatorname{Vol}_\Omega}\,,\qquad
 \langle f\rangle_\mu=\int_X\dd\mu\,f\,.
\ee
A dimensionless closure diagnostic is
\be\label{eq:Kappa}
 \kappa(g)=\ell_0^2\left\langle |\dd\omega|_{g_0}^2\right\rangle_\mu\,.
\ee
If $F_{\alpha\beta}$ is the coordinate change from chart $\alpha$ to chart $\beta$, compatible local predictions obey $g_\alpha=F_{\alpha\beta}^*g_\beta$ on their overlap.
We can therefore define
\be\label{eq:Gamma}
 \gamma(g)=\left\langle
 \frac{1}{N_{\mathrm{tr}}(p)}
 \sum_{(\alpha,\beta)\in\mathcal P(p)}
 \left|g_\alpha-F_{\alpha\beta}^*g_\beta\right|_{g_0}^2
 \right\rangle_\mu\,.
\ee
Here, $\mathcal P(p)$ contains the tested overlapping chart pairs, $N_{\mathrm{tr}}(p)$ is their number, and the pullback includes the Jacobians acting on both metric indices.
These formulas give a convenient normalization of the closure and transition penalties used in~\cite{Jejjala:2020wcc,Anderson:2020hux}; the tensor norms use $g_0$.
Monte Carlo estimates require the weights appropriate to $\dd\mu$.

The more general objective introduced in~\cite{Larfors:2021pbb} combines five geometric conditions:
\be\label{eq:loss}
 \mathcal L=
 \alpha_{\mathrm{MA}}\mathcal L_{\mathrm{MA}}
 +\alpha_{\mathrm{K}}\kappa
 +\alpha_{\mathrm{tr}}\gamma
 +\alpha_{\mathrm{Ric}}\mathcal L_{\mathrm{Ricci}}
 +\alpha_{\mathrm{vol}}\mathcal L_{\mathrm{vol}}\,.
\ee
The nonnegative coefficients balance the Monge--Amp\`ere residual, closure, chart compatibility, Ricci curvature, and agreement with the prescribed volume.
For example, one can use the $\sigma$-measure for $\mathcal L_{\mathrm{MA}}$ and $|\operatorname{Vol}_g/V_0-1|$ for $\mathcal L_{\mathrm{vol}}$.
Once global K\"ahler geometry is ensured, the Monge--Amp\`ere equation supplies the Ricci-flatness condition without differentiating the predicted metric.
The more expensive Ricci term can therefore be omitted during training and evaluated independently for validation.

Volume matching does not determine the K\"ahler class when $h^{1,1}>1$.
Further period constraints or an ansatz that preserves the class are needed.
This distinction applies also to K3, which has $h^{1,1}=20$: restricting attention to a quartic in projective space selects an ambient polarization, rather than removing its other K\"ahler directions.
The potential construction below incorporates closure and class preservation directly.

\subsection{Learning the K\"ahler potential}\label{sec:potential}

The additive $\Phi$ model, first implemented in~\cite{Larfors:2021pbb} and developed further in~\cite{Larfors:2022nep}, replaces the metric components by a real scalar correction to a reference K\"ahler potential.
This is a particular neural potential ansatz; the contemporaneous holomorphic and bihomogeneous constructions of~\cite{Douglas:2020hpv} are discussed in Section~\ref{subsec:holomorphic-networks}.
Although the reference potential is defined locally, the difference between two K\"ahler forms in the same class on compact $X$ can be expressed through a global function $\Phi$:
\be\label{eq:potential-metric}
 \omega_\Phi=\omega_0+\text{i}\partial\bar\partial\Phi
 \,,\qquad
 (g_\Phi)_{a\bar b}=(g_0)_{a\bar b}+\partial_a\partial_{\bar b}\Phi \,.
\ee
If $\Phi$ is smooth and globally defined, closure, chart compatibility, and preservation of $[\omega_0]$ follow from the ansatz.
Positive definiteness of $g_\Phi$ must still be checked or enforced.
A network acting on local coordinates does not automatically define a global scalar, so the choice of inputs matters.

\begin{figure}[tbp]
 \centering
 \includegraphics[width=0.9\linewidth]{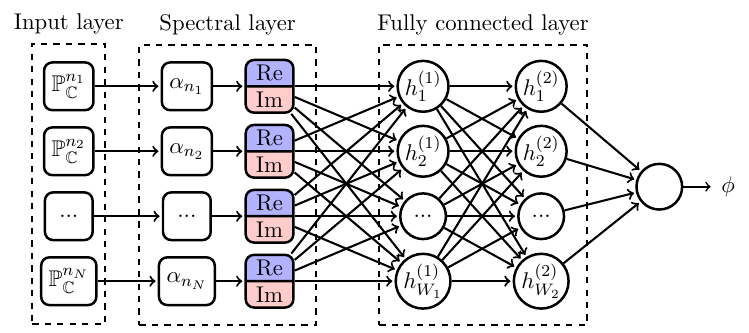}
 \caption{The spectral network produces a globally well defined potential.
 The fixed maps $\alpha_{n_i}$ supply the projective features of~\eqref{eq:projective-features}, separately for each ambient factor, to the trainable network. Reproduced from Figure~27 of~\cite{Berglund:2022gvm}.}
 \label{fig:spectral-network}
\end{figure}

\paragraph{Spectral networks:}
Berglund and collaborators~\cite{Berglund:2022gvm} ensure globality by supplying the network with functions that are invariant under projective rescaling.
For a hypersurface in $\mathbb P^N$, these are the real and imaginary components of
\be\label{eq:projective-features}
 P_{A\bar B}(z)=\frac{z_A\bar z_{\bar B}}{\sum_{C=0}^N|z_C|^2}
 \,,\qquad A,B=0,\ldots,N \,.
\ee
The matrix $P$ is unchanged by $z\mapsto\lambda z$ for any nonzero complex $\lambda$.
Its Hermitian symmetry and unit trace allow redundant inputs to be omitted.
For a product of projective spaces, the construction is applied separately to each factor.
The term \emph{spectral} refers to the projective function basis used in spectral approximations; a fixed feature layer is followed by an ordinary smooth neural network whose real output is $\Phi_\theta$.
This gives a globally defined scalar on the ambient space and hence on $X$, as illustrated in Figure~\ref{fig:spectral-network}.
The resulting model needs no separate closure or chart transition penalty, and its exact correction fixes the K\"ahler class throughout training.

In this subsection, let $\nu$ denote the volume matched form 
\be
 \nu=\left(\int_X\omega_0^n\right)\dd\nu\,,
\ee
where $\dd\nu$ is the probability measure defined in~\eqref{eq:sampling-target}.
Thus $\int_X\nu=\int_X\omega_0^n$; the distinction between the volume form $\nu$ and the probability measure $\dd\nu$ fixes the normalization in the following formulas.
Define the dimensionless volume ratio and normalized average by
\be\label{eq:potential-residual}
 \eta(\Phi)=\frac{\omega_\Phi^n}{\nu}
 \,,\qquad
 \langle f\rangle_\nu=\frac{\int_X\nu\,f}{\int_X\nu} \,.
\ee
The Monge--Amp\`ere equation is $\eta=1$, and a possible training objective is $\langle|\eta-1|^2\rangle_\nu$.
The normalization of $\nu$ is fixed by the prescribed K\"ahler class.
Automatic differentiation of $\Phi_\theta$ supplies the metric, while Monte Carlo evaluation uses the integration weights appropriate to the point sampling distribution, as in Section~\ref{sec:points}.

\begin{figure}[tbp]
 \centering
  \includegraphics[width=0.66\linewidth]{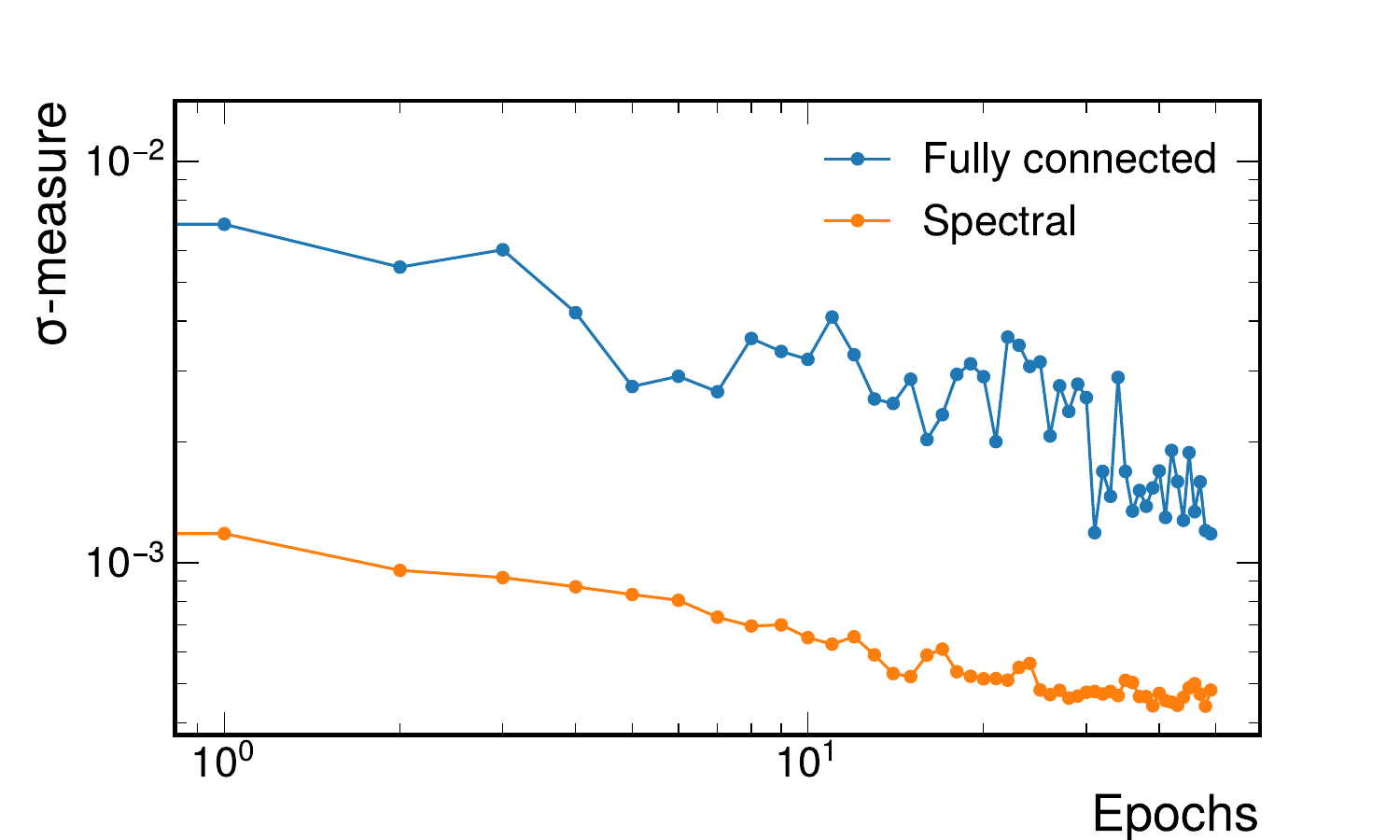}
 \caption{
 Comparison of the validation $\sigma$-measure for an ordinary fully connected network and the spectral construction on the Fermat K3 surface; the latter reaches values below $10^{-3}$ in the reported experiment.
 Reproduced from Figure~31 of~\cite{Berglund:2022gvm}.}
 \label{fig:spectral-network-performance}
\end{figure}

\paragraph{Curvature and topological validation:}
A small Monge--Amp\`ere residual ensures matching the volume form, but curvature depends on further derivatives of the learned potential.
An independent check is therefore to evaluate a topological quantity through its curvature representative.
Writing the Euler form as $e(g)=\mathcal E(g)\,\dd V_g$, the Chern--Gauss--Bonnet theorem on a smooth K3 surface gives
\be\label{eq:k3-euler-validation}
 \int_X\dd V_{g_\Phi}\,\mathcal E(g_\Phi)
 =\int_X c_2(TX,g_\Phi)=\chi(X)=24 \,.
\ee
For a potential model this requires derivatives through fourth order and probes the consistency of the metric, curvature computation, and numerical integration.
The full curvature expression must be retained for an approximate metric, including terms that vanish only at exact Ricci-flatness.
Since~\eqref{eq:k3-euler-validation} holds for every smooth global K\"ahler metric, it supplements the Ricci-flatness residual and does not replace it.

The tests in~\cite{Berglund:2022gvm} use the Cefal\`u family of quartic surfaces,
\be\label{eq:cefalu-family}
 X_\lambda=\left\{\sum_{A=0}^3z_A^4-\frac{\lambda}{3}
       \left(\sum_{A=0}^3z_A^2\right)^2=0\right\}
       \subset\mathbb P^3 \,.
\ee
Here, $\lambda$ varies the complex structure while the ambient polarization is held fixed.
The Fermat K3 surface occurs at $\lambda=0$, giving the benchmark in Figure~\ref{fig:spectral-network-performance}(b).
At $\lambda=3/4,1,3/2,3$, the surface develops respectively $8,16,12,4$ isolated $A_1$ singularities; the $\lambda=1$ member is the Kummer quartic.

At these singular fibers the Euler integral requires more care.
Write $X_s=X_\lambda\setminus\operatorname{Sing}(X_\lambda)$, let $N=|\operatorname{Sing}(X_\lambda)|$, and denote the minimal resolution by $\widetilde X_\lambda$.
The induced Fubini--Study metric obeys the identity underlying the observation in~\cite{Berglund:2022gvm,Aluffi:2019pfaffian}:
\be\label{eq:cefalu-euler-singular}
 \chi(\widetilde X_\lambda)
 =\chi(X_s)+2N
 =\int_{X_s}\dd V_{g_{\mathrm{FS}}}\,\mathcal E(g_{\mathrm{FS}})+2N
 =24 \,.
\ee
Each node is replaced by an exceptional $\mathbb P^1$ of Euler characteristic two.
The identification of the bulk curvature integral with $\chi(X_s)$ uses the behavior of the ambient induced metric near the deleted points; it is not a general Gauss--Bonnet formula without boundary terms for every metric on $X_s$.\footnote{For an orbifold metric that is smooth on the local $\mathbb C^2/\mathbb Z_2$ covers, the bulk Euler integral is instead $24-\tfrac32N$~\cite{Roulleau:2017kummer}; in particular, it vanishes for the flat Kummer orbifold with sixteen nodes.}
In the numerical experiments, spectral networks substantially improve the Euler integrals relative to the ordinary network, but appreciable discrepancies remain at some singular values, particularly near the Kummer quartic.
Thus the analogous equality for a learned metric is an observed consistency test within the numerical setup, rather than an independent proof that the singular Ricci-flat metric has been recovered.
Concentrated curvature near a degeneration also makes this test sensitive to sampling and the removal of the cutoff.

\begin{figure}[tbp]
 \centering
 \begin{subfigure}{0.78\textwidth}
  \centering
  \includegraphics[width=\linewidth]{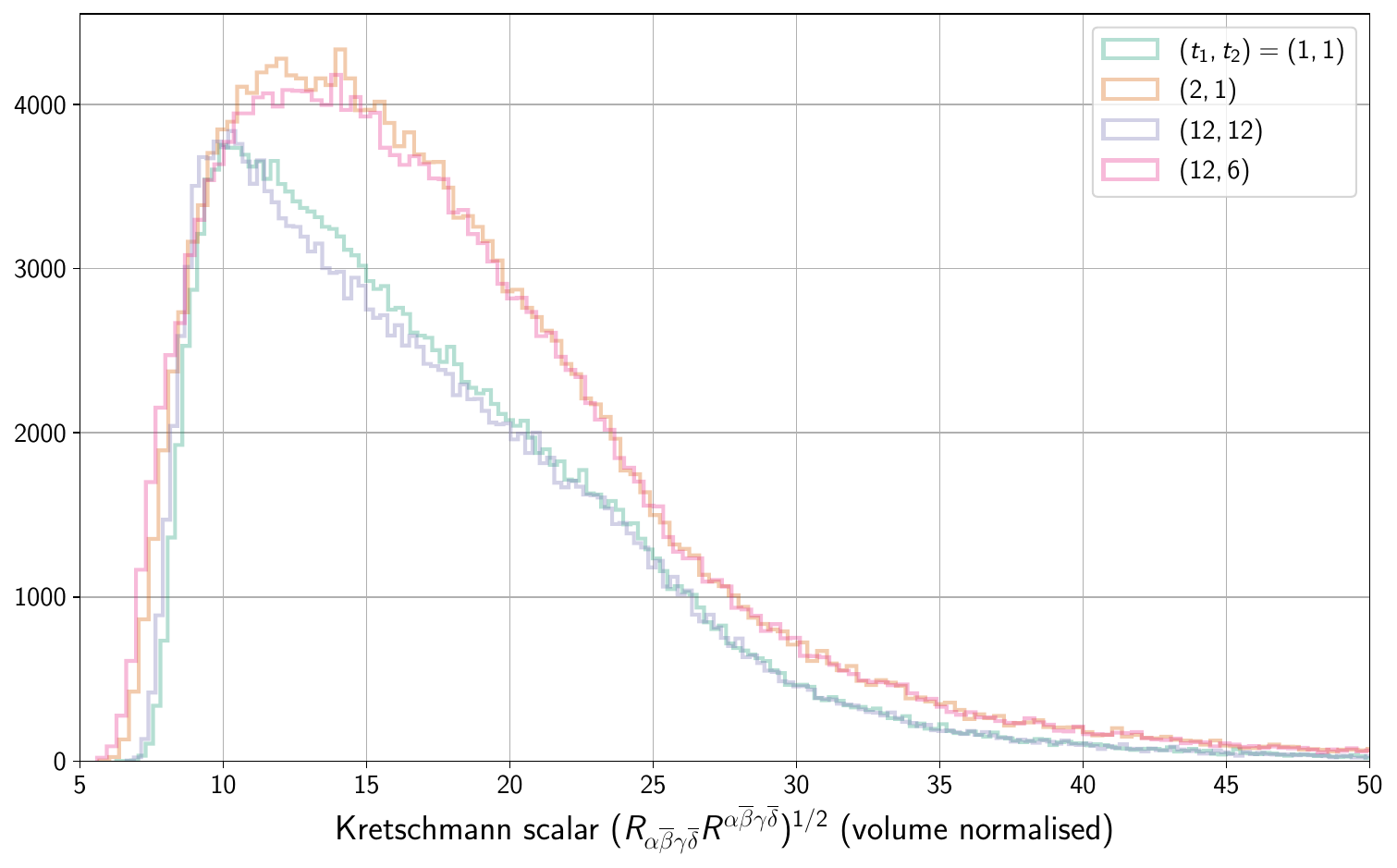}
  \caption{Four K\"ahler classes on two rays of the cone.}
 \end{subfigure}
 \par\medskip
 \begin{subfigure}{0.94\textwidth}
  \centering
  \includegraphics[width=\linewidth]{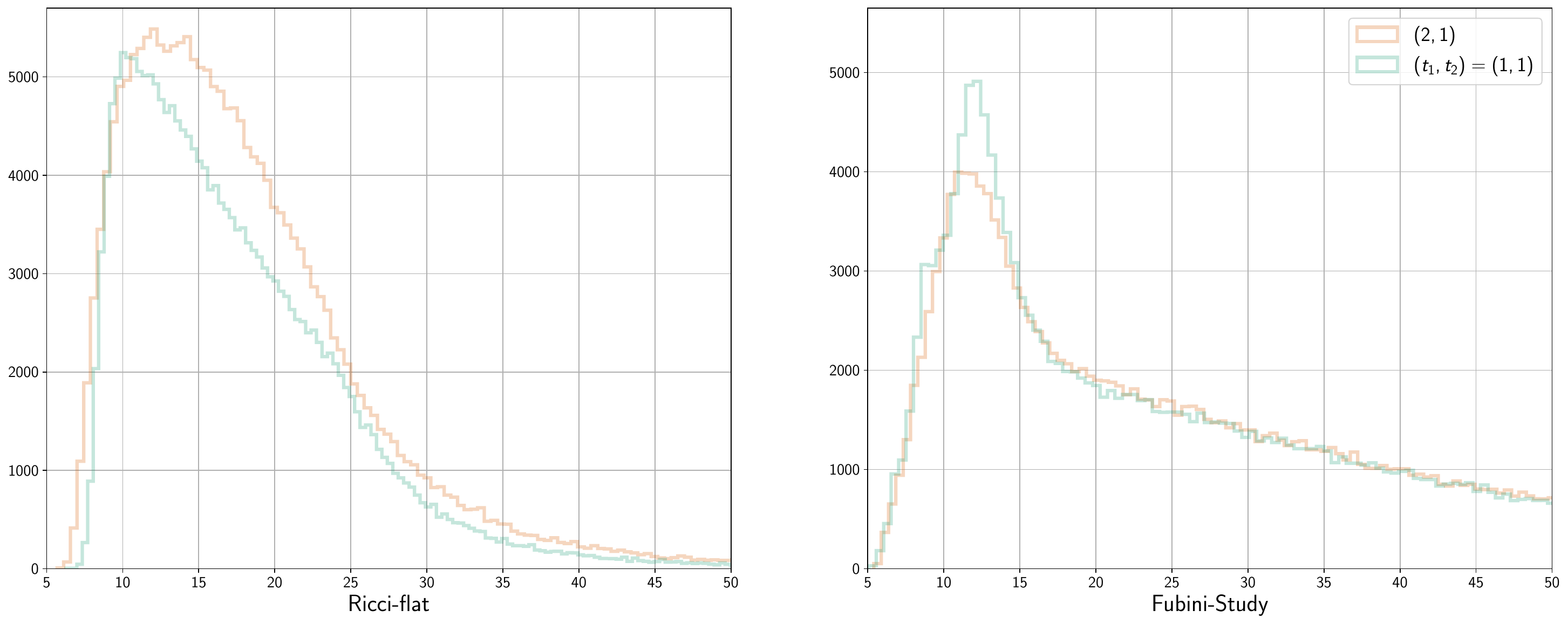}
  \caption{Learned Ricci-flat and induced Fubini--Study metrics.}
 \end{subfigure}
 \caption{Numerical Ricci-flat metrics on a bidegree $(4,2)$ hypersurface in $\mathbb P^3\times\mathbb P^1$ at fixed complex structure $\psi=0.2$.
 The histograms show the Riemann curvature norm, the square root of the contraction displayed on the horizontal axes, with the overall volume normalized; the vertical axes give counts.
 In (a), approximately coincident distributions for equal $t_1/t_2$ reflect overall metric rescaling, while the ratios $1$ and $2$ give different shapes.
 Panel (b) compares the learned metric (left) with the induced Fubini--Study metric (right) at $(t_1,t_2)=(1,1)$ and $(2,1)$.
 Reproduced from Figures~8 and~9 of~\cite{Butbaia:2024xgj}.}
 \label{fig:kahler-class-metrics}
\end{figure}

\paragraph{Varying the K\"ahler class:}
The global potential ansatz permits a separate numerical Ricci-flat metric to be learned in each prescribed K\"ahler class on the same complex manifold.
For example,~\cite{Butbaia:2024xgj} studies the bidegree $(4,2)$ hypersurface $X_\psi\subset\mathbb P^3\times\mathbb P^1$, with $(h^{1,1},h^{2,1})=(2,86)$ and $\chi=-168$.
The reference form $\omega_0=t_1\omega_1+t_2\omega_2$, where the two forms are induced from the ambient factors, specifies the class before training.
Figure~\ref{fig:kahler-class-metrics} holds the complex structure fixed at $\psi=0.2$ and compares the choices $(t_1,t_2)=(1,1),(2,1),(12,12),(12,6)$.
Changing the overall scale moves along a ray of the K\"ahler cone; changing $t_1/t_2$ changes the metric shape.
Accordingly, after volume normalization the curvature distributions approximately coincide along each ray, while they distinguish the two ratios.
The comparison with the induced Fubini--Study metrics shows how learning the Ricci-flat representative changes the local curvature within each class.
The training and residual checks establish the numerical approximation; the curvature distributions illustrate the resulting geometric differences.

\paragraph{Iterative neural corrections:}
L\"ust, Ruehle, and Schreyer demonstrate successive potential corrections~\cite{Lust:2026mys}.
Using the normalized objective above, we can formulate this refinement, starting with $\Phi^{(0)}=0$, as
\begin{align}\label{eq:iterative-potential}
 \Phi^{(m)}&=\Phi^{(m-1)}+a_m\phi_{\theta_m}\,,\nonumber\\*
 \mathcal L_m(\theta_m)
 &=\left\langle\left|\eta\!\left(\Phi^{(m-1)}+a_m\phi_{\theta_m}\right)-1\right|^2\right\rangle_\nu \,.
\end{align}
At each stage the earlier weights are frozen and only the new network is trained, using the full nonlinear residual of the accumulated metric.
The authors find $a_m=1$ sufficient; their two-stage benchmark reduces the test $\sigma$-measure from approximately $2\times10^{-2}$ to $2\times10^{-3}$, with some overfitting.
This demonstrates refinement, not a convergence theorem for arbitrarily many stages.

For implementation, freezing parameters must preserve the earlier networks' coordinate derivatives, which still enter the accumulated metric.
Each new correction should be checked for positivity and evaluated on independent validation points.
If necessary, its amplitude can be reduced before acceptance; this is a safeguard we propose, not part of the reported benchmark.
The additive constant in each potential is irrelevant and can be fixed by imposing zero mean.
An iteration should stop when validation ceases to improve, even if training loss continues to fall.

The linearization provides a useful interpretation and suggests a separate extension.
Donaldson proposed an iterative refinement of approximate Calabi--Yau metrics in which a correction to the K\"ahler potential is obtained by applying the inverse Laplacian to the volume form discrepancy~\cite{Donaldson:2005hvr}.
For a positive $\omega_m=\omega_{\Phi^{(m)}}$ and $\eta_m=\omega_m^n/\nu$, a small correction $u$ obeys
\begin{align}\label{eq:metric-linearization}
 \Delta_m u&=\frac{n\,\text{i}\partial\bar\partial u\wedge\omega_m^{n-1}}{\omega_m^n}\,,\nonumber\\*
 \frac{(\omega_m+\text{i}\partial\bar\partial u)^n}{\nu}
 &=\eta_m\left(1+\Delta_m u+O(\|\text{i}\partial\bar\partial u\|_{\omega_m}^2)\right) \,.
\end{align}
Here, the norm measures the pointwise size of the correction relative to the current Hermitian metric.
Thus, a Newton-type neural correction would solve
\be\label{eq:neural-newton}
 \Delta_m u=\eta_m^{-1}-1\,,
 \qquad
 \int_X \omega_m^n\,(\eta_m^{-1}-1)=0 \,.
\ee
The second identity follows from the fixed volume normalization and supplies the compatibility condition for this Poisson equation on compact $X$.
Training a correction network on this linear equation, then accepting a damped update only after checking positivity and the nonlinear residual, is a possible Newton-type variant.
It should be distinguished from the nonlinear training in~\eqref{eq:iterative-potential}.

\subsection{Metrics across moduli space}\label{sec:metric-moduli}

Applications often require a family of Ricci-flat metrics rather than a metric at a single point in moduli space.
Constantin, Lukas, and Nutricati~\cite{Constantin:2026bky} construct approximate K\"ahler potentials with explicit dependence on the K\"ahler parameters.
Starting from numerical metrics learned with \texttt{cymetric}, they fit the correction to a reference potential using a finite set of globally defined functions built from sections of an auxiliary ample line bundle.
They then use symbolic regression to express the fitted coefficients as functions of ratios of K\"ahler parameters, factoring out the overall volume scaling.
The examples are the bicubic in $\mathbb P^2\times\mathbb P^2$ and a hypersurface of bidegree $(2,4)$ in $\mathbb P^1\times\mathbb P^3$, both with $h^{1,1}=2$ and with the complex structure fixed at a symmetric member.
For the bicubic, the final symbolic potential has a reported average relative disagreement of approximately $2.2\%$ with the learned potential, in the potential normalization and error convention of~\cite{Constantin:2026bky}.
Relative potential errors depend on the additive and K\"ahler gauge conventions, so this percentage is not an invariant relative error of the metric.
It also does not by itself bound errors in derivatives or in physical observables.

Constantin, Lee, Lukas, and Nutricati~\cite{Constantin:2026rgm} extend this construction to simultaneous dependence on complex structure and K\"ahler parameters.
Their demonstration uses a symmetry restricted family of bicubics with one complex structure parameter and both K\"ahler parameters.
Thus ``full moduli dependence'' refers to both types of moduli within this family, rather than all $83$ complex structure moduli of a general bicubic.
Using $2{,}037$ sampled moduli points, they report an average potential disagreement of $3.2\%$, in their chosen potential normalization and error convention, together with a Monge--Amp\`ere $\sigma$ loss of approximately $0.06$~\cite{Constantin:2026rgm}.
Such constructions turn sampled numerical solutions into explicit expressions that can be evaluated within their domain of validity in moduli space; their relation to other analytic approximations is discussed in Section~\ref{sec:analytics}.
For heterotic applications, these metrics provide one ingredient in following physical couplings as the moduli vary.
Neither paper computes the associated HYM connections, harmonic matter forms, or normalized Yukawa couplings, which remain additional requirements.
The auxiliary line bundle used to parameterize a Calabi--Yau metric should in particular be distinguished from the physical gauge bundle.

\subsection{Holomorphic and bihomogeneous networks}
\label{subsec:holomorphic-networks}

The packages \href{https://github.com/yidiq7/MLGeometry}{\texttt{MLGeometry}}, \href{https://github.com/pythoncymetric/cymetric}{\texttt{cymetric}}, \href{https://github.com/ml4physics/cyjax}{\texttt{CYJAX}}, and \href{https://github.com/Justin-Tan/cymyc}{\texttt{cymyc}}, reviewed in Section~\ref{sec:calabi-yau-software}, share the strategy of optimizing a parameterized metric against geometric equations, usually through a Monge--Amp\`ere loss.
The distinction made by holomorphic networks concerns the representation of the trial metric and the constraints built into it.
In particular, the algebraic ansatz used by \texttt{CYJAX} is closely related to this construction: it learns a Hermitian form on a prescribed space of sections, whereas a holomorphic network also learns the sections entering the potential~\cite{Gerdes:2022nzr,Douglas:2021zdn}.

The construction starts from holomorphic sections of a line bundle $L$, represented by homogeneous coordinates in the simplest projective example.
Complex linear maps without biases are alternated with homogeneous polynomial activations, such as componentwise squaring.
Squaring takes sections of $L^m$ to sections of $L^{2m}$; after $r$ such activations the outputs $F_{\theta,\alpha}$ are sections of $L^k$, with $k=2^r$.
Depth therefore produces high degree sections through a structured, nonlinear parameterization, without assigning an independent coefficient to every monomial.
The available approximation space depends on both depth and width~\cite{Douglas:2021zdn}.

The network outputs determine a real potential,
\be
  K_\theta=\frac{1}{k}\log\!\left(F_\theta^\dagger H F_\theta\right)\,,
  \qquad
  g_{i\bar\jmath}=\partial_i\partial_{\bar\jmath}K_\theta\,,
  \label{eq:holomorphic-network-potential}
\ee
where $H$ is a positive definite Hermitian matrix and an overall conventional normalization is understood.
Homogeneity ensures that changes of projective representative shift $K_\theta$ only by a K\"ahler transformation.
When the sections define an embedding, the metric is the pullback of a Fubini--Study metric.
Holomorphy thus characterizes the constituent sections; the potential itself depends on both the coordinates and their conjugates~\cite{Douglas:2021zdn}.

The numerical implementation underlying \texttt{MLGeometry} favors a related \emph{bihomogeneous} network.
It first forms the real and imaginary parts of $z_i\bar z_j$, then applies real linear maps and homogeneous polynomial activations, taking the logarithm of the resulting scalar to obtain the potential.
This preserves the required projective transformation law while allowing a flexible parameterization.
Positive definiteness must still be checked for a general bihomogeneous ansatz.
In both constructions, Ricci flatness is imposed by training the metric volume form to match the appropriately normalized volume form determined by $\Omega$.
The architectural distinction therefore complements, rather than replaces, the common loss based approach~\cite{Douglas:2020hpv}.

\subsection{K\"ahler--Ricci flow}
\label{sec:kahler-ricci-flow}

Hamilton~\cite{Hamilton:1982ThreeManifoldsPositive} introduced Ricci flow, $\partial_t g_{ij}=-2R_{ij}$, as a means of relating the geometry and topology of manifolds.
A central innovation of Perelman~\cite{Perelman:2006un} was to express Ricci flow, modulo diffeomorphisms, as the gradient flow of
\be
    \mathcal F(g,f)=\int_M \dd V_g\, (R+|\nabla f|^2)e^{-f}\,,
\ee
where, with fixed weighted measure $\dd m=e^{-f}\dd V_g$, its Euler--Lagrange equation is $R_{ij}+\nabla_i\nabla_j f=0$, and the associated gradient evolution is
\be
 \partial_t g_{ij}=-2(R_{ij}+\nabla_i\nabla_j f)
 \,,\qquad
 \partial_t f=-\Delta f-R \,.
\ee
After a time dependent diffeomorphism, these equations become ordinary Ricci flow coupled to $\partial_t f=-\Delta f+|\nabla f|^2-R$, or equivalently the backward conjugate heat equation $\partial_t u=-\Delta u+Ru$ for $u=e^{-f}$.
With $f=2\Phi$, the functional takes the familiar metric--dilaton form
\be
    \mathcal F=\int_M \dd V_g\, e^{-2\Phi}(R+4|\nabla\Phi|^2) \,,
\ee
where the target space scalar $\Phi$ is the dilaton coupled to the worldsheet curvature.
Perelman~\cite{Perelman:2006un,Perelman:2006up} developed monotone entropy functionals and reduced volume, proved the absence of local collapsing at scales with controlled curvature, and used these tools to analyze singularities and continue the flow by surgery, cutting along suitable necks and capping the resulting boundaries.
His analysis of the flow with surgery and its long time behavior established Thurston's geometrization conjecture: after prime decomposition and cutting along suitable incompressible tori, every closed orientable three manifold decomposes into pieces admitting one of Thurston's eight model geometries.
For a closed simply connected three manifold, Perelman's finite extinction theorem implies that the manifold is diffeomorphic to $S^3$, proving the Poincar\'e conjecture~\cite{Perelman:2003uq}.
A physical counterpart of this geometric evolution had already appeared in Friedan's study of the renormalization group for nonlinear sigma models~\cite{Friedan:1980jf,Friedan:1980jm}.
In the two-dimensional sigma model describing the string worldsheet, with the target space metric as the only background coupling, its one loop beta function is proportional to the target space Ricci tensor, so evolution toward the infrared reproduces Ricci flow after a rescaling of flow time.
This correspondence holds at leading order in $\alpha'$, with higher orders introducing curvature corrections.

K\"ahler--Ricci flow provides a dynamical construction of the Ricci-flat metric in a prescribed K\"ahler class.
Let $X$ be a smooth connected compact Calabi--Yau manifold of complex dimension $n$, with a fixed complex structure and initial K\"ahler form $\omega_0$.
For a projective embedding, a natural initial condition is the restriction of a Fubini--Study form.
With $\omega_t=i g_{a\bar b}(t)\,\dd z^a\wedge \dd\bar z^{\bar b}$, our conventions are
\be
 \frac{\partial\omega_t}{\partial t}=-\rho(\omega_t)
 \,,\qquad
 \rho(\omega_t)=-i\partial\bar\partial\log\det g(t)
 \,,\qquad \omega_{t=0}=\omega_0 \,.
 \label{eq:krf-geometric}
\ee
Since $[\rho]=2\pi c_1(X)=0$, the class $[\omega_t]=[\omega_0]$ and the volume $V=\int_X\omega_0^n/n!$ remain fixed.
No additional volume normalization term is needed.
Cao's theorem ensures that this flow exists smoothly for all $t\geq0$ and converges to the unique Ricci flat K\"ahler metric in $[\omega_0]$~\cite{Cao:1985DeformationKahlerMetrics}.
Ricci flatness does not imply vanishing of the full Riemann curvature tensor; the limiting metric is flat on a complex torus, but generally curved on a K3 surface or a Calabi--Yau threefold.

The fixed class allows the evolution to be expressed through a global real potential, $\omega_t=\omega_0+i\partial\bar\partial\varphi_t$.
For a nowhere vanishing holomorphic $n$ form $\Omega$, define the probability measure
\be
 \dd\nu=\frac{i^{n^2}\Omega\wedge\bar\Omega}
 {\int_X i^{n^2}\Omega\wedge\bar\Omega}
 \,,\qquad
 \eta_t=\frac{\omega_t^n/n!}{V\,\dd\nu} \,.
 \label{eq:krf-volume}
\ee
The flow then becomes a parabolic complex Monge--Amp\`ere equation~\cite{Tosatti:2015KAWA},
\be
 \frac{\partial\varphi_t}{\partial t}
 =\log\eta_t+c(t)
 \,,\qquad
 \varphi_0=0
 \,,\qquad
 \omega_0+i\partial\bar\partial\varphi_t>0 \,.
 \label{eq:krf-potential}
\ee
The function $c(t)$ fixes the freedom to add a spatial constant to the potential and has no effect on the metric.
Applying $i\partial\bar\partial$ recovers~\eqref{eq:krf-geometric}, because $\Omega$ is holomorphic and nowhere zero.
At a stationary metric, $\log\eta$ is pluriharmonic and hence constant on compact $X$; the volume constraint then gives $\eta=1$.
Thus, the endpoint solves precisely the Calabi--Yau equation.
The parabolic character is also explicit: in the gauge $c(t)=0$, differentiating once gives $\partial_t\dot\varphi=\Delta_t\dot\varphi$, where $\Delta_t=g^{a\bar b}(t)\partial_a\partial_{\bar b}$.
This heat equation describes the smoothing of the volume discrepancy, while the global convergence theorem requires uniform geometric estimates in addition to the maximum principle.

Noting that neural networks minimize loss functions via gradient flow, the idea of using K\"ahler--Ricci flow to construct the Ricci-flat Calabi--Yau metric was advanced in~\cite{Jejjala:2020wcc}.
After fitting the initial Fubini--Study metric, each step constructs the Euler target
\be
 g^{(k+1)}_{a\bar b}
 =g^{(k)}_{a\bar b}-\Delta t\,R_{a\bar b}[g^{(k)}] \,.
 \label{eq:krf-euler}
\ee
The network is retrained against this target, with losses for first and second derivatives as well as metric components.
Such derivative control matters because curvature depends on second derivatives of the metric.
Their explicit flow illustration uses a conventional differential equation solver on a real torus; neural flow studies on the complex torus and K3 are described as preliminary.
The principal Calabi--Yau results in that paper instead use direct minimization of geometric losses.
Consequently, the proposed flow algorithm and the demonstrated accuracy of those learned metrics should be assessed separately.

For a numerical implementation, the potential formulation offers useful structural control.
A smooth global network correction $\varphi_\theta$ preserves the K\"ahler class and closure of $\omega_\theta$ exactly, while positivity must still be enforced or checked.
Equation~\eqref{eq:krf-potential} involves only second spatial derivatives of the potential, whereas evaluating the Ricci tensor from it requires fourth derivatives.
Finite time steps and imperfect network fits introduce errors not contemplated in the continuum theorem.
Reducing the step size, controlling the fit between successive times, and checking positive definiteness are therefore essential to establish that a computation approximates the intended flow.
The geometric time $t$ is distinct from the optimizer's training time.

Halverson and Ruehle make the latter distinction precise by studying the metric evolution induced by gradient descent~\cite{Halverson:2023ndu}.
Writing $A,B$ for independent real metric component labels and $\tau$ for training time gives
\be
 \begin{aligned}
 \frac{\partial g_A(x)}{\partial\tau}
 &=-\int_X \dd\nu(y)\,\Theta_{AB}(x,y;\theta)
 \frac{\delta\mathcal L}{\delta g_B(y)} \,,\\*
 \Theta_{AB}(x,y;\theta)
 &=\sum_I\frac{\partial g_A(x)}{\partial\theta_I}
 \frac{\partial g_B(y)}{\partial\theta_I} \,.
 \end{aligned}
 \label{eq:krf-ntk}
\ee
The functional derivative here is defined relative to $\dd\nu$.
The metric neural tangent kernel $\Theta$ generally evolves during training, couples different points, and mixes metric components.
An appropriate infinite width limit can freeze the kernel, but obtaining a local geometric flow requires further architectural assumptions.
With locality, suitable component structure, and an appropriate functional, the authors recover Perelman's formulation of Ricci flow.
In that formulation, an auxiliary scalar $f$ enters $\mathcal F(g,f)=\int_X \dd V_g\,(R+|\nabla f|^2)e^{-f}$.
Its gradient flow at fixed weighted measure is equivalent to Ricci flow through a time dependent diffeomorphism~\cite{Perelman:2006un}.
Ordinary descent on a volume residual therefore need not trace a K\"ahler--Ricci flow trajectory, even when both procedures seek the same endpoint.

On the Fermat quintic, the fixed kernels tested by Halverson and Ruehle fit training points well but generalize less effectively than finite networks whose kernels evolve~\cite{Halverson:2023ndu}.
This supports the importance of feature learning in those experiments, rather than a general obstruction to numerical Ricci flow.
For either approach, validation should combine the volume discrepancy $\sigma=\int_X \dd\nu\,|1-\eta|$ with curvature and positivity checks on independent points.
A small integrated volume error alone does not control its derivatives and hence does not guarantee small Ricci curvature.
When claiming to follow the flow, one should evaluate its flow equation residual and show that the results converge as the time step is reduced and the sampling density is increased. 

\subsection{Point sampling revisited}

\subsubsection{From complete intersections to toric hypersurfaces}

For a complete intersection in a product of projective spaces, one intersects its defining equations with $n$ independent random hyperplanes, distributed among the factors so that the intersection is finite and nonempty.
The resulting measure is a mixed wedge product of restricted Fubini--Study forms.
Larfors, Lukas, Ruehle, and Schneider extend this construction to toric Calabi--Yau hypersurfaces in \texttt{cymetric}~\cite{Larfors:2022nep}.
Base point free nef line bundles $L_\alpha$ on the toric ambient space supply monomial sections and projective maps
\be
 \Phi_\alpha=[s_0^{(\alpha)}:\cdots:s_{r_\alpha}^{(\alpha)}] \,,
 \qquad
 J_\alpha=\frac{\text{i}}{2\pi}\partial\bar\partial
       \log\sum_j|s_j^{(\alpha)}|^2 \,.
 \label{eq:toric-sampling-forms}
\ee
Intersecting the hypersurface with $n$ random sections having independent complex Gaussian coefficients produces
\be
 \dd\rho_{\boldsymbol\alpha}
  =\frac{J_{\alpha_1}\wedge\cdots\wedge J_{\alpha_n}|_X}
  {\int_XJ_{\alpha_1}\wedge\cdots\wedge J_{\alpha_n}} \,.
 \label{eq:toric-sampling-measure}
\ee
The chosen intersection must have positive degree.
Relations among section coordinates must be retained: the projective image is generally not a complete intersection.
Toric sampling therefore involves a polynomial system with the appropriate quotient and patch data.
Individual $J_\alpha$ can be semipositive, and the mixed measure need not equal the volume of $J=\sum_\alpha t^\alpha J_\alpha$.
Integration still uses the ratio of the target measure to the actual sampling measure.

The established random section implementation in \texttt{cymetric} combines Sage preprocessing of the toric data with Mathematica point generation, producing the local pullbacks and weights needed downstream~\cite{Cymetric:ndCode}.
The geometric extension thus preserves the random section principle while changing the equations, charts, and sampling density.
For a mixture of sampling prescriptions, write $\dd\rho=\sum_A\pi_A \dd\rho_A$, where $\pi_A$ is the fraction of points drawn from prescription $A$.
A pooled estimator can use $\dd\nu/\dd\rho$; using a single Fubini--Study density for this mixture would generally give the wrong integral.

\subsubsection{Improving the sampling distribution}

Correct weights remove bias from a mismatched sampling measure, but cannot compensate for regions represented by very few points at finite sample size.
To see the statistical issue, write $\dd\nu=p\,\dd\lambda$ and $\dd\rho=q\,\dd\lambda$ relative to a common reference measure.
For a real valued $f$, let $I=\int_X \dd\nu\,f$ and $\widehat I=N^{-1}\sum_i f(p_i)p(p_i)/q(p_i)$.
For independent points, adequate proposal support, and finite second moment,
\be
 \operatorname{Var}(\widehat I)
 =\frac{1}{N}\left[\int_X \dd\lambda\,\frac{f^2p^2}{q}-I^2\right] \,.
 \label{eq:sampling-variance}
\ee
Large density ratios can therefore cause noisy quadrature and stochastic gradients.
Sampling directly from $\dd\nu$ makes the volume weights constant, although the proposal that minimizes the variance for a particular integrand is proportional to $|f|p$.
There is consequently no universally optimal sampler for all observables and all stages of training.

Sampling with respect to $\Omega\wedge\bar\Omega$ is possible because its density is known analytically.
The current \texttt{cymetric} repository~\cite{Cymetric:ndCode} provides MCMC alternatives for both CICYs and toric hypersurfaces.
These routines target the holomorphic volume measure using tangent space proposals, Newton projection onto the defining equations, and a Metropolis acceptance step based on the holomorphic volume density.
Pilot chains tune the step size and thinning interval, followed by an initial equilibration period and sampling.
The dataset interface assigns a common integration weight.
This changes the generated point distribution, whereas the random section workflow retains its proposal distribution and supplies variable integration weights.
MCMC samples are correlated, so convergence and autocorrelation must be assessed; sensitivity to the finite proposal step and numerical projection should also be checked before interpreting the samples as target volume data.

An alternative is rejection sampling: given a proposal $\dd\rho$ and a bound $M\geq\sup_X(\dd\nu/\dd\rho)$, accepting a proposed point with probability $(\dd\nu/\dd\rho)/M$ gives the exact target law.
A large bound makes this inefficient; replacing the supremum by an unchecked sample maximum can bias the result.
Resampling a finite weighted cloud is another option, but duplicates points and cannot supply regions absent from the original cloud.

Headrick and Nassar~\cite{Headrick:2009jz} pursue target volume sampling using a thin region around the hypersurface: they approximate its delta function measure by a finite width region $|Q|<\epsilon$ in coordinate patches, accept points in that region, and project them onto the hypersurface.
Their construction has an $O(\epsilon^2)$ systematic error, with acceptance also scaling as $O(\epsilon^2)$; it illustrates the tradeoff between distributional accuracy and point generation cost.
Cui and Gray~\cite{Cui:2019uhy} use this approach in numerical metric computations.
Such finite width sampling should be distinguished from exact rejection on points already lying on $X$.

An alternative is to improve the proposal while retaining random section sampling.
Keller and Lukic~\cite{Keller:2009vj} construct improved point sets from several Fubini--Study metrics.
Their partition assigns a region to the metric whose normalized ratio of target volume to proposal volume is closest to one, and retains points generated by that metric only within its assigned region.
Further metrics are introduced near extrema of the volume ratio.
This adapts the sampling density to the target geometry without first solving for the Ricci flat metric.

L\"ust, Ruehle, and Schreyer~\cite{Lust:2026mys} implement this strategy with potentials proportional to $\log(s^\dagger\lambda^{-1}s)$. Starting with samples from the standard Fubini--Study measure, they identify points $x$ (represented by homogeneous coordinates) with unusually large or small masses $m(x)$ (defined as the normalized volume ratio relative to the reference Fubini--Study measure), and construct additional sampling proposals that assign unit mass to those points.
For linear sections their adjustment is
\be
 \lambda_x=\frac{I+\epsilon P_x}{1+\epsilon} \,,\qquad
 P_x=\frac{xx^\dagger}{x^\dagger x} \,,\qquad
 \epsilon=m(x)^{1/n}-1 \,,
 \label{eq:adapted-fs-sampling}
\ee
where $x$ is the homogeneous coordinate vector and $m$ is the normalized volume ratio relative to the reference Fubini--Study measure.
Cholesky transformations of Gaussian draws generate the adjusted ensembles.
They discuss regional rejection and independent pilot samples for choosing the regions.
These proposals approximate the target volume distribution; they do not make it exactly uniform.
Their calculations near a conifold also show that denser sampling of difficult regions can reveal numerical cancellation errors in curvature observables, so better coverage must be accompanied by accuracy checks.

For regional rejection, the retained density includes the region indicator and its acceptance normalization.
For pooled, unrejected ensembles it is the mixture density.
These give different quadrature weights.
Reducing the spread of weights is useful, but does not by itself establish correct integration if the density of the retained sample has been misidentified.

\paragraph{A torus benchmark:}
The Fermat cubic $Z_0^3+Z_1^3+Z_2^3=0$ provides a useful test because its Ricci flat metric is explicitly flat after uniformization.
If $u$ is the coordinate on a fundamental cell of $\mathbb C/(\mathbb Z+\tau\mathbb Z)$, then $\Omega=C\,\dd u$.
Drawing $u=a+b\tau$ with independent uniform $a,b\in[0,1)$ and mapping back to the cubic therefore samples $\dd\nu$.
We compare this \emph{flat sampling} with \emph{Fubini--Study sampling}, obtained by intersecting random lines in $\mathbb P^2$ with the cubic~\cite{Berglund:unpubPointSelection}.

In both experiments we use $100{,}000$ points and a spectral network with three layers of $500$ neurons each, starting from the Fubini--Study metric.
Changing the point distribution is a separate choice from changing the initial metric: the run using flat sampling also starts from the same reference geometry.

\begin{figure}[tbp]
\centering
\begin{minipage}[b]{.43\textwidth}
 \centering\textsf{Fubini--Study sampling}\\[3pt]
 \includegraphics[width=\linewidth]{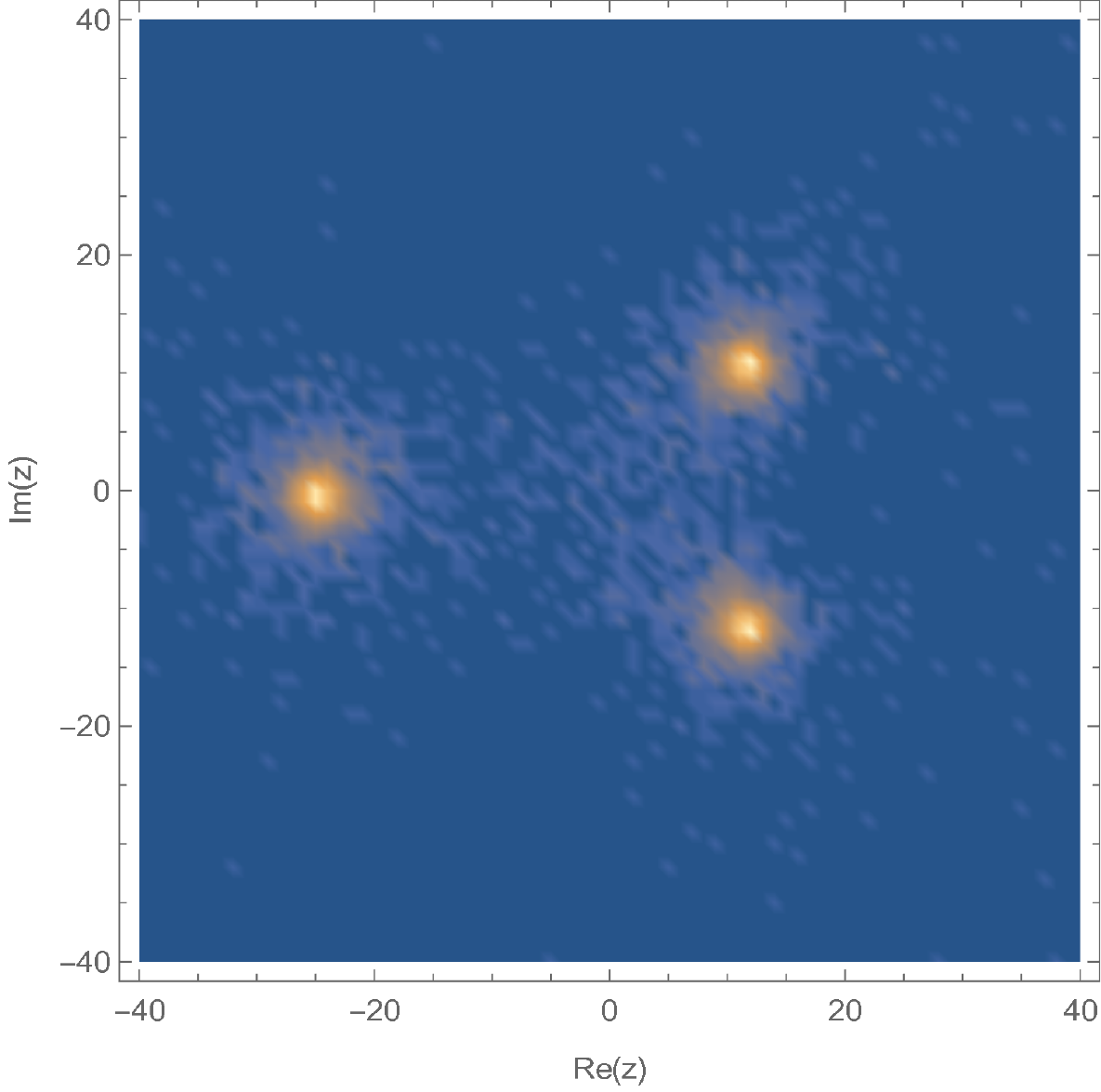}
\end{minipage}\hfill
\begin{minipage}[b]{.43\textwidth}
 \centering\textsf{Flat sampling}\\[3pt]
 \includegraphics[width=\linewidth]{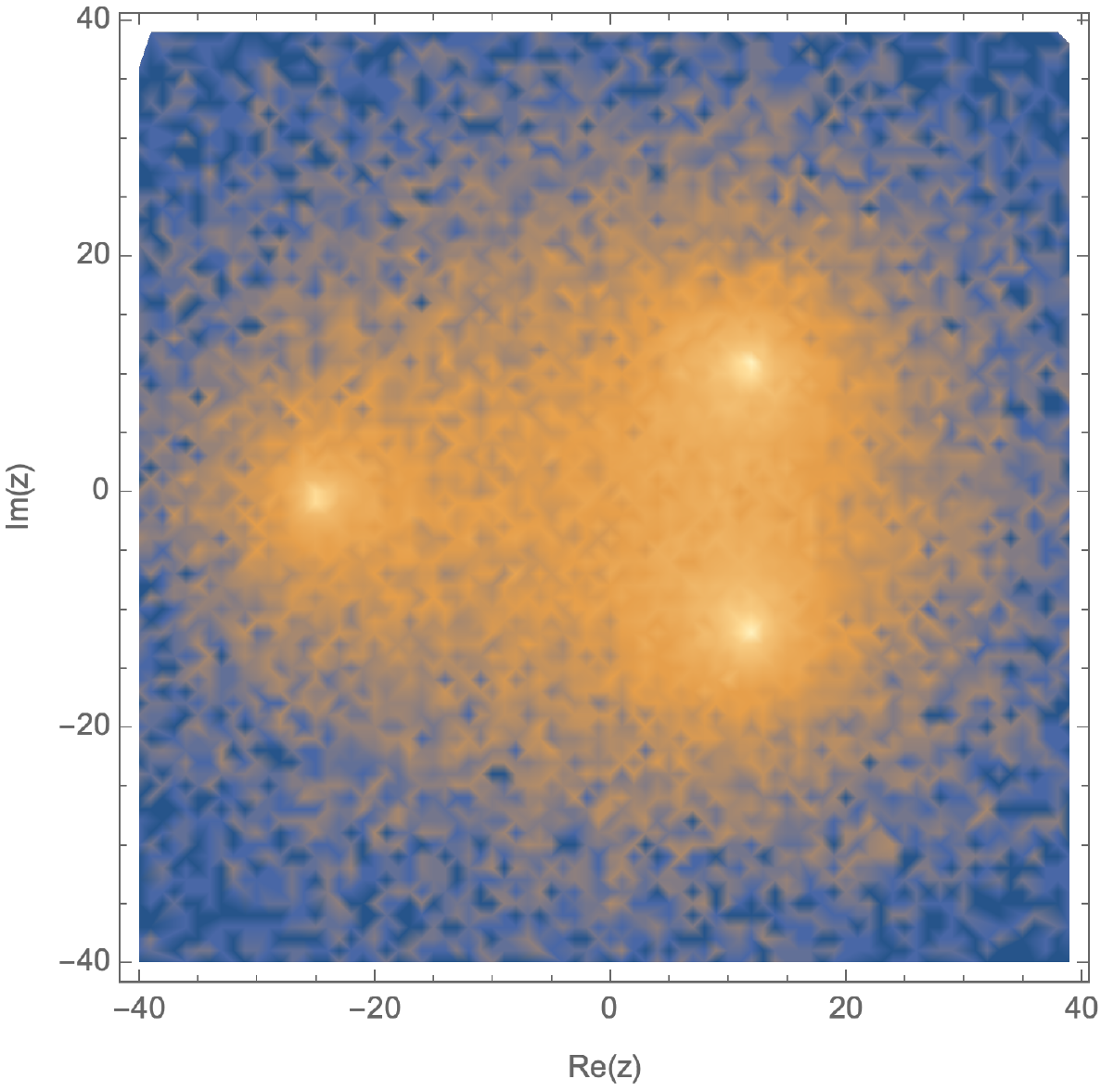}
\end{minipage}\hfill
\includegraphics[width=.065\textwidth]{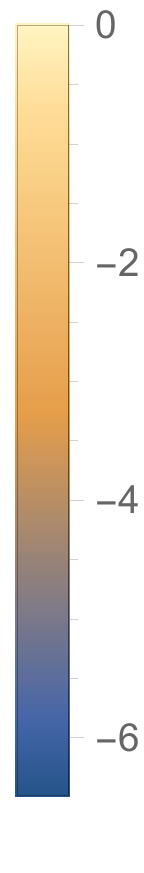}
\caption{\small Point density on a coordinate patch of the Fermat cubic for Fubini--Study sampling (left) and flat sampling (right).
The common color key encodes density, with warmer colors indicating larger values.
Fubini--Study points are obtained by random line intersections in $\mathbb P^2$; flat points are sampled uniformly in the fundamental cell and mapped back to the cubic.
Uniformity with respect to either intrinsic measure does not imply uniform coordinate density.
Reproduced from~\cite{Berglund:unpubPointSelection}.}
\label{fig:torus-sampling-density}
\end{figure}

Figure~\ref{fig:torus-sampling-density} shows the resulting point densities on a common coordinate patch of the cubic.
The Fubini--Study sample is strongly concentrated near three locations in this patch, whereas flat sampling populates a broader part of the displayed region.
Flat sampling need not look uniform in this coordinate: the uniform coordinate is $u$, and its density in a local coordinate $z$ includes the Jacobian $|\dd u/\dd z|^2$.
The maps thus visualize the distinction between intrinsic volume measures and coordinate density.

\begin{figure}[tbp]
\centering
\begin{minipage}[b]{.47\textwidth}
 \centering\textsf{Fubini--Study sampling}\\[3pt]
 \includegraphics[width=\linewidth]{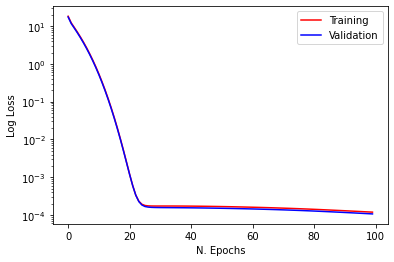}
\end{minipage}\hfill
\begin{minipage}[b]{.47\textwidth}
 \centering\textsf{Flat sampling}\\[3pt]
 \includegraphics[width=\linewidth]{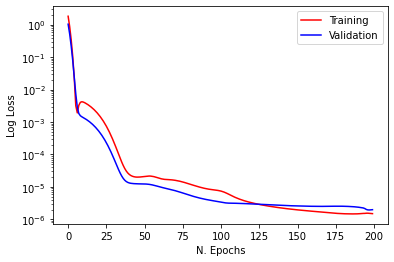}
\end{minipage}
\caption{\small Reported training (red) and validation (blue) $\sigma$ losses for the two torus sampling schemes, with both runs weighted and validated against the same measure.
Both experiments use $100{,}000$ points, three spectral network layers with $500$ neurons per layer, and Fubini--Study initialization.
The vertical axes are logarithmic; the horizontal axes cover different numbers of epochs.
Flat sampling attains a lower loss late in training in these experiments.
Reproduced from~\cite{Berglund:unpubPointSelection}.}
\label{fig:torus-sampling-loss}
\end{figure}

The training and validation curves in Figure~\ref{fig:torus-sampling-loss} show the reported $\sigma$ loss used in these experiments.
Both runs are weighted and validated against the same measure.
Flat sampling reaches a smaller loss late in training: the terminal values are of order $10^{-4}$ for Fubini--Study sampling and $10^{-6}$ for flat sampling, in the convention of the supplied curves.
The displayed runs extend to approximately $100$ and $200$ epochs, respectively, so the terminal comparison is not a benchmark at equal computational cost.
The lower loss with flat sampling illustrates its potential benefit in this benchmark, with the architecture, initial metric, and weighting measure held fixed.
Improved volume form matching does not by itself bound curvature errors.

The torus is unusually favorable because uniformization provides an exact target volume sampler.
On higher dimensional Calabi--Yau manifolds, the corresponding practical objective is to improve coverage and control weight fluctuations using the algebraically known $\Omega$ measure.
Adaptive proposals may additionally emphasize regions of large residual or curvature variation, while retaining coverage of the whole manifold.
If the intended loss remains an integral against $\dd\nu$, the weights must follow every change of sampling proposal.
Independent validation against a common target measure then separates a better metric approximation from a change in the sampling error.

\section{Bundle computations and Yukawa couplings}
\label{sec:bundle-computations}

\subsection{General remarks on Yukawa couplings}
\label{sec:yukawa}

Consider the $E_8\times E_8$ heterotic string compactified on a Calabi--Yau threefold $X$.
We work at leading order in the Calabi--Yau approximation and focus on the gauge background in the visible $E_8$ factor.
Its structure group $H\subset E_8$ determines the surviving four-dimensional gauge group, the commutant $G=C_{E_8}(H)$.
If the background in the second $E_8$ factor is trivial, the resulting gauge group is $G\times E_8$.
The internal components of the ten-dimensional gauge field supply charged scalars, while the gaugino supplies their fermionic partners.
Their Yukawa interactions descend from the gauge interaction of these fields.
Thus both the matter representations and their couplings are inherited from the $E_8$ gauge theory~\cite{Candelas:1985en,Strominger:1985ks}.

\paragraph{Holomorphic bundles and supersymmetry:}
For $H=SU(n)$, let $V$ denote the associated rank $n$ vector bundle in the fundamental representation, with trivial determinant.
Unbroken $\mathcal N=1$ supersymmetry requires its unitary connection $A$ to obey the Hermitian Yang--Mills (HYM) equations,
\be
 F_A^{0,2}=0 \,,\qquad \Lambda_\omega F_A=0 \,,
 \label{eq:standard-HYM}
\ee
where $\Lambda_\omega$ contracts the curvature with the K\"ahler form; unitarity also gives $F_A^{2,0}=0$.
The first equation makes $\bar\partial_A^2=0$.
It therefore equips $V$ with a holomorphic structure, equivalently local holomorphic frames related by holomorphic transition functions.
This is why a holomorphic bundle is required: it implements part of the supersymmetry conditions and provides the Dolbeault complex whose cohomology describes the massless chiral fields.
An infinitesimal mode satisfies $\bar\partial_A a=0$, and changes $a\mapsto a+\bar\partial_A s$ are gauge equivalences.
The resulting classes lie in $H^1(X,V)$, or in the corresponding cohomology of another associated bundle (\textit{e.g.}, $V^*$, $\wedge^2 V$).

Holomorphicity alone does not solve the second equation in~\eqref{eq:standard-HYM}.
For a fixed K\"ahler class, a holomorphic bundle admits a HYM connection precisely when it is slope polystable, \textit{i.e.}, it is a direct sum of stable bundles with the same slope~\cite{Donaldson:1985zz,Uhlenbeck:1986ntc}.
Here, stability means that every holomorphic subbundle $W\subset V$ with
$0<\operatorname{rk}W<\operatorname{rk}V$ satisfies
$\mu_\omega(W)<\mu_\omega(V)$, where
\be
 \mu_\omega(W)
 =\frac{1}{\operatorname{rk}W}
   \int_X c_1(W)\wedge\omega^2 \,.
\ee
For an $SU(n)$ bundle the slope is zero, so all stable summands must have zero slope.
The HYM connection determines the bundle Laplacian and selects the harmonic representatives of the matter classes.
Its Hermitian bundle metric is therefore needed for their kinetic normalization.

\paragraph{From the bundle to the GUT:}
For the $SU(n)$ structure groups considered here, decomposing the adjoint of $E_8$ under $G\times H$ gives
\be
 \bm{248}\longrightarrow
 (\operatorname{Adj}_G,\bm{1})\oplus(\bm{1},\operatorname{Adj}_H)
 \oplus\bigoplus_i(R_i,r_i) \,.
 \label{248->...}
\ee
If $V_i$ is the vector bundle associated to the representation $r_i$ of $H$, then $H^1(X,V_i)$ counts chiral multiplets transforming in the representation $R_i$.
The same decomposition supplies infinitesimal bundle deformations from $H^1(X,\operatorname{End}_0V)$; their extension to finite moduli can be obstructed.
For full structure group $SU(n)$ in the usual embeddings, the principal choices are~\cite{Slansky:1981yr,Blesneag:2015pvz}
\be
 \begin{array}{c|c|c}
 \operatorname{rk}V & H & G\\ \hline
 3 & SU(3) & E_6\\
 4 & SU(4) & SO(10)\\
 5 & SU(5) & SU(5)
 \end{array} \,.
 \label{eq:bundle-gut-groups}
\ee
We use the conventional GUT name $SO(10)$; its spinor $\bm{16}$ is a representation of the covering group $\operatorname{Spin}(10)$.
The rank of $V$ in this table is its fiber dimension, and the two $SU(5)$ factors in the last row have distinct roles.
The actual structure group determines the commutant: reducing it, for example by splitting the bundle, can enlarge the surviving gauge symmetry.

For $E_6\times SU(3)\subset E_8$, the branching is
\be
 \bm{248}\longrightarrow
 (\bm{78},\bm{1})\oplus(\bm{1},\bm{8})
 \oplus(\bm{27},\bm{3})
 \oplus(\overline{\bm{27}},\overline{\bm{3}}) \,.
 \label{eq:standard-branching}
\ee
The $\bm{27}$ admits a symmetric cubic invariant.
Under $SO(10)\times U(1)\subset E_6$, it decomposes as
$\bm{27}=\bm{16}_{1}\oplus\bm{10}_{-2}\oplus\bm{1}_{4}$, so its cubic includes the $\bm{16}\,\bm{16}\,\bm{10}$ matter coupling and the $\bm{10}\,\bm{10}\,\bm{1}$ coupling~\cite{Slansky:1981yr}.
The conjugate representation admits the corresponding conjugate invariant.
On the bundle side, the $SU(3)$ invariant contracts three factors of $V$ into $\wedge^3V\simeq\mathcal O_X$, permitting the $\bm{27}^3$ coupling; the dual contraction permits $\overline{\bm{27}}^3$.

Changing to a rank four bundle with structure group $SU(4)$ gives $SO(10)$ matter in $\bm{16}$, $\overline{\bm{16}}$, and $\bm{10}$, counted by $H^1(X,V)$, $H^1(X,V^*)$, and $H^1(X,\wedge^2V)$.
The contraction $V\otimes V\otimes\wedge^2V\to\wedge^4V\simeq\mathcal O_X$ produces the $\bm{16}\,\bm{16}\,\bm{10}$ Yukawa coupling.
For a rank five bundle with structure group $SU(5)$, $H^1(X,V)$ supplies $\bm{10}$ multiplets and $H^1(X,\wedge^2V)$ supplies $\overline{\bm{5}}$ multiplets of the visible $SU(5)$; the dual bundles supply their conjugates.
The invariant contractions then allow the familiar $\bm{10}\,\bm{10}\,\bm{5}$ and $\bm{10}\,\overline{\bm{5}}\,\overline{\bm{5}}$ couplings~\cite{Blesneag:2015pvz}.

\paragraph{Holomorphic couplings and physical normalization:}
Let $\Phi^I$ denote the massless chiral superfields and $a_I$ their internal $(0,1)$ forms.
Their effective $\mathcal N=1$ interactions are organized by the K\"ahler potential $K$, superpotential $W$, and gauge kinetic function $f_{AB}$.
Suppressing gravitational couplings, the action can be written in flat superspace as
\be
 \begin{aligned}
 S_\text{rigid}={}&\int\mathrm d^4x\,
 \bigg[\int\mathrm d^4\theta\,K(\Phi,\bar\Phi;\mathscr V)\\
 &+\left\{\int\mathrm d^2\theta\,
 \left(W(\Phi)+\frac14 f_{AB}(\Phi)\,
 \mathcal W^{A\alpha}\mathcal W^B_\alpha\right)
 +\mathrm{h.c.}\right\}\bigg] \,.
 \end{aligned}
 \label{eq:action}
\ee
Here, $\mathscr V$ denotes the vector superfields and $\mathcal W^A_\alpha$ their field strengths; dependence on the neutral moduli is implicit.
The K\"ahler potential is real and includes the gauge completion of the matter kinetic terms, while $W$ and $f_{AB}$ are holomorphic.
Gauge invariance constrains $W$; if an $R$ symmetry is imposed with $R(\theta)=1$, then $W$ must have $R$ charge $2$.
The gauge Chern--Simons contribution to the heterotic superpotential yields cubic terms
\be
 W\supset\frac{1}{6}\kappa_{IJK}\Phi^I\Phi^J\Phi^K \,,
 \label{eq:suppot}
\ee
which encode the interactions of two matter fermions with a scalar.
Suppressing a common normalization, their coefficients are
\be
 \kappa_{IJK}
 =\int_X\Omega\wedge
   \mathcal C\big(a_I\wedge a_J\wedge a_K\big) \,,
 \label{eq:yuk1}
\ee
where $\mathcal C$ is the invariant contraction of the bundle indices inherited from the $E_8$ Lie bracket.
It maps the product of associated bundles to $\mathcal O_X$, while the corresponding visible representations must combine to a gauge singlet.
These conditions allow a coupling but do not guarantee that its overlap integral is nonzero.
The integral depends only on the cohomology classes of the three forms, so it can be evaluated without finding harmonic representatives.
It can nevertheless vary with complex structure and bundle moduli~\cite{Strominger:1985ks,Candelas:1987se,Blesneag:2015pvz,Blesneag:2016yag}.

The coefficients $\kappa_{IJK}$ are unnormalized: the fields in~\eqref{eq:suppot} need not have canonical kinetic terms.
At vanishing vector superfields, the expansion
\be
 K=K_0+Z_{\bar I J}\bar\Phi^{\bar I}\Phi^J+\cdots
 \label{eq:matter-kahler-expansion}
\ee
identifies their Hermitian kinetic matrix $Z$; $K_0$ is the moduli K\"ahler potential.\footnote{
In the worldsheet description, matter kinetic coefficients are determined by two-point functions of the corresponding vertex operators.
For exactly marginal deformations, the analogous two-point functions define the Zamolodchikov metric~\cite{Zamolodchikov:1986gt,Dixon:1989fj}.}
This matrix is obtained geometrically from harmonic representatives,
\be
 Z_{\bar I J}\ \propto\
 \frac{1}{\operatorname{Vol}_g(X)}
 \int_X\mathrm d\operatorname{vol}_g\,
 \langle a_{I,\mathrm{harm}},a_{J,\mathrm{harm}}\rangle_{g,h} \,,
 \label{eq:norm}
\ee
with the appropriate Hermitian metric $h$ on each associated bundle.
The omitted common factor is fixed by dimensional reduction and the field normalization conventions and can depend on neutral moduli; the standard embedding convention is made explicit below.
The Ricci-flat metric $g$ and the HYM bundle metric determine both the harmonic equations and this inner product~\cite{Douglas:2015aga,Hubsch:2025qhd}.
Writing $\Phi^I=(E_Z)^I{}_a\Phi^a_\text{can}$, choose $E_Z^\dagger Z E_Z=\bm{1}$.
Including the supergravity normalization and working in four-dimensional Planck units, the physical couplings are then
\be
 Y_{abc}=e^{K_0/2}\kappa_{IJK}
 (E_Z)^I{}_a(E_Z)^J{}_b(E_Z)^K{}_c \,,
 \label{eq:tripleYuk}
\ee
In an orthogonal basis this amounts to dividing by the square root of each field's kinetic eigenvalue, in addition to the common supergravity factor.
The distinction between the cohomological cubic and its geometric normalization is central to numerical Yukawa calculations.
For the $\bm{27}^3$ sector in the standard embedding, the latter can be recovered from complex structure geometry, as we now explain.

\subsection{Yukawa couplings in the standard embedding}
\label{sec:standard-embedding}

For the \emph{standard embedding} on a Calabi--Yau threefold, one sets $V=T_X$ and identifies the visible gauge connection with the $SU(3)$ connection induced by the Ricci-flat K\"ahler metric~\cite{Candelas:1985en}.
The tangent bundle curvature has type $(1,1)$, and its contraction with $\omega$ vanishes because the Ricci tensor vanishes.
It therefore solves the HYM equations~\eqref{eq:standard-HYM}.
The same metric supplies the geometry of $X$ and the Hermitian metric on the bundle fibers, so there is no independent HYM problem to solve.
Identifying the two curvatures also cancels their contributions to the heterotic Bianchi identity.
The visible gauge group is $E_6$, with matter representations given by~\eqref{eq:standard-branching}.
This specialization relates physical Yukawa couplings to complex structure geometry and underlies the numerical methods of~\cite{Butbaia:2024tje,Berglund:2024uqv,Butbaia:2024xgj}.

\paragraph{The spectrum and the Hodge numbers:}
The bundle cohomologies in Section~\ref{sec:yukawa} now become Hodge cohomologies: contraction with the nowhere vanishing holomorphic three-form identifies $T_X$ with $\Omega_X^2$, giving
\begin{align}
 H^1(X,T_X)&\simeq H^{2,1}(X) \,,&
 H^2(X,T_X)&\simeq H^{2,2}(X) \,,\nonumber\\
 h^1(X,T_X)&=h^{2,1}(X) \,,&
 h^2(X,T_X)&=h^{1,1}(X) \,.
 \label{eq:standard-spectrum}
\end{align}
In the standard embedding, $\bm{27}$ and $\overline{\bm{27}}$ multiplets are counted by $h^1(X,V)$ and $h^1(X,V^*)=h^2(X,V)$, respectively, where the last equality follows from Serre duality. Consequently, $N_{\bm{27}}-N_{\overline{\bm{27}}}=h^{2,1}-h^{1,1}=-\chi(X)/2$, and the absolute net generation number is $|\chi(X)|/2$.
The same space $H^1(X,T_X)$ describes infinitesimal complex structure deformations.
This is the reason that complex structure geometry controls the normalization of the $\bm{27}$ fields.
Infinitesimal deformations of the holomorphic bundle $T_X$ are a different set of singlets, counted by $H^1(X,\operatorname{End}T_X)$.
Explicitly, writing
$\Omega=\frac{1}{3!}\Omega_{ijk}\,\mathrm dz^i\wedge\mathrm dz^j\wedge\mathrm dz^k$,
the correspondence between an ordinary $(2,1)$-form and a tangent bundle valued $(0,1)$-form is
\begin{align}
 \chi
 &=\frac12\chi_{jk\bar\ell}\,
   \mathrm dz^j\wedge\mathrm dz^k\wedge\mathrm d\bar z^{\bar\ell}
 \;\longleftrightarrow\;
 a=a^i{}_{\bar\ell}\,\mathrm d\bar z^{\bar\ell}
   \otimes\frac{\partial}{\partial z^i} \,,\nonumber\\
 \chi_{jk\bar\ell}
 &=\Omega_{ijk}a^i{}_{\bar\ell} \,,\qquad
 a^i{}_{\bar\ell}=\frac12\Theta^{ijk}\chi_{jk\bar\ell} \,,
 \label{eq:standard-form-map}
\end{align}
where $\Theta$ is the inverse holomorphic volume trivector, normalized by
$\frac1{3!}\Theta^{ijk}\Omega_{ijk}=1$.
Equivalently, with a Hermitian metric,
$\Theta^{ijk}=\bar\Omega^{ijk}/\|\Omega\|_g^2$, where
$\|\Omega\|_g^2=\frac1{3!}\Omega_{ijk}\bar\Omega^{ijk}$ and the indices on $\bar\Omega$ are raised using $g$.
For unit normalized $\Omega$, the inverse map is simply
$a^i{}_{\bar\ell}=\frac12\bar\Omega^{ijk}\chi_{jk\bar\ell}$.
We write the forward map as $\chi=a\lrcorner\Omega$: the symbol $\lrcorner$ contracts the tangent vector index of $a$ with one holomorphic index of $\Omega$, retaining the antiholomorphic differential form index.

\paragraph{Topological and quasi-topological couplings:}
The distinction concerns the holomorphic, unnormalized coefficients of the superpotential~\cite{Strominger:1985ks,Candelas:1987se,Butbaia:2024tje}.
Ignoring worldsheet instantons, the $\overline{\bm{27}}^3$ coefficients are triple intersection numbers,
\be
 d_{ABC}=\int_X\omega_A\wedge\omega_B\wedge\omega_C \,,
 \label{eq:standard-topological}
\ee
for an integral basis $[\omega_A]\in H^{1,1}(X)$.
They are \emph{topological}: in a fixed transported cohomology basis, they do not change under continuous changes of the metric or complex structure.
The $\bm{27}^3$ coefficients instead involve representatives $a_i$ of the corresponding Kodaira--Spencer classes,
\be
 \kappa_{ijk}
 =\int_X\Omega\wedge a_i^\mu\wedge a_j^\nu\wedge a_k^\rho\,
   \Omega_{\mu\nu\rho} \,.
 \label{eq:standard-holomorphic-yukawa}
\ee
The cohomological invariance of~\eqref{eq:yuk1} still holds, but the classes and holomorphic form vary with complex structure, and so do the couplings.
They are therefore \emph{quasi-topological}, rather than invariants of the underlying differentiable manifold~\cite{Strominger:1985ks,Candelas:1987se,Blesneag:2015pvz}.

The kinetic normalization in~\eqref{eq:tripleYuk} is still required.
In particular, quasi-topological $\overline{\bm{27}}^3$ coefficients need not give K\"ahler independent \emph{physical} couplings.
For the $\bm{27}^3$ sector, the standard embedding identifies the necessary matter metric with a rescaling of the Weil--Petersson metric.

\paragraph{Weil--Petersson geometry:}
Let $t^a$, $a=1,\ldots,h^{2,1}(X)$, be local complex structure coordinates, and choose a holomorphic family of three-forms $\Omega(t)$.
Define
\be
 \mathcal S(t)=\text{i}\int_X\Omega(t)\wedge\overline{\Omega(t)}>0 \,,\qquad
 K_\text{cs}=-\log\mathcal S \,,\qquad
 G^\text{WP}_{a\bar b}=\partial_a\partial_{\bar b}K_\text{cs} \,.
 \label{eq:wp_metric}
\ee
This is the conventional, volume normalized Weil--Petersson metric.
To see its relation to the Ricci-flat metric $g$, let $a_g$ and $b_g$ be harmonic tangent bundle valued $(0,1)$-forms representing the corresponding deformations, and set $\chi_a=a_g\lrcorner\Omega$.
Then
\be
 G^\text{WP}_{a\bar b}
 =\frac{1}{\operatorname{Vol}_g(X)}
   \int_X\mathrm d\operatorname{vol}_g\,\langle a_g,b_g\rangle_g
 =-\frac{\displaystyle\int_X\chi_a\wedge\overline{\chi_b}}
         {\displaystyle\int_X\Omega\wedge\overline\Omega} \,.
 \label{eq:WPH}
\ee
The inner product uses $g$ both on the cotangent index and on the tangent bundle index.
Ricci-flatness makes $\Omega$ parallel and relates its norm to the Riemannian volume form, yielding the equality with the complex structure expression~\cite{Tian:1987Smoothness,Todorov:1989umc,Butbaia:2024tje}.
An unnormalized $L^2$ pairing contains the additional factor $\operatorname{Vol}_g(X)$.
One therefore needs the Ricci-flat metric to evaluate the harmonic form expression directly, but can obtain the same answer from~\eqref{eq:wp_metric} without constructing that metric.

To normalize~\eqref{eq:standard-holomorphic-yukawa}, choose a frame $E$ satisfying
$E^i{}_aG^\text{WP}_{i\bar j}\overline{E^j{}_b}=\delta_{a\bar b}$.
The normalized complex structure Yukawa tensor is then
\be
 Y_{abc}=e^{K_\text{cs}}\kappa_{ijk}E^i{}_aE^j{}_bE^k{}_c \,,
 \label{eq:NormYuk}
\ee
up to a common dilaton/gauge normalization, held fixed here.
In a basis orthogonal with respect to $G^\text{WP}$, with squared norms $N_i$, this reduces to
\be
 Y_{ijk}=\frac{\kappa_{ijk}}
 {\mathcal S\,\sqrt{N_iN_jN_k}} \,.
 \label{eq:standard-orthogonal-yukawa}
\ee
The three norm factors are the field rescalings required for canonical kinetic terms; the factor $\mathcal S^{-1}=e^{K_\text{cs}}$ completes the normalization of the holomorphic cubic.
This is why the relative sizes of the physical couplings can differ substantially from those of the unnormalized coefficients.
At tree level, the matter metric $Z$ is proportional to $e^{(K_\text{K}-K_\text{cs})/3}G^\text{WP}$.
Here, $K_\text{K}$ and $K_\text{S}$ denote the K\"ahler potentials of the K\"ahler moduli and dilaton, respectively.
The three field rescalings cancel the $K_\text{K}$ dependence of $e^{K_0/2}=e^{(K_\text{S}+K_\text{K}+K_\text{cs})/2}$ in~\eqref{eq:tripleYuk}, leaving $e^{K_\text{cs}}$ up to the fixed dilaton and gauge factor~\cite{Dixon:1989fj}.
Thus the frame $E$ here orthonormalizes $G^\text{WP}$, while $E_Z$ in Section~\ref{sec:yukawa} orthonormalizes the full matter metric.
Equation~\eqref{eq:NormYuk} makes the remaining complex structure dependence explicit.

\paragraph{Three ways to compute the couplings:}
The first two methods below use special geometry to compute the normalization; the third computes the harmonic wavefunctions themselves.
They target the same tensor~\eqref{eq:NormYuk}, with a consistent choice of cohomology basis and field normalization.

\noindent\textbf{\textit{Periods:}}
Choose a symplectic basis $(A^I,B_I)$ of three-cycles, with $I=0,\ldots,h^{2,1}(X)$, and calculate
$Z^I=\int_{A^I}\Omega$ and $\mathcal F_I=\int_{B_I}\Omega$.
Their symplectic pairing determines $\mathcal S$ and hence $G^\text{WP}$; their variation also determines the holomorphic Yukawa tensor.
Picard--Fuchs equations provide a practical route to these periods~\cite{Candelas:1990rm,Morrison:1991cd}.
This approach is particularly effective for families with few complex structure moduli.
There is no mathematical restriction to $h^{2,1}=1$, although constructing and solving a multivariable period system can become difficult.

\noindent\textbf{\textit{Kodaira--Spencer integration:}}
The Kodaira--Spencer map sends a deformation direction to its tangent bundle cohomology class:
\be
 \rho:T_t\mathcal M_\text{cs}\longrightarrow H^1(X_t,T_{X_t}) \,,
 \qquad \partial_a\longmapsto[\Phi_a] \,.
 \label{eq:ks_map}
\ee
For an explicitly presented family, variations of its defining equations and Poincar\'e residue expressions for $\Omega$ give the derivatives needed in~\eqref{eq:wp_metric}.
The resulting integrals over $X$ can be evaluated by Monte Carlo sampling, without solving for periods or constructing a Ricci-flat metric~\cite{Keller:2009vj,Butbaia:2024tje}.
This calculation retains the full variation
$\partial_a\Omega=\Omega'_a+\Phi_a\lrcorner\Omega$, including its $(3,0)$ part.
It is not equivalent to inserting arbitrary nonharmonic representatives into the last expression of~\eqref{eq:WPH}.
The holomorphic cubic is computed using the same cohomology representatives.
This method applies for arbitrary $h^{2,1}$: it computes the matrix of pairings for the supplied deformation directions, rather than relying on a one modulus reduction.

\noindent\textbf{\textit{Machine learned harmonic forms:}}
\label{sec:mlmetricsandharmonics}
First approximate the Ricci-flat metric in a chosen K\"ahler class.
Then, for each reference representative $\Phi_a$, learn a smooth global section $s_{a,\theta}$ of $T_X$ and set
\be
 \eta_{a,\theta}=\Phi_a+\bar\partial_{T_X}s_{a,\theta} \,,\qquad
 \mathcal L_H=\sum_{a=1}^{h^{2,1}}
 \big\|\bar\partial_{T_X}^{\dagger}\eta_{a,\theta}\big\|_{L^2(g)}^2 \,.
 \label{eq:harmonicNN}
\ee
The ansatz is closed and remains in the prescribed cohomology class; minimizing the loss imposes co-closure and approximates the harmonic representative.
The adjoint depends on the learned metric.
Global sections with the correct bundle transition laws are essential: independently specified local network outputs need not define a valid cohomologous correction.
For projective presentations, one can pull back the ambient twisted forms $z^i\,\mathrm dz^j-z^j\,\mathrm dz^i$, conjugate and dualize them with the inverse Hermitian metric, and compensate their projective weights to construct tangent sections with learned scalar coefficients~\cite{Ashmore:2020ujw,Berglund:2024uqv}.
Related constructions of globally defined learned forms appear in~\cite{Douglas:2024pmn}.
The resulting forms supply the kinetic matrix through~\eqref{eq:WPH} and the normalized couplings through~\eqref{eq:NormYuk}.
The construction and its implementation in \texttt{cymyc} allow simultaneous learning of all $h^{2,1}$ representatives~\cite{Berglund:2024uqv,Butbaia:2024xgj}.
Thus this approach, like Kodaira--Spencer integration, applies for arbitrary $h^{2,1}$, with computational cost increasing as more deformation directions are included.
Learned harmonic representatives also enter Yukawa calculations for line bundle models~\cite{Constantin:2024yxh} and for general non-Abelian bundles, where the HYM metric must be determined separately~\cite{Mishra:2025xkr}.

\begin{figure}[tbp]
 \centering
 \includegraphics[width=\linewidth]{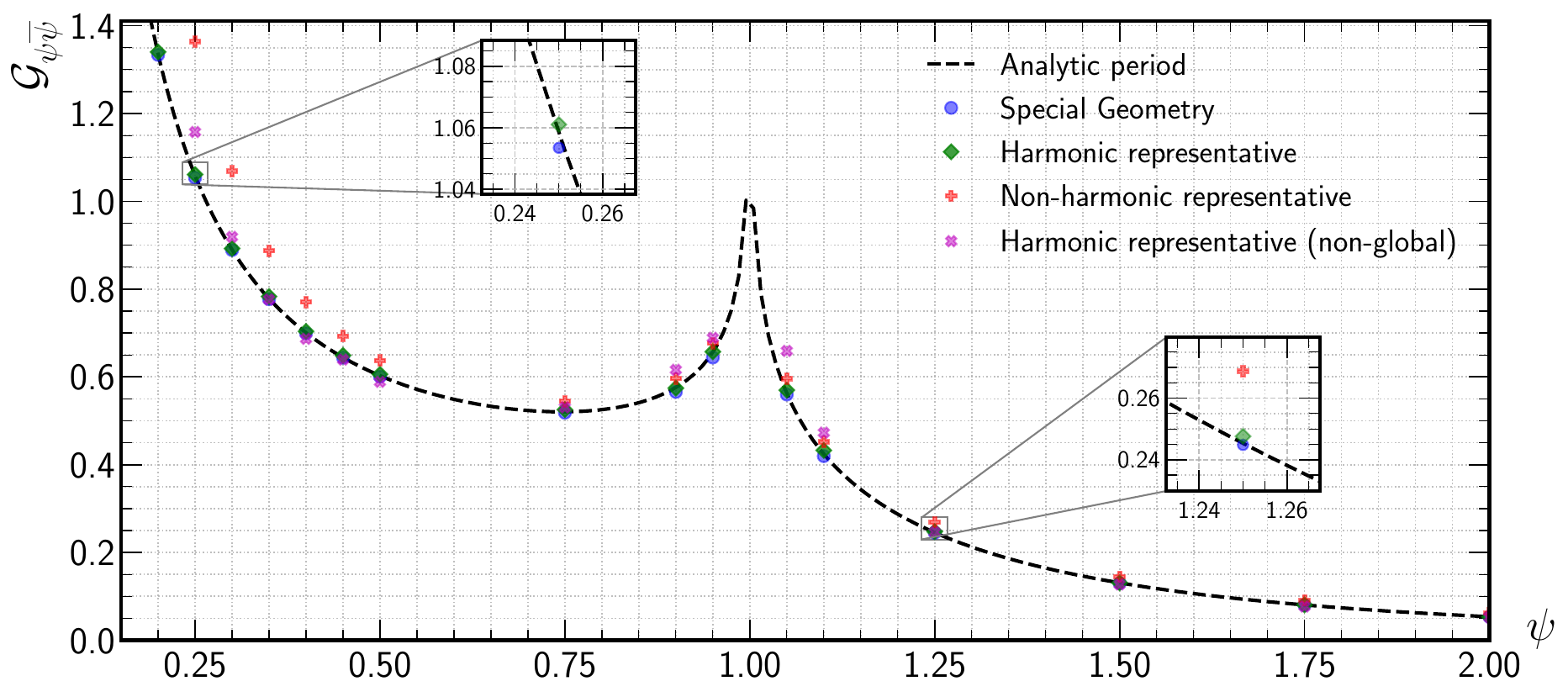}
 \par\smallskip
 \includegraphics[width=\linewidth]{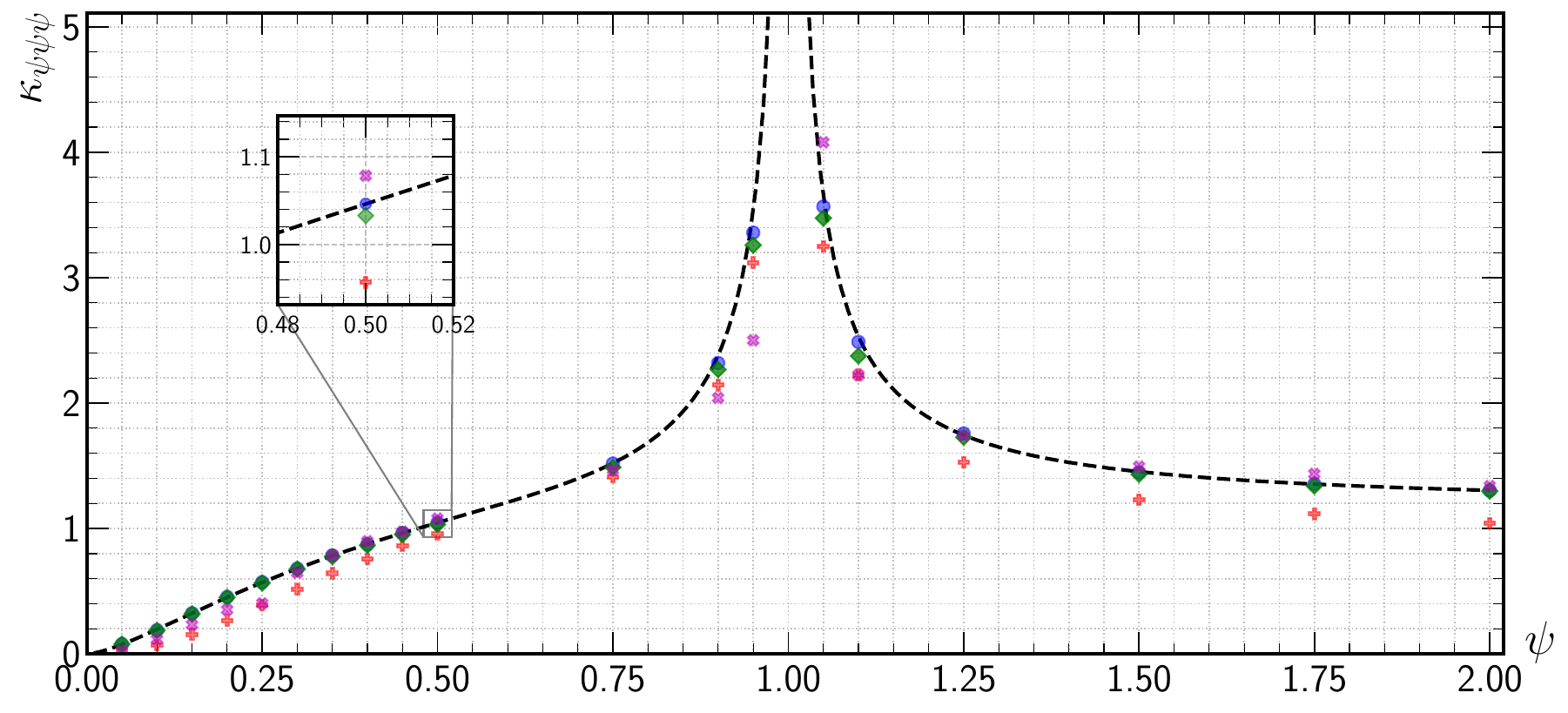}
 \caption{\small Comparison on the mirror of $\mathbb P^5[3,3]$ along real $\psi$.
 The upper and lower panels show the Weil--Petersson metric and normalized Yukawa coupling, respectively.
 Blue points use the full special geometry calculation and green diamonds use learned harmonic representatives.
 The red and magenta diagnostics respectively omit harmonic projection in the harmonic pairing and use a correction that is not globally defined.
 The plotted $\kappa_{\psi\psi\psi}$ denotes the normalized coupling, called $Y_{\psi\psi\psi}$ here.
 Reproduced from~\cite{Berglund:2024uqv}.}
 \label{fig:standard-bicubic}
\end{figure}

\paragraph{Agreement and the effect of normalization:}
The mirror of the complete intersection $\mathbb P^5[3,3]$ provides a direct comparison of all three methods.
Its Hodge numbers are $(h^{1,1},h^{2,1})=(73,1)$, so a period calculation is available.
Figure~\ref{fig:standard-bicubic} compares the Weil--Petersson metric and normalized coupling from periods, Kodaira--Spencer integration, and globally defined learned harmonic forms~\cite{Butbaia:2024tje,Berglund:2024uqv,Butbaia:2024xgj}.
The agreement tests the metric and harmonic form calculations together.
The unsuccessful nonharmonic and non-global prescriptions also show why the hypotheses of the normalization formula matter.
Errors grow in high curvature regions; the comparison is a numerical validation of these observables, not a proof of an exact local metric.

\begin{figure}[tbp]
 \centering
 \includegraphics[width=\linewidth]{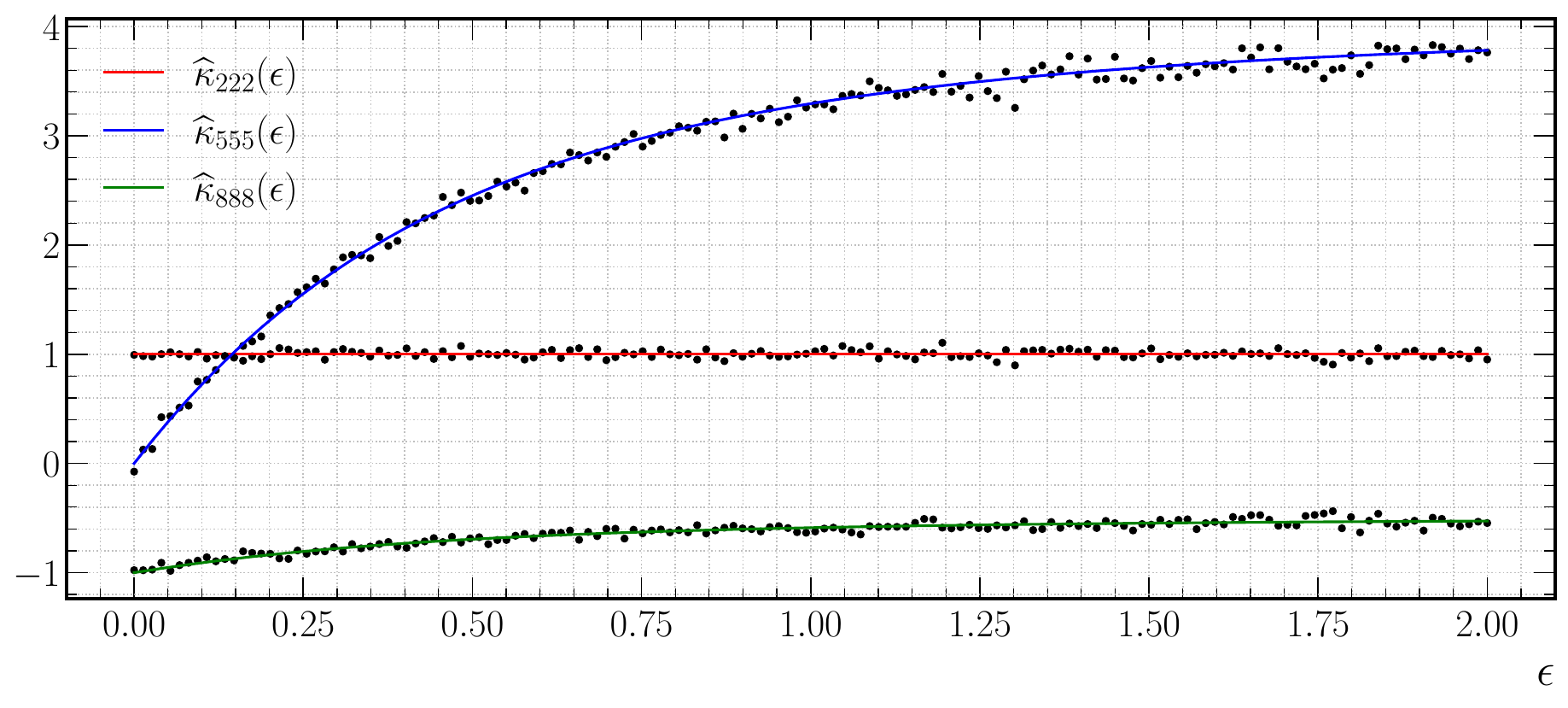}
 \par\smallskip
 \includegraphics[width=\linewidth]{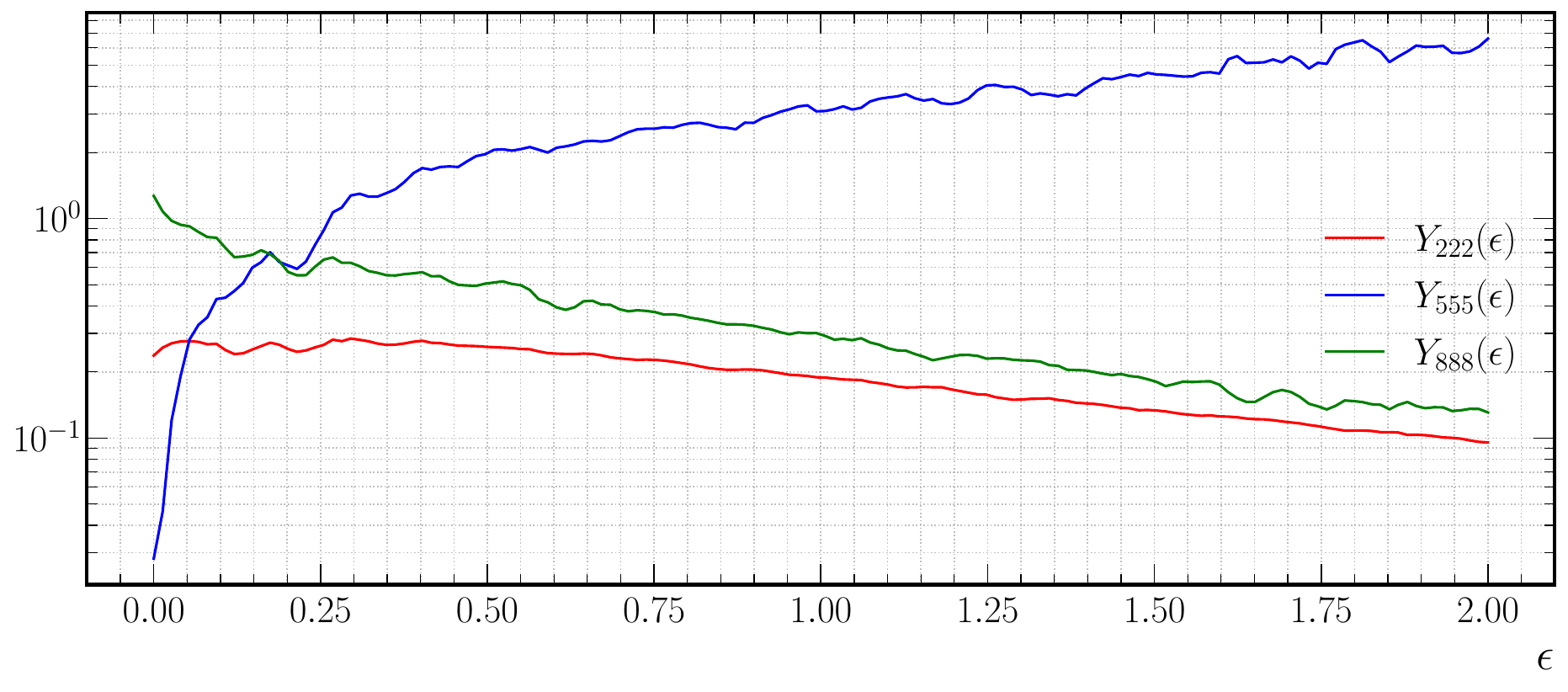}
 \caption{\small Cubic Yukawa couplings on the $\mathbb Z_3$ quotient of Tian--Yau along~\eqref{eq:standard-TY-deformation}, reproduced from Figures~8 and~9 of~\cite{Butbaia:2024tje}.
 Top: holomorphic ratios $\widehat\kappa_{aaa}=\kappa_{aaa}/\kappa_{111}$, with analytic curves and numerical points.
 Bottom: magnitudes of the canonically normalized couplings on a logarithmic scale.
 Moving away from the symmetric member at $\epsilon=0$ generates a hierarchy between $Y_{555}$ and the other displayed couplings.
 The lower curves were smoothed with a Savitzky--Golay filter, as described in the original paper.}
 \label{fig:standard-tian-yau}
\end{figure}

A Tian--Yau quotient illustrates how Yukawa hierarchies develop as the complex structure moves away from a symmetric point.
The smooth free $\mathbb Z_3$ quotient has $(h^{1,1},h^{2,1})=(6,9)$ and $\chi=-6$, giving three net generations in the standard embedding.
Following~\cite{Butbaia:2024tje}, keep the two cubic equations in $\mathbb P^3\times\mathbb P^3$ fixed and vary the bilinear equation as
\be
 p_2=x_0y_0+x_1y_1+(1+\epsilon)(x_2y_2+x_3y_3)=0 \,,\qquad
 \epsilon\in\mathbb R \,.
 \label{eq:standard-TY-deformation}
\ee
The member at $\epsilon=0$ has enhanced discrete symmetry.
This one parameter path lies in a moduli space with $h^{2,1}=9$; it does not reduce the normalization problem to one matter field.
Figure~\ref{fig:standard-tian-yau} compares selected holomorphic cubic ratios with the magnitudes of the corresponding normalized couplings.
As $\epsilon$ increases, $|Y_{555}|$ grows relative to the other displayed couplings, producing a hierarchy of approximately two orders of magnitude over the plotted range.
The physical hierarchy is therefore visible upon moving away from the symmetric point in complex structure moduli space, and its size requires the kinetic normalization.
The logarithmic panel shows absolute values: crossings of magnitudes need not be equalities of the signed couplings.

Finally, the K\"ahler class is not a parameter of the complex structure geometry in~\eqref{eq:wp_metric}.
At tree level, at fixed complex structure and with the normalization above, the standard embedding $\bm{27}^3$ couplings are therefore the same when evaluated using the Ricci-flat metric in \emph{any K\"ahler class}~\cite{Berglund:2024uqv}.
The local metric and harmonic wavefunctions change, but their normalized pairings recover the same Weil--Petersson metric and Yukawa tensor, up to the usual unitary choice of canonical field basis.
This holds for arbitrary $h^{2,1}$ and supplies a further consistency check for the numerical methods.
This statement only concerns the standard embedding; general bundle compactifications do not have this reduction to complex structure geometry.

\subsection{Yukawa couplings in non-standard embeddings}
\label{sec:non-standard}

The textbook relation $N_{\text{gen}}=\frac12|\chi(X)|$ is a consequence of the standard embedding, rather than a universal prediction of heterotic compactification.
It follows from the Hodge number identification in~\eqref{eq:standard-spectrum} when $V=T_X$.
Here, $N_{\text{gen}}$ denotes the absolute net chiral multiplicity.

In a non-standard embedding, the gauge bundle $V$ is an additional choice whose topology need not coincide with that of $T_X$.
Consider a stable holomorphic $SU(n)$ bundle on $X$, with $n>1$.
Since $c_1(V)=c_1(T_X)=0$, the Hirzebruch--Riemann--Roch theorem gives~\cite{Anderson:2008uw}
\be
\operatorname{ind}(V) = \sum_{q=0}^{3}(-1)^q h^q(X,V)
= \int_X \operatorname{ch}(V)\wedge\operatorname{Td}(T_X)
= \frac12\int_X c_3(V) \,.
\ee
Stability implies $H^0(X,V)=H^3(X,V)=0$, while Serre duality identifies $H^2(X,V)$ with the dual of $H^1(X,V^*)$.
For the usual $SU(3)$ and $SU(4)$ constructions, generations and antigenerations are counted by $H^1(X,V)$ and $H^1(X,V^*)$, respectively, so
\be
N_{\text{gen}} = \left|h^1(X,V)-h^1(X,V^*)\right| = |\operatorname{ind}(V)|
= \frac12\left|\int_X c_3(V)\right| \,.
\ee
More generally, the multiplicity of each charged representation is determined by the cohomology of its associated bundle.
The index fixes the chiral excess but does not determine the number of vectorlike pairs.

Anomaly cancellation does not restore the relation between the generation number and the Euler characteristic.
The cohomological Bianchi identity requires~\cite{Anderson:2012yf}
\be
c_2(T_X)-c_2(V)-c_2(V_{\text{hid}}) = [W] \,,
\ee
where $V_{\text{hid}}$ is the hidden sector bundle and $[W]\in H^4(X,\mathbb{Z})$ is the Poincar\'e dual of the effective curve class wrapped by five-branes.
This constrains the second Chern classes, whereas the chiral index depends on the third Chern class.
Even when the hidden bundle is trivial and five-branes are absent, the condition $c_2(V)=c_2(T_X)$ does not imply $c_3(V)=c_3(T_X)$.
Consequently, consistent bundle choices can produce different net generation numbers on the same Calabi--Yau threefold.

Changing the net index does require a change in the relevant bundle topology: a smooth deformation of the holomorphic structure of $T_X$ preserves its Chern classes and therefore preserves the standard embedding value of $N_{\text{gen}}$.
An explicit illustration of the greater freedom is provided by Braun, He, Ovrut, and Pantev~\cite{Braun:2005bw}.
Their quotient threefold $X=\widetilde X/(\mathbb Z_3\times\mathbb Z_3)$ has $h^{1,1}=h^{2,1}=3$, hence $\chi(X)=0$, and carries an $SU(4)$ bundle with $\int_X c_3(V)=-6$.
The corresponding Chern number on the ninefold cover is $-54$.
The resulting theory has three net generations despite the vanishing Euler characteristic.
Thus, the chiral spectrum depends on the topology of the gauge bundle as well as that of the compactification manifold.

\paragraph{Line bundle sums and more general bundles:}
A particularly tractable setting is
\be
 V=\bigoplus_{a=1}^{n}L_a \,,\qquad
 \bigotimes_{a=1}^{n}L_a\simeq\mathcal O_X \,,\qquad
 \mu_\omega(L_a)=0 \,,\quad\text{for all }a \,.
 \label{eq:split-bundle}
\ee
The tensor product condition makes the determinant trivial, while the common zero slope condition makes the sum polystable at the chosen K\"ahler class.
Its structure group lies in $S(U(1)^n)\subset SU(n)$, and the Abelian charges impose useful selection rules on the superpotential, even when the corresponding gauge bosons acquire St\"uckelberg masses~\cite{Anderson:2012yf,Constantin:2024yaz}.
At a split locus, the HYM problem reduces to finding metrics on the individual line bundles.
However, sums of line bundles are not the most general setting: stable bundles with non-Abelian structure group need not split, and can be constructed, for example, as nontrivial extensions or as the cohomology of monads.
A presentation using line bundles does not mean that the resulting bundle is their direct sum.
Moreover, expectation values of charged bundle moduli can deform a split sum into an indecomposable bundle, subject to the supersymmetry conditions and stability in the chosen K\"ahler chamber~\cite{Constantin:2025vyt}.

\paragraph{Direct computations of normalized couplings:}
For a non-standard embedding, the Ricci-flat metric must be supplemented by a Hermitian bundle metric whose Chern connection solves the HYM equations~\eqref{eq:standard-HYM}, and by harmonic representatives of the matter cohomology classes.
There is generally no reduction of the matter metric to the Weil--Petersson metric as in the standard embedding.
ML supplies numerical approximations to these geometric fields at specified values of the moduli~\cite{Constantin:2024yxh,Mishra:2025xkr,Chervonyi:2026tjx}.
Neural networks parameterize corrections to a reference K\"ahler potential, bundle metrics, and matter forms; training enforces the Monge--Amp\`ere, HYM, and harmonic equations at sampled points, following the physics informed strategy of Section~\ref{sec:pinn}.
The bundle metric ansatz must be positive definite and compatible with the transition functions of $V$; an arbitrary matrix valued function on each coordinate patch is insufficient.
For the matter forms, the learned $\bar\partial_V$ exact corrections preserve the cohomology classes supplied by the algebraic construction.
The geometric equations supply the training signal, while Monte Carlo integration of the matter kinetic terms supplies the matrices needed for canonical normalization.
The holomorphic Yukawa coefficients can be evaluated separately by cohomological methods, avoiding numerical integration errors in small superpotential couplings~\cite{Chervonyi:2026tjx}.
Residuals in the geometric equations, cohomologically predicted zeros, symmetry constraints, and convergence of the integrals supply complementary checks.

Constantin and collaborators~\cite{Constantin:2024yxh} carried out this calculation for up-quark Yukawa couplings in a model with a sum of five line bundles on a $\mathbb Z_2\times\mathbb Z_2$ quotient of the tetraquadric.
The Abelian selection rules restrict the perturbative up-quark Yukawa matrix to rank at most two, leaving at least one mass zero.
In the one parameter complex structure family examined, a symmetry forces the two nonzero masses to be degenerate, a feature reproduced within numerical errors.
The authors also study a point away from that family, where the degeneracy is lifted.
For this example, the reported masses are
\begin{align}
 (m_u,m_c,m_t)_{\text{full}}
 &\simeq e^{-\phi}|\langle H_u\rangle|\,(0,\;0.009,\;0.016) \,,\nonumber\\*
 (m_u,m_c,m_t)_{\text{reference}}
 &\simeq e^{-\phi}|\langle H_u\rangle|\,(0,\;0.007,\;0.013) \,,\nonumber\\*
 (m_u,m_c,m_t)_{\text{canonical}}
 &\simeq e^{-\phi}|\langle H_u\rangle|\,(0,\;0.004,\;0.008) \,.
 \label{eq:nonstandard-masses}
\end{align}
Here, $\phi$ is the dilaton and $\langle H_u\rangle$ is the up-type Higgs expectation value.
The second line uses analytic reference quantities, while the third assumes canonical matter kinetic terms without computing their geometric normalization.
Reference quantities give a better approximation than simply assuming canonical kinetic terms, but the full normalization still changes the predicted masses appreciably.
The authors estimate a numerical accuracy of roughly $10\%$ for the full calculation.

The phenomenological conclusion is more restrained than the computational achievement.
The larger mass remains too small to account for the top quark mass, and the splitting is insufficient to reproduce the observed top to charm hierarchy.
The example was selected from five randomly chosen invariant defining polynomials, rather than an extensive search of moduli space.
The paper thus demonstrates a calculation of physical couplings and the importance of normalization, while leaving a realistic flavor fit for further work.

\paragraph{Geometric hierarchies and model scans:}
Chervonyi, Ruehle, and Tomasev~\cite{Chervonyi:2026tjx} identify two mechanisms for generating quark mass hierarchies at tree level.
The first tunes the holomorphic mass matrix toward a rank deficient locus in the smooth interior of complex structure moduli space.
Near a simple rank drop, with regular kinetic metrics and nonzero heavier masses, a transverse displacement gives $m_{\mathrm{light}}/m_{\mathrm{heavy}}\propto|\delta C|$.
This permits parametrically large hierarchies without degenerating the geometry, in a regime where the supergravity approximation remains applicable.
The second mechanism uses localization near polynomial factorization loci to produce unequal kinetic norms, requiring competing harmonic norms to remain nonzero in the limit.
Moderate localization can complement holomorphic tuning, although extreme degeneration produces high curvature.
The tetraquadric analysis illustrates why harmonic representatives matter: an apparent hierarchy in reference quantities disappears after harmonic projection.

The authors scan $63{,}325$ upstairs $SU(5)$ models and an earlier catalog yielding $2{,}112$ downstairs Standard Models, including compatible quotient actions and Wilson-line GUT breaking.
Within these catalogs and their selection conditions, they find one candidate supporting a full-rank up-type Yukawa matrix: Model~500 on a freely acting $\mathbb Z_2$ quotient of CICY~5273.
The covering threefold has codimension five in the ambient space $\mathbb{P}^1\times\mathbb{P}^1\times\mathbb{P}^1\times\mathbb{P}^1\times\mathbb{P}^2\times\mathbb{P}^2$ and $h^{1,1}=6$, while the quotient has $h^{1,1}=5$.
Three distinct matter line bundles and three Higgs sectors on a cohomology jumping locus permit rank three; the Wilson line distinguishes quark and antiquark wavefunctions, avoiding the hierarchy obstruction of the symmetric upstairs matrix with vanishing diagonal.

A central methodological advance is exact evaluation of the holomorphic overlaps through ambient integrals, extending Koszul--Dolbeault residue methods across the higher steps of the codimension five resolution.
This removes integration noise that would overwhelm small couplings or mislead the moduli search.
Numerical calculations supply the kinetic terms and common volume normalization.
Harmonic projection minimizes the kinetic norm within a cohomology class, while pointwise comparison of physical and reference inner products bounds their minimum norms.
These estimates constrain normalization changes given metric comparison data, without assuming that the corrections are small.

At selected geometries, a sequential computation uses $17$ networks: one for the Ricci-flat metric, three for independent HYM bundle metrics, and thirteen for harmonic matter and Higgs forms.
The largest network predicts $21{,}600$ real coefficients in an ambient section frame.
Parallel training within the latter stages, implemented in JAX/Equinox on NVIDIA B200 GPUs, completes one training seed in approximately $90$ minutes.

The resulting smooth examples exhibit an up-quark hierarchy spanning approximately five orders of magnitude.
The comparison with low energy masses uses two-loop RG evolution, assuming MSSM running down to a supersymmetry matching scale $M_{\mathrm{SUSY}}=3\,\mathrm{TeV}$ and Standard Model running below it.
This also assumes that two of the three Higgs pairs decouple near unification; the required bundle deformation and its corrected Yukawas have not been computed.
No model in the scan has both Yukawa matrices of full rank at the split bundle locus: the selected model leaves the down and strange quarks massless and tests only the bottom column of CKM mixing.
The moduli and Higgs directions are chosen phenomenologically, numerical uncertainties are not complete error bounds, and model dependent supersymmetric threshold corrections are omitted.
Whether moduli stabilization dynamically favors the hierarchy loci remains open.

\paragraph{Flavor textures and explicit compactifications:}
A complementary computational task is to identify flavor textures allowed by the bundle charges and singlet fields.
Constantin and collaborators~\cite{Constantin:2024yaz} search for Froggatt--Nielsen models motivated by split rank five heterotic bundles.
These splittings include higher rank summands as well as sums of line bundles.
Exhaustive searches and a genetic algorithm explore singlet charge assignments and patterns of singlet expectation values, selecting candidate theories according to their ability to reproduce masses and mixing.
Singlet expectation values and nonperturbative factors suppress selected operators, generating hierarchical Yukawa textures consistent with the Abelian charges.
The authors fit quark and charged lepton masses and CKM mixing magnitudes using these suppression parameters and coefficients of order one, with leading kinetic normalization effects absorbed into the latter.
This is a search within effective theories compatible with heterotic charge patterns; the fits are not tied to explicit Calabi--Yau realizations, and their coefficients are not computed from normalized overlap integrals.

The subsequent study by Constantin, Leung, Lukas, and Nutricati~\cite{Constantin:2025vyt} supplies two explicit line bundle constructions on a $\mathbb Z_2$ quotient of the complete intersection $X_{6770}$.
These are obtained through a systematic model scan and parameter fitting, providing model building targets for future geometric coupling calculations rather than a machine learning determination of the operator coefficients.
They have three families, one Higgs doublet pair, and no exotic charged matter.
The cohomology and charge assignments determine which singlet insertions and nonperturbative terms can contribute to the Yukawa matrices.
Charged singlet expectation values also describe deformations away from the split bundle locus.
With suitable choices of these expectation values and operator coefficients, the resulting textures fit the charged fermion masses and CKM mixing magnitudes.
The coefficients are fitted effective parameters that absorb matter normalization and renormalization group effects; their dependence on the geometric moduli is not calculated.
The result therefore establishes compatibility of explicit bundle constructions with the observed flavor hierarchies under the stated assumptions, including nonvanishing instanton prefactors and unbroken $R$ parity, rather than deriving the fitted coefficients at an identified point in moduli space.
Moduli stabilization and supersymmetry breaking remain to be addressed.

\paragraph{Beyond line bundle sums:}
Mishra and Tan~\cite{Mishra:2025xkr} develop an HYM computation for general holomorphic vector bundles, using generating sections of a sufficiently positive twist to construct globally compatible neural ans\"atze.
The networks learn a scalar potential for the determinant part and a positive Hermitian endomorphism for the trace free part, with alternating optimization enforcing the corresponding HYM conditions.
This construction is not intrinsically restricted to a particular rank or to Abelian structure groups.
Its practical applicability depends on obtaining the generating sections, and large twists can make the computation expensive.
The examples include monad bundles on the Fermat quartic and quintic.
For the Yukawa calculation they use the rank three bundle on the quintic defined by
\be
 0\longrightarrow V\longrightarrow\mathcal O_X(1)^{\oplus4}
 \longrightarrow\mathcal O_X(4)\longrightarrow0 \,.
 \label{eq:nonstandard-monad}
\ee
Although the defining sequence involves sums of line bundles, its kernel $V$ is a non-Abelian $SU(3)$ bundle, giving an $E_6$ gauge theory.
Further networks parameterize $\bar\partial_V$ exact corrections to explicit cohomology representatives, minimizing a Rayleigh quotient for the bundle Laplacian to obtain approximate harmonic forms.
The calculation treats selected subsets of the $50$ dimensional space $H^1(X,V)$, using the restricted matter metric to normalize couplings within each selected subspace, rather than computing the full Yukawa tensor or a three family model.
It reproduces algebraically predicted zeros and shows how normalization changes the relative strengths of the nonzero couplings.
This extends direct physical Yukawa computations beyond split sums of line bundles.

Together, these studies distinguish three tasks: identifying the allowed holomorphic couplings, fitting their coefficients within an effective theory, and calculating their physical normalization from the compactification geometry.
For a general bundle, physical couplings can depend on complex structure, K\"ahler, and bundle moduli; the K\"ahler independence of the standard embedding $\bm{27}^3$ sector does not extend automatically.
The families of background metrics discussed in Section~\ref{sec:metric-moduli} provide one ingredient for such studies of moduli dependence.

\subsection{From bundle metrics to axion--matter couplings}\label{sec:axion-matter}

Bundle valued matter wavefunctions enter both physical Yukawa normalization and other four-dimensional couplings.
Benabou \textit{et al.}\ compute leading perturbative axion--matter couplings in two heterotic line bundle models on tetraquadric quotients~\cite{Benabou:2026lcu}.
Neural networks approximate the Ricci-flat metric, line bundle metrics, and harmonic matter forms by imposing the geometric equations described in Section~\ref{sec:non-standard}.
The learned fields supply the integrands for direct Monte Carlo evaluation of both the matter kinetic matrix~\eqref{eq:norm}, denoted $G_{I\bar J}$, and the bosonic axion overlap matrices $\Lambda_{i,I\bar J}$.
In the paper's holomorphic matter coordinates, dimensional reduction gives the analytic relation $\Lambda_{i,I\bar J}=\tfrac12\partial_{t^i}G_{I\bar J}$, expressing the axion couplings through the K\"ahler moduli dependence of the matter metric.

Fermionic couplings include an additional universal supergravity term and require canonical normalization~\cite{Benabou:2026lcu}.
Near the Kaluza--Klein matching scale, the reported parametric behavior is
\be\label{eq:axion-matter-scaling}
 \mathcal L\supset c_\psi\frac{\partial_\mu a}{f_a}
 \psi^\dagger\bar\sigma^\mu\psi\,,
 \qquad
 c_\psi^{\text{MD}}\sim\frac{1}{4\pi t}\,,
 \qquad
 c_\psi^{\text{MI}}\sim\frac{\alpha_{\text{YM}}}{8\pi} \,.
\ee
Here, MD and MI denote model dependent and model independent axions, and $t$ represents the K\"ahler modulus scale.
The MI fermion coupling is loop sized but includes a string tree level supergravity contribution.
These are perturbative matching results; omitted nonperturbative Yukawa corrections can induce order-one shifts in physical axion couplings after the field rotations to the fermion mass basis~\cite{Benabou:2026lcu}.

\section{Machine learning further differential geometry}
\label{sec:differential-geometry}

\subsection{Harmonic fluxes and warped backgrounds}\label{sec:warped}

A learned metric also supplies the differential operators needed to calculate other quantities on the compactification space.
L\"ust, Ruehle, and Schreyer apply this idea to the one-parameter Dwork family, learning the Ricci-flat metric, harmonic flux representatives, and a warp factor in sequence~\cite{Lust:2026mys}.
Schematically, the last step is
\be\label{eq:warp-pipeline}
 \tilde g\longrightarrow G_3\longrightarrow h=\text{e}^{-4A}\,,
 \qquad
 -\tilde\nabla^2h=S_{\text{flux}}+S_{\text{loc}} \,,
\ee
where the sources depend on the flux and localized charge.
On a compact space their sum must integrate to zero.
The Poisson equation fixes $h$ only up to an additive constant, and physical use also requires $h>0$.
The implementation uses adapted Fubini--Study sampling and a weighted Huber loss for large sources.
Its benchmark smears negative charge because the chosen geometry lacks the required orientifold involution; it is a numerical GKP-type toy background, not a localized orientifold solution~\cite{Lust:2026mys}.

\subsection{$\alpha'$ corrections}
\label{sec:ml-alpha-prime}

The Ricci-flat Calabi--Yau metric is the leading approximation to the internal string geometry at large volume.
Nemeschansky and Sen~\cite{Nemeschansky:1986yx} showed that higher loop corrections to the worldsheet beta function can be compensated by deforming the metric, preserving conformal invariance without preserving Ricci-flatness.
Becker, Becker, and Robbins~\cite{Becker:2015bra} related this construction to the target space description in terms of $SU(3)$ structures.
In type IIB string theory, the leading bulk higher derivative correction involves $\alpha'^3 R^4$ terms, whose reduction produces the Euler characteristic dependent correction to the four-dimensional moduli K\"ahler potential~\cite{Becker:2002nn,Bonetti:2016dqh}.

Fraser-Taliente, Harvey, and Kim~\cite{Fraser-Taliente:2024etl} implemented the leading tree-level metric correction numerically on the Fermat quintic.
At fixed order in $\alpha'$, the corrected metric depends on field redefinition conventions; physical observables require the other fields and fluctuation operators to be treated consistently.
For a smooth compact threefold $X$, their K\"ahler potential description gives, through order $\alpha'^3$,
\be
 g_{i\bar j}=\widetilde g_{i\bar j}
 +\alpha'^3\partial_i\partial_{\bar j}\delta K \,,\qquad
 \widetilde\Delta\delta K
 =\frac{(2\pi)^3\zeta(3)}{4}
 \left(\frac{\chi(X)}{V}-Q[\widetilde g]\right)
 \equiv\rho \,.
 \label{eq:ml-alpha-poisson}
\ee
Here, $\widetilde g$ is the Ricci-flat seed, $\widetilde\Delta=\widetilde g^{i\bar j}\partial_i\partial_{\bar j}$, $V=\int_X dV_{\widetilde g}$, and the Euler density is normalized by $\int_XQ\,dV_{\widetilde g}=\chi(X)$.
Subtracting its average ensures that the source integrates to zero, while a globally defined $\delta K$ preserves the K\"ahler class and is determined up to an additive constant.
Starting from a metric learned with \texttt{cymetric}, they evaluate the curvature source and train a second network $\delta K_\theta$ on sampled points $\mathcal P\subset X$ with the residual loss
\be
 \mathcal L_{\alpha'}
 =\sum_{p\in\mathcal P}
 \left|\widetilde\Delta\delta K_\theta(p)-\rho(p)\right| \,.
 \label{eq:ml-alpha-loss}
\ee
They report a residual of order one percent for approximately $10^4$ network parameters and compute induced shifts in the scalar Laplacian spectrum.

As discussed in~\cite{Berglund:2026hdh}, this provides a route from numerical Ricci flat seeds to corrected backgrounds, with local curvature tests supplementing global volume estimates.
Control requires accurate curvature and perturbatively small corrections; complete physical spectra additionally require the corrected fluctuation operators and other background fields.

\subsection{\texorpdfstring{$SU(3)$}{SU(3)} structures with intrinsic torsion}
\label{sec:su3-structure-metrics}

The distinction between $SU(3)$ structure and $SU(3)$ holonomy is essential when extending metric learning to flux compactifications.
An $SU(3)$ structure on a six-dimensional manifold is specified by a compatible real two-form $J$ and decomposable complex three-form $\Omega$, with a positive associated metric and normalization
\be
 J\wedge\Omega=0 \,,\qquad
 \frac{1}{3!}J^3=\frac{i}{8}\Omega\wedge\overline\Omega \,. 
 \label{eq:su3-structure-normalization}
\ee
These algebraic conditions do not make the forms parallel under the Levi-Civita connection.
Their exterior derivatives determine five intrinsic torsion classes $W_1,\ldots,W_5$; all five vanish precisely for a torsion free structure, whose metric has holonomy contained in $SU(3)$~\cite{Chiossi:2002tw}.
Thus, a manifold that admits a Calabi--Yau metric can also support other $SU(3)$ structures whose metrics are neither K\"ahler nor Ricci-flat.
Learning such a structure requires differential constraints adapted to its torsion, rather than the Calabi--Yau Monge--Amp\`ere equation alone.

Larfors, Lukas, and Ruehle~\cite{Larfors:2018nce} constructed explicit torsional structures on complete intersection Calabi--Yau threefolds, providing analytic benchmarks for this problem.
The heterotic supersymmetry conditions allow the torsion pattern
\be
 W_1=W_2=0 \,,\qquad W_4=\frac12 W_5=\dd\phi \,,
 \label{eq:su3-heterotic-torsion}
\ee
in their conventions, where $\phi$ is the dilaton and these relations do not require $W_3$ to vanish.
Their quintic example has $J=aJ_{\text{FS}}$ and $\Omega=a^2\Omega_0$, with a known positive function $a$, and satisfies $W_3=0$ and $W_4=\dd\log a$.
This is a conformal rescaling of the restricted Fubini--Study metric, which is itself not Ricci-flat.

Anderson \textit{et al.}~\cite{Anderson:2020hux} used this example to demonstrate neural approximation of a metric with prescribed nonzero torsion.
They fixed $\Omega$ to its analytic expression and trained the components of $J$, using losses for the volume normalization, agreement on coordinate overlaps, and the differential residual $\dd J-\dd\log a\wedge J$.
The experiment used the quintic at complex structure parameter $\psi=10$.
Comparison with the known metric, allowing the relevant coordinate transformations, reduced their $L^1$ metric discrepancy from $0.511$ for the initial Fubini--Study metric to $0.025$ after training.
These are errors in their chosen diagnostic, not percentages or coordinate independent bounds.
The paper also formulated a more general approach in which networks represent both defining forms and the dilaton and are trained against torsion constraints; the implemented benchmark learned $J$ with $\Omega$ fixed.

Completing this program for a heterotic vacuum requires more than fitting the supersymmetry torsion classes.
The gauge connection must obey its instanton equations.
In the absence of localized sources, the flux must also satisfy the anomaly cancellation condition~\cite{Strominger:1986uh}
\be
 \dd H=\frac{\alpha'}{4}\left[\operatorname{tr}(R\wedge R)-\operatorname{tr}(F\wedge F)\right] \,.
 \label{eq:su3-bianchi}
\ee
The demonstrated neural benchmark does not solve this coupled system.
For new examples, independent checks of positivity, global compatibility, torsion residuals, and the Bianchi identity are therefore needed before interpreting a learned metric as a string background.

\subsection{\texorpdfstring{$G_2$}{G2} structures and metrics}
\label{sec:g2-structure-metrics}
A positive three-form $\varphi$ on a seven-manifold determines a Riemannian metric $g_\varphi$ and its Hodge operator $\ast_\varphi$.
The structure is torsion free precisely when
\be\label{eq:learned-g2-torsion-free}
 \dd\varphi=0\,,\qquad
 \dd(\ast_\varphi\varphi)=0\,.
\ee
These equations imply Ricci flatness and holonomy contained in $G_2$; identifying the holonomy with all of $G_2$ requires additional global information~\cite{Karigiannis:2009FlowsG2}.
A neural representation must therefore control both differential conditions while maintaining positivity and consistency between the learned form, metric, and coordinate patches.
Small Ricci curvature by itself does not verify the defining $G_2$ equations.

Beyond the prediction of topological invariants considered in~\cite{Aggarwal:2023swe}, Heyes, Hirst, S\'a Earp, and Silva learn $G_2$ structures on the compact seven-dimensional link over the Fermat quintic~\cite{Heyes:2026rch}.
Their \href{https://github.com/TomasSilva/LearningG2}{\texttt{LearningG2}} implementation uses a neural Calabi--Yau metric to generate training data for the link's three-form and associated metric, then fits their independent components by supervised learning.
For the contact form $\eta$, transverse K\"ahler form $\omega=\dd\eta$, and normalized transverse holomorphic volume form $\Upsilon$, the target satisfies
\be\label{eq:learned-contact-g2}
 \varphi=\eta\wedge\omega+\operatorname{Re}\Upsilon\,,\qquad
 \dd\varphi=\omega\wedge\omega\neq0\,,\qquad
 \dd(\ast_\varphi\varphi)=0\,.
\ee
Thus, the target geometry is a coclosed $G_2$ structure with intrinsic torsion, rather than a torsion free compactification metric.
The reported component mean squared errors are of order $10^{-6}$ after normalization, but numerical differentiation produces appreciable residuals in the expected torsion identities.
Accurate tensor interpolation consequently provides a weaker validation than accurate differential geometry.
Evolution toward vanishing torsion using a Laplacian coflow is proposed as subsequent work~\cite{Heyes:2026rch}.
For this fixed compact link, however, standard topological facts imply an obstruction to a smooth torsion free endpoint.
The radial normal and the gradients of the real and imaginary parts of the defining polynomial give a global normal frame in $\mathbb R^{10}$, so $TL\oplus\mathbb R^3$ is trivial and $p_1(L)=0$; this is the stable parallelizability of an isolated hypersurface link~\cite{Durfee:1975BranchedCovers}.
The link is also $2$-connected.
A compact torsion free $G_2$ structure on such a manifold would have full $G_2$ holonomy and require $\langle p_1(L)\smile[\varphi],[L]\rangle<0$, contradicting $p_1(L)=0$~\cite{Corti:2015SemiFanoG2}.
This deduction rules out a smooth compact torsion free limit on the unchanged link, while leaving singular, collapsing, or noncompact limits as different questions.
For the natural contact Calabi--Yau initial condition, the analytic coflow in fact develops an infinite time singularity and collapses after volume normalization~\cite{Lotay:2022ContactCYFlows}.

A complementary approach learns geometric inputs to an analytic construction.
Douglas, Platt, Qi, and Barbosa approximate Calabi--Yau metrics and harmonic one-forms on their real loci~\cite{Douglas:2024pmn}.
The motivation is the Joyce--Karigiannis construction, which resolves suitable $G_2$ orbifolds by gluing families of Eguchi--Hanson spaces, with a nowhere vanishing harmonic one-form on the singular locus determining the resolution data~\cite{Joyce:2021EguchiHanson}.
For a complete intersection whose real locus is diffeomorphic to $S^1\times S^2$, their numerical results near a singular limit support the existence of such a form.
The calculation supplies evidence for the required input, rather than the metric on the completed seven-manifold.
The authors also formulate an analytic route to a computer assisted proof but do not establish the required quantitative bounds~\cite{Douglas:2024pmn}.
In particular, a positive minimum over sampled points does not certify that the exact harmonic form is everywhere nonzero.

\subsection{Special Lagrangian submanifolds}
\label{sec:special-lagrangians}

Metric learning also gives access to calibrated submanifolds, whose defining equations involve both the K\"ahler form and the holomorphic volume form.
For an oriented real three-dimensional submanifold $\iota:L\hookrightarrow X$ of a Calabi--Yau threefold, the special Lagrangian conditions at constant phase $\theta$ are
\be
 \iota^*J=0\,,\qquad
 \iota^*\operatorname{Im}(\text{e}^{-i\theta}\Omega)=0\,,\qquad
 \iota^*\operatorname{Re}(\text{e}^{-i\theta}\Omega)=\operatorname{vol}_L\,,
 \label{eq:learned-special-lagrangian}
\ee
with $\Omega$ normalized relative to the metric and the orientation chosen as indicated by the last equality~\cite{Joyce:2001xt}.
These cycles are relevant to supersymmetric branes and to the special Lagrangian torus fibrations proposed in the SYZ description of mirror symmetry~\cite{Strominger:1996it}.

Unpublished work by Ahmed, Qi, and Ruehle explores special Lagrangian submanifolds using genetic algorithms and quality-diversity optimization~\cite{Ahmed:2025SearchingNewSpecial}.
(See also the accompanying \href{https://github.com/yidiq7/SLagSearch}{code}~\cite{SLagSearch:ndCode}.)
Candidate three-cycles in a quintic are cut out by three real bihomogeneous constraints $z^\dagger H_a z=0$, with Hermitian matrices $H_a$ and homogeneous coordinates $z$.
The search evaluates geometric residuals on sampled points, while a diversity criterion encourages exploration of distinct candidates instead of repeatedly finding the same solution.
The public presentation uses a Fubini--Study approximation and identifies replacement by a numerical Calabi--Yau metric as a subsequent step~\cite{Ahmed:2025SearchingNewSpecial}.
This work recovers known examples, including $\mathbb{RP}^3$ in the Fermat quintic and $T^3$ examples in the large complex structure limit of the Dwork family.
Establishing new special Lagrangians requires control of smoothness and dimension, as well as the pullback conditions for the desired Calabi--Yau metric; small sampled residuals alone do not supply such a result.
The distinction is especially relevant to the SYZ application, which requires a compatible family of torus fibers and its singular fibers, beyond isolated candidate cycles.

\subsection{Towards analytics}
\label{sec:analytics}

Analytic constructions and numerical approximations provide complementary access to Ricci-flat Calabi--Yau metrics.
An analytic description can specify the metric through integral equations or a convergent series, even when a finite expression in familiar functions is unavailable.
Numerical calculations can also reveal simple approximate formul\ae\ and motivate conjectures about the exact geometry.
These possibilities should be distinguished from a rigorous error estimate proving that a computed metric lies near the Ricci-flat solution.
Recent work illustrates all three directions.
In particular, Douglas, G\'omez-Serrano, Lehmann, Platt, and Qi~\cite{Douglas:2026jop} combine an explicit analytic approximation on the smooth Fermat quartic K3 with a computer assisted proof that an exact Ricci-flat metric lies nearby.
Their result connects numerical approximation with constructive existence away from a degenerating limit; it does not give a finite closed form expression for the exact metric.

\paragraph{Analytic constructions of Ricci-flat metrics:}
Kachru, Tripathy, and Zimet~\cite{Kachru:2018van} approach the K3 metric through the moduli space of a six-dimensional heterotic little string theory compactified on $T^3$.
Compactification first on $T^2$ supplies a four-dimensional theory whose BPS states become instanton corrections after a further circle compactification.
The Gaiotto--Moore--Neitzke construction~\cite{Gaiotto:2010okc,Gaiotto:2009hg} encodes these corrections in integral equations for twistor Darboux coordinates, from which the hyper-K\"ahler metric can be recovered.
The inputs are central charges and BPS indices; wall crossing ensures compatibility between descriptions in different chambers.
This gives a systematic reconstruction of the metric from protected physical data, beginning with a semi-flat approximation.
The authors argue for convergence at sufficiently large circle radius, where the exponential suppression of massive states can overcome the growth of the BPS spectrum.
The construction is presented at the level of physical reasoning, and its description away from singular fibers does not by itself establish a global extension across them.
Its practical implementation also requires the relevant BPS degeneracies.

Their subsequent work~\cite{Kachru:2020tat} develops a complementary description using D-geometry and an infinite dimensional hyper-K\"ahler quotient.
Starting near the $T^4/\mathbb{Z}_2$ orbifold, Fayet--Iliopoulos parameters deform the moment map equations and resolve the singularities.
Expanding in these parameters reduces the construction to successive linear problems for Fourier modes, with higher order sources determined by lower-order solutions.
The authors compute the first nontrivial metric corrections and verify Ricci flatness to that order.
Poisson resummation and the relation between Higgs branch and Coulomb branch descriptions connect this expansion to the little string construction, allowing part of the required BPS spectrum to be extracted.
Thus, the two papers~\cite{Kachru:2018van,Kachru:2020tat} relate an analytic expansion of the geometry to a protected state counting problem.
The explicitly evaluated perturbation uses coordinates appropriate to generic points near the orbifold limit; global control near the exceptional loci and convergence of the full expansion require further analysis.

A different construction is given by Fu, Liu, Xiao, and Yau~\cite{Fu:2026ExplicitKummer}, who obtain a globally convergent representation on a fixed Kummer K3 surface resolving a square torus quotient.
They consider a one-parameter family of K\"ahler classes with equal exceptional curve areas.
An explicit background $\omega_a$ joins Eguchi--Hanson metrics to the flat exterior and has normalized volume residual $f_a=O(a^4)$.
With $A_a=-\operatorname{tr}_{\omega_a}(\text{i}\partial\bar\partial)$, its inverse $G_a$ on functions of zero mean, and the polarized quadratic Monge--Amp\`ere term $Q_a$, the correction satisfies
\begin{align}
 -A_a u+Q_a(u,u)=f_a \,,\qquad
 U_{a,1}=-G_af_a \,,\qquad
 U_{a,m}=G_a\sum_{j=1}^{m-1}Q_a(U_{a,j},U_{a,m-j}) \,,
 \label{eq:analytic-kummer-recursion}
\end{align}
for $m\geq2$.
The Green operator is specified through periodic kernels, separated radial equations, and interface matching.
An $O(a^{-1})$ inverse estimate and a Catalan number bound prove absolute convergence of $u=\sum_{m\geq1}U_{a,m}$ for sufficiently small fixed $a$, including at the exceptional curves.
The resulting positive form $\omega_a+\text{i}\partial\bar\partial u$ is the Ricci-flat metric in the prescribed class.
This is a convergent expansion in the nonlinear source degree; its coefficients themselves depend on $a$, so it is not a Taylor expansion at $a=0$.

A direct numerical test of this analytic construction is not currently controlled by The results of Berglund \textit{et al.}~\cite{Berglund:2022gvm} do not constitute a numerical test of  this analytic construction. Over the Cefal\`u family, one see that the numerical results are poorest at the $\lambda=1$ Kummer quartic, where one attains the highest validation $\sigma$-loss among the singular cases studied, and the truncated Euler density integral does not show the convergence observed at the other sampled singular values. There is also a difference between the geometries being compared: the numerical model is a singular quartic with sixteen nodes, whereas the analytic construction describes a smooth resolution with positive exceptional curve areas. A meaningful comparison would therefore require matching the underlying torus complex structure and K\"ahler class.

The numerical results of Berglund \textit{et al.}~\cite{Berglund:2022gvm} are poorest at the $\lambda=1$ Kummer member of the Cefal\`u family.
Among the singular cases studied, this point has the highest validation $\sigma$-loss, and the truncated Euler density integral fails to exhibit the convergence observed at the other sampled singular values.
A direct comparison using these data therefore fails to provide a controlled test of the analytic proposal.
Moreover, the numerical model is a singular quartic with sixteen nodes, while the analytic construction describes a smooth resolution with positive exceptional curve areas.
A comparison would also require matching the underlying torus complex structure and K\"ahler class.

Fu, Liu, Xiao, and Yau~\cite{Fu:2026ExplicitSchoen} then extend their strategy to a special Schoen threefold obtained by resolving a shifted square torus quotient with sixteen disjoint elliptic singular curves.
The local building blocks are products of Eguchi--Hanson spaces with elliptic curves.
In complex dimension three the Monge--Amp\`ere equation contains both quadratic and cubic terms, which enter the recursive correction.
An explicit inverse assembled from separated modes, radial kernels, and a convergent boundary transmission series supplies the required analytic estimates.
The authors prove convergence and give potential and relative metric remainder bounds for $0<a\leq2^{-2260}$.
This very small interval is a sufficient quantitative bound for their expansion on a fixed model, rather than the range in which Ricci-flat metrics exist.
The result supplies an exact series representation in a restricted family of K\"ahler classes; it does not cover arbitrary Schoen complex structures or establish an efficient numerical implementation throughout moduli space.
Together, these constructions make explicit the additional work needed to turn a perturbative geometric description into a convergent representation of a compact metric.

\paragraph{Analytic structure in numerical approximations:}
For numerical metrics, a useful goal is to extract a compact expression that retains the accuracy of a much larger learned representation.
Mirjani\'c and Mishra~\cite{Mirjanic:2024gek} pursue this by studying symmetries of ambient representatives of the K\"ahler potential correction on Fermat hypersurfaces and related deformations.
Symmetric functions of normalized coordinate moduli provide inputs for small neural models and for symbolic regression, yielding short approximate potentials whose Monge--Amp\`ere residuals can be evaluated directly.
Their proposed propagation of additional ambient, or ``extrinsic,'' symmetries is a conjecture: such symmetries of a representative need not be intrinsic isometries of the hypersurface.
The resulting representation in terms of coordinate magnitudes alone and some formul\ae\ on restricted loci are conditional on this conjecture.
Other pointwise statements follow directly from discrete isotropy and volume normalization.
The main numerical achievement is therefore an interpretable analytic approximation, accompanied by structural predictions that can be investigated independently.

Lee and Lukas~\cite{Lee:2025pue} instead fit neural K\"ahler potential data to symmetry reduced versions of Donaldson's ansatz, as in~\eqref{eq:FSk}.
They study the Dwork quintic family and a one-parameter family of bicubics in $\mathbb{P}^2\times\mathbb{P}^2$, restricting the latter to equal ambient K\"ahler parameters.
At low section degree, symmetry leaves only a few independent coefficients, whose dependence on the complex structure parameter $\psi$ can itself be fitted by simple functions.
The resulting expressions are inexpensive to evaluate and expose how the approximation varies over the sampled family.
The coefficients are optimized against learned potentials, so they need not be the balanced coefficients obtained from Donaldson iteration.
Their observed dependence predominantly on $|\psi|$ is an empirical simplification within numerical accuracy, rather than an exact phase symmetry of the Ricci-flat metric.
Likewise, agreement of potential values at the percent level is not a certified bound on the metric, which depends on second derivatives.

Analytic constraints also improve the reliability of the approximation itself.
Mirjani\'c and Mishra~\cite{Mirjanic:2026fpq} explain how uniqueness in a fixed K\"ahler class identifies class preserving holomorphic automorphisms with holomorphic isometries of the Ricci-flat metric.
This gives a principled symmetry constraint on learned and symbolic ans\"atze.
They derive a global expression for the holomorphic volume to reference volume ratio on complete intersections in products of projective spaces, and distinguish this quantity from weights associated with a particular sampling distribution.
Their graph network represents ambient coordinates by vertices and projectively invariant features by edges, incorporating permutation symmetry into the potential.
Experiments on the Fermat quintic demonstrate favorable training behavior, while comparisons with other symmetry prescriptions show that a small Monge--Amp\`ere loss can coexist with poor curvature diagnostics.
This work supports analytic approximation through geometric constraints, explicit volume identities, and careful validation of derivatives.

The ambient balanced metric analysis of Berglund \textit{et al.}~\cite{Berglund:2026hel}, discussed in Section~\ref{sec:ambient-balanced}, illustrates how numerical coefficient asymptotics can suggest analytic descriptions of SYZ degeneration.
These remain conjectural until convergence is controlled as the section degree increases and the geometry degenerates.

Douglas \textit{et al.}~\cite{Douglas:2026jop} provide an explicit analytic approximation with a rigorous error certificate on
\be
 X=\left\{[Z_0:Z_1:Z_2:Z_3]\in\mathbb{P}^3:
 \sum_{a=0}^{3}Z_a^4=0\right\} \,.
 \label{eq:certified-fermat-quartic}
\ee
Their potential is
\be
 \Phi=\log p_0+\sum_{L=13}^{24}N_Lp_0^{-L/16} \,.
 \label{eq:certified-quartic-potential}
\ee
Here $p_0$ is a Hermitian polynomial of bidegree $(16,16)$ that is strictly positive on $X$, while $N_L$ has bidegree $(L,L)$; the correction terms are therefore homogeneous of degree zero.
The forms respect the discrete symmetries of the quartic, and their stored coefficients are treated as exact numbers in the verification.
A constant multiple of $\text{i}\partial\bar\partial\Phi$ defines the volume one form $\widetilde\omega$ in the normalized hyperplane class.
The construction fits a low degree base and then corrects it through linear least squares in numerically orthonormalized bases, using extended precision to control the severe conditioning of high degree monomials.
This yields a finite, explicitly evaluable approximation whose positivity and residual are subsequently verified on the whole surface.
The authors identify the accompanying \href{https://github.com/yidiq7/machine_percision_cy_metric}{code repository} for the computer assisted bounds on the Fermat quartic.
The exact metric is obtained by the convergent fixed point construction described below, rather than by identifying the finite ansatz with a Ricci-flat solution.

\paragraph{From numerical evidence to a certified metric:}
For the Fermat quartic, a K3 manifold, this step is now achieved by Douglas \textit{et al.}~\cite{Douglas:2026jop}.
Writing $\widetilde g$ for the metric associated with~\eqref{eq:certified-quartic-potential}, their Theorem~1.1 proves that there is a smooth real function $u$ such that
\begin{align}
 \omega_{\mathrm{CY}}&=\widetilde\omega+\text{i}\partial\bar\partial u \,,
 &\|u\|_{\mathscr{X}}&\leq1.551\times10^{-4} \,,\nonumber\\
 \|\text{i}\partial\bar\partial u\|_{C^0(X,\widetilde g)}
 &\leq1.553\times10^{-4} \,.
 \label{eq:certified-quartic-error}
\end{align}
The $\mathscr{X}$-norm is the paper's explicitly weighted combination of H\"older $C^{2,\alpha}$ and Sobolev $H^2$ control, with $\alpha=1/2$.
The second inequality is a uniform bound on the K\"ahler form correction measured using $\widetilde g$; it is not a residual loss or a bound on the curvature difference.
The proof constructs the exact metric without assuming Yau's theorem.

There are two distinct verification steps.
First, Taylor models with rigorous remainder and rounding bounds, together with a verified covering of the surface, establish positivity and global bounds for the volume residual.
With $\eta=\mu_{\widetilde\omega}/\widehat\mu$, where $\widehat\mu$ is the volume one holomorphic volume measure, and $f=1-\eta^{-1}$, Proposition~2.4 gives
\be
 \|f\|_{C^0(X)}\leq5.0\times10^{-20} \,,\qquad
 \sup_X|\nabla f|_{\widetilde g}\leq3.37\times10^{-16} \,.
 \label{eq:certified-quartic-residual}
\ee
The much smaller sampled mean residual, approximately $1.7\times10^{-22}$, is a numerical diagnostic and is not the certified uniform bound.
Second, explicit geometric and elliptic estimates control the inverse Laplacian on the symmetry invariant functions of zero mean and the nonlinear Monge--Amp\`ere term, allowing a contraction argument to turn the residual bounds into~\eqref{eq:certified-quartic-error}.
A small residual, even if globally certified, does not by itself supply a metric error bound without this second step.

Platt's treatment of the Nirenberg problem~\cite{Platt:2026Nirenberg} illustrates the same general strategy in another setting.
For prescribed Gaussian curvature on $S^2$, he uses interval arithmetic, rigorous residual estimates, and bounds on the inverse linearized operator to promote numerical candidates to exact solutions through a contraction argument.
The inverse estimate controls both the computed finite dimensional part and the remaining modes.
The Fermat quartic result supplies the corresponding quantitative estimates for a compact Calabi--Yau example.

Related computer assisted existence proofs illustrate the broader scope of this program.
Buttsworth and Hodgkinson~\cite{Buttsworth:2024Einstein} construct a non-round Einstein metric on $S^{12}$ by rigorously controlling an approximate solution and proving that it can be perturbed to an exact one.
The existence of a spherically symmetric, non-Schwarzschild black hole in Einstein--Weyl gravity is certified by combining a convergent horizon expansion, validated integration across a compact exterior region, and a fixed point construction of the asymptotic tail~\cite{Goldstein:2026svk}.
The matching step uses the Poincar\'e--Miranda theorem and proves existence without uniqueness of the matching zero.
The black hole solution has vanishing Ricci scalar but nonvanishing Ricci tensor.
The relevance of this work here is the method of promoting numerical evidence to an existence theorem.

To explain the structure of the certification argument in general, fix the complex structure and K\"ahler class on a smooth compact connected $X$, and suppose that $\widetilde\omega$ is a smooth approximation whose global K\"ahler property and positivity have been verified.
Let $\mu$ be the positive volume form proportional to $\text{i}^{n^2}\Omega\wedge\bar\Omega$, normalized by $\int_X\mu=\int_X\widetilde\omega^n$.
For a real potential correction $u$, define
\be
 F(u)=\frac{(\widetilde\omega+\text{i}\partial\bar\partial u)^n-\mu}
 {\widetilde\omega^n} \,,\qquad
 \int_X u\,\widetilde\omega^n=0 \,.
 \label{eq:analytic-certification-map}
\ee
Stokes' theorem ensures that $F(u)$ also has zero $\widetilde\omega^n$-mean.
Its linearization at $u=0$ is
$L=\Delta_{\widetilde\omega}$, with
$\Delta_{\widetilde\omega}u=\operatorname{tr}_{\widetilde\omega}(\text{i}\partial\bar\partial u)$.
This equals $\tfrac12\Delta_{\widetilde g}u$ when the real Laplacian is defined as $\Delta_{\widetilde g}=\operatorname{div}_{\widetilde g}\nabla_{\widetilde g}$, as in~\cite{Douglas:2026jop}.
Removing constants makes this operator invertible.
For a schematic formulation, work from $C^{2,\alpha}_0(X)$ to $C^{0,\alpha}_0(X)$, where $0<\alpha<1$ and the subscript denotes zero mean.
The Fermat quartic proof uses specially weighted norms to obtain effective constants; its numerical bounds should not be identified with constants in these unweighted spaces.
Writing $F(u)=F(0)+Lu+N(u)$, suppose verified estimates give
\begin{align}
 \|F(0)\|_{C^{0,\alpha}}&\leq\varepsilon \,,\qquad
 \|L^{-1}\|_{C^{0,\alpha}_0\to C^{2,\alpha}_0}\leq M \,,\nonumber\\
 \|N(u)-N(v)\|_{C^{0,\alpha}}
 &\leq Cr\,\|u-v\|_{C^{2,\alpha}}
 \quad\text{for }\|u\|_{C^{2,\alpha}},\|v\|_{C^{2,\alpha}}\leq r \,.
 \label{eq:analytic-certification-bounds}
\end{align}
Then the sufficient inequalities
\be
 M\varepsilon+MC r^2\leq r \,,\qquad MC r<1
 \label{eq:analytic-certification-radius}
\ee
make $u\mapsto-L^{-1}(F(0)+N(u))$ a contraction of this ball, provided the ball remains uniformly K\"ahler.
The fixed point has $\|u\|_{C^{2,\alpha}}\leq r$, and elliptic regularity for the smooth background data then gives a smooth Ricci-flat metric.
The potential bound gives a corresponding $C^{0,\alpha}$ bound on the metric correction, with the constant determined by the chosen norms.
Yau's theorem already guarantees existence in the chosen K\"ahler class, but this argument establishes existence independently and also certifies proximity to the computed metric.
The Fermat quartic construction realizes this program with explicit constants and controlled rounding and covering errors.
A small loss measured on finitely many sample points does not supply these bounds.

Earlier work by Douglas, Platt, Qi, and Barbosa~\cite{Douglas:2024pmn} used Monge--Amp\`ere residuals and inverse Laplacian estimates to formulate conditions ensuring proximity to the exact Ricci-flat metric.
The new result verifies the necessary quantitative bounds for a particular smooth K3 and K\"ahler class~\cite{Douglas:2026jop}.
Extending such certificates to numerical metrics on Calabi--Yau threefolds, across moduli space, or near degenerations requires new bounds for each geometry, including the higher order Monge--Amp\`ere nonlinearity in higher dimension.
Certified errors for bundle metrics, harmonic representatives, and physical observables require further analysis as well.

\section{Software for numerical Calabi--Yau geometry}
\label{sec:calabi-yau-software}

The public packages \texttt{MLGeometry}, \texttt{cymetric}, \texttt{CYJAX}, and \texttt{cymyc}, discussed here in order of first public code release, provide complementary implementations of numerical Ricci flat Calabi--Yau geometry.
All four are principally written in Python, with automatic differentiation and neural network optimization supplied by \href{https://docs.jax.dev/en/latest/}{JAX} or \href{https://www.tensorflow.org/}{TensorFlow}.
Their common workflow starts with an algebraic description of a manifold, generates points and integration weights, and optimizes a parameterized metric or K\"ahler potential against geometric equations, rather than precomputed Ricci flat metrics.
The underlying ans\"atze and losses are discussed in Sections~\ref{sec:pinn} and~\ref{sec:potential}; here we compare the supplied implementations, supported geometries, treatment of moduli, and calculations available after training~\cite{Anderson:2023viv}.

\texttt{MLGeometry} implements the neural potential approach developed by Douglas, Lakshminarasimhan, and Qi.
The inspected \texttt{MLGeometry-JAX 2.2.2} source uses JAX, \href{https://flax.readthedocs.io/en/stable/}{Flax}, and \href{https://optax.readthedocs.io/en/latest/}{Optax}, whereas the original publication and earlier releases used TensorFlow/\href{https://keras.io/}{Keras}.
Its bihomogeneous network representation, reviewed in Section~\ref{subsec:holomorphic-networks}, provides a structured algebraic approximation without requiring an imposed discrete symmetry of the manifold~\cite{MLGeometry:ndCode,Douglas:2020hpv}.
The package includes point sampling, local coordinate and integration data, several volume form matching losses, Adam and L-BFGS optimization, and evaluation of the learned metric.
Its tutorial demonstrates independent training and test samples and geometric diagnostics.
The inspected geometry classes support hypersurfaces and intersections of two hypersurfaces in a single projective space, together with sampling of their real loci.
More general complete intersections or ambient spaces require extensions to the supplied geometry routines~\cite{MLGeometry:ndCode}.

The \texttt{cymetric} package offers an integrated workflow for complete intersection Calabi--Yau manifolds in products of projective spaces and for anticanonical hypersurfaces in toric varieties, including Kreuzer--Skarke constructions.
The original implementation considered here uses TensorFlow/Keras and additionally contains \href{https://www.wolfram.com/language/}{Wolfram Language} code and a Mathematica interface; the established toric workflow uses \href{https://www.sagemath.org/}{SageMath} for algebraic geometric preprocessing and Mathematica for point generation.
Given defining equations and prescribed complex structure and K\"ahler parameters, \texttt{cymetric} constructs samples, quadrature weights, holomorphic volume data, and coordinate pullbacks.
Its five principal neural ans\"atze include direct prediction of metric components, several corrections to the Fubini--Study metric, and a scalar potential correction, allowing comparisons of these parameterizations within one package~\cite{Cymetric:ndCode,Larfors:2022nep}.
Explicit K\"ahler class control is supplied by matching numerical line bundle slopes to intersection theoretic values.
Training combines the Monge--Amp\`ere residual with the K\"ahler, coordinate transition, and class constraints required by the chosen ansatz, with curvature based diagnostics also available.
The source includes Donaldson balanced metric routines and supporting calculations for line bundle Hermitian--Yang--Mills problems.
The implemented reference forms determine the accessible K\"ahler classes, and standard training is performed at specified moduli.
A separate, updated \href{https://github.com/ruehlef/cymetric}{\texttt{cymetric} fork} adds \href{https://pytorch.org/}{PyTorch} and JAX backends; these are distinct from the TensorFlow implementation described above~\cite{Ruehle:ndCymetricFork,Cymetric:ndCode}.

\texttt{CYJAX} uses JAX, Flax neural networks, and Optax optimization, implementing both Donaldson iteration and neural optimization of the algebraic K\"ahler potential ansatz~\eqref{eq:FSk}, up to normalization conventions.
In its supplied learning workflow, a network maps complex structure parameters to a Cholesky parameterization of the positive Hermitian matrix defining the potential, allowing one model to represent a family of metrics.
Training uses a normalized Monge--Amp\`ere residual estimated by Monte Carlo integration, without requiring a preliminary dataset of Donaldson metrics~\cite{Gerdes:2022nzr,CYJAX:ndCode}.
The package exposes modular routines for polynomial evaluation, projective coordinate transformations, random line point sampling, integration, and curvature evaluation.
Its examples include the Dwork family and a two parameter quintic family.
The supplied implementation concerns varieties defined by one hypersurface equation in one projective space; the section degree fixes the finite dimensional approximation space~\cite{CYJAX:ndCode}.

The \texttt{cymyc} package is built on JAX, using Flax for neural networks and Optax for optimization.
Its metric workflow treats complete intersections in products of complex projective spaces, including hypersurfaces as a special case.
It implements the global scalar correction of Section~\ref{sec:potential} using projectively invariant inputs for each ambient factor, preserving closedness and the reference cohomology class.
Point generation, holomorphic volume data, pullbacks, and a weighted Monge--Amp\`ere loss complete the metric learning workflow.
The available reference forms select K\"ahler classes inherited from the ambient projective factors~\cite{Cymyc:ndCode}.
A distinctive strength is its integration of metric approximation with subsequent differential geometry: differentiable routines evaluate connections, Riemann and Ricci curvature, Chern forms, and Euler characteristic integrals.
It also approximates tangent bundle valued harmonic forms and provides calculations of the Weil--Petersson metric and canonically normalized Yukawa couplings in the standard embedding.
The independent Kodaira--Spencer route described in Section~\ref{sec:standard-embedding} computes the complex structure moduli space metric by numerical integration without first learning a Ricci flat metric, providing a cross check.
The demonstrated Yukawa workflows specify the relevant bundle and cohomology data; they should not be interpreted as automatic solutions for arbitrary heterotic compactifications~\cite{Butbaia:2024xgj}.

\begingroup
\emergencystretch=2em
Several smaller projects complement these four libraries.
The code \href{https://github.com/harveyThomas4692/jax_metric_CalabiYau}{\texttt{jax\_\allowbreak metric\_\allowbreak CalabiYau}} is a Python/JAX implementation of projectively invariant scalar potential learning, with quintic and tetraquadric examples, particularly suited to illustrating the construction~\cite{JaxMetricCalabiYau:ndCode}.
The code \href{https://github.com/jake997/invariant_layers_cymetric}{\texttt{invariant\_\allowbreak layers\_\allowbreak cymetric}} supplies Python/TensorFlow extensions that impose group invariance through nontrainable canonicalization layers; the associated study by Hendi, Larfors, and Walden applies these to metric learning~\cite{InvariantLayersCymetric:ndCode,Hendi:2024yin}.
The code \href{https://github.com/kitft/heteroticyukawas}{\texttt{heteroticyukawas}} extends the Python/TensorFlow \texttt{cymetric} workflow to Calabi--Yau metrics, line bundle metrics, harmonic representatives, and physical Yukawa couplings on the tetraquadric~\cite{HeteroticYukawas:ndCode}.
The toric and topological tools reviewed in Section~\ref{sec:cy-databases}, including \texttt{CYTools}, complement these local Ricci flat metric solvers~\cite{Demirtas:2022hqf}.
\par
\endgroup

Package choice depends on the geometric scope and required observables, with the numerical diagnostics of Section~\ref{sec:don} and the distinction from rigorous certification in Section~\ref{sec:analytics} applying throughout.
Comparisons of speed or accuracy require common geometries, precision, hardware, sampling budgets, and error criteria~\cite{Douglas:2020hpv,Anderson:2023viv}.
Mirjani\'c and Mishra~\cite{Mirjanic:2024gek} compare the accuracy of several machine learning approaches on the same manifold, the Fermat quintic.
Their Tables~1 and~2 report $\sigma$ measures and, for selected methods, curvature diagnostics, providing a common geometric benchmark.

\section{Conclusions and outlook}
\label{sec:conclusion}

The Standard Model of particle physics has strikingly elegant features.
It is a renormalizable chiral gauge theory based on a compact gauge group.
Local gauge symmetry constrains the allowed interactions, while the Yukawa matrices and Higgs potential contain independent parameters.
The matter representations cancel gauge and mixed gauge--gravitational anomalies.
Together, these features give a mathematically coherent structure.
The Lagrangian, however, is full of seemingly arbitrary details: three generations for no obvious reason, a zoo of Yukawa couplings spanning many orders of magnitude, an effective QCD $\theta$-angle whose smallness is unexplained in the minimal Standard Model, a Higgs mass whose radiative stability cries out for an explanation, and a handful of mixing angles and CP-violating phases that we mostly just measure and plug in.
From an aesthetic viewpoint, this looks more like a catalog of numbers than a uniquely compelling structure.

Whether the Standard Model is beautiful or ugly is, in the end, a matter of taste.
We must find a realization of this theory with moduli stabilized as a proof of principle that string theory connects to the real world.

In a string compactification, particle physics is encoded in the topology and geometry of the extra dimensions.
The internal space, together with the bundles, branes, fluxes and other data appropriate to the construction, determines the four-dimensional effective theory.
Questions about the observed matter content and interactions thereby become concrete questions about string compactification.
The motivation for studying Calabi--Yau manifolds is consequently physical as well as mathematical: understanding the topology and computing the geometry of the extra dimensions are essential steps towards explaining the structure and parameters of particle physics.

The program to model the real world via compactification of a consistent theory of quantum gravity was initiated in 1985~\cite{Candelas:1985en}.
This article therefore serves as one status update on aspects of this challenge some 40 years later.
In 2026, following two and a half superstring revolutions, the theory and our ability to compute have matured enough that we can justly preface string phenomenology with the word \emph{precision}.

The methods we have reviewed address successive parts of the global agenda.
Learning topological invariants, identifying equivalent presentations and generating new geometries improve our ability to explore candidate compactifications.
Numerical metric and bundle calculations then make it possible to investigate selected candidates in greater detail.
The remarkable technical progress we have made is owed to the transformative leaps in artificial intelligence during the past decade.
Machine learning contributes both to the search through discrete possibilities and to the approximation of functions constrained by geometric equations.
These tasks become especially useful when connected: physical requirements can guide the search, and detailed calculations can determine which candidates warrant further attention.
The algebraic and differential geometry underlying each construction provides the conditions against which the learned results must be tested.

Several priorities follow for future work.
Databases should record equivalences and the scope of their coverage so that counts of presentations, geometries and vacua can be distinguished.
Search algorithms should make their selection biases explicit, especially when their output is used to infer statistical properties of compactifications.
Metric solvers should be tested on common geometries and extended to families, with particular attention to the deterioration of numerical control near degenerations.
Compatible conventions and reusable software would allow the output of one calculation to become a checked input to the next.
Such comparisons can also reveal which structures learned by a model admit an algebraic or analytic explanation, turning numerical observations into precise conjectures.
For an account of the physical and mathematical calculations that become accessible once a Ricci-flat metric is available, we refer to a companion review~\cite{Berglund:2026hdh}.

These questions place the study of Calabi--Yau and related geometries within the broader framework of the string landscape and the swampland~\cite{Susskind:2003kw,Vafa:2005ui}.
The landscape comprises consistent string vacua, while the swampland consists of apparently consistent effective theories that admit no completion in quantum gravity.
A database of Calabi--Yau manifolds organizes ingredients for compactification; constructing vacua also requires the appropriate bundles, branes and fluxes, together with an analysis of moduli stabilization and consistency.
Vacuum selection asks why our Universe realizes a particular admissible vacuum, connecting compactification dynamics and cosmology with observational constraints.
Machine learning can advance this program by identifying promising geometries, accelerating the calculation of effective theory data and directing expensive computations towards candidates with specified physical properties~\cite{He:2017aed,Carifio:2017bov}.
Building on the statistical approach to string vacua developed by Denef and Douglas~\cite{Denef:2004ze}, machine learning can also be used to search for correlations among masses, couplings and other parameters of the low energy effective theory across ensembles of Calabi--Yau compactifications.
The same tools can help test proposed swampland constraints across families of compactifications and reveal correlations that suggest new conjectures, provided that their interpretation respects the domain of validity of the effective theory.
Such calculations must distinguish established consistency conditions from conjectural bounds and patterns inferred from finite datasets.
A learned ranking of candidates becomes a statement about physical vacuum selection only when tied to a justified dynamical mechanism or probability measure.

The Standard Model of particle physics together with the cosmological Standard Model supplies a demanding target for this program.
If dark energy is a positive cosmological constant, an outstanding challenge is to construct a demonstrably controlled Calabi--Yau compactification that simultaneously admits a metastable de Sitter vacuum and reproduces the observed spectrum and interactions of the Standard Model.
The KKLT proposal~\cite{Kachru:2003aw} and the Large Volume Scenario with an appropriate uplift~\cite{Balasubramanian:2005zx} provide important starting points, complemented by explicit candidate de Sitter vacua developed by McAllister and collaborators~\cite{Kachru:2019dvo,McAllister:2024lnt}.
Progress toward combining cosmology and particle physics within a single construction is illustrated by Cicoli and collaborators~\cite{Cicoli:2021dhg}, who embed a Standard Model quiver into a global type IIB Calabi--Yau orientifold compactification with stabilized closed string moduli, a candidate de Sitter minimum, and an MSSM-like spectrum in the low energy.
A de Sitter vacuum necessarily breaks the four-dimensional $\mathcal{N}=1$ supersymmetry, so the mechanism responsible for positive vacuum energy must also be compatible with the particle physics sector.
The de Sitter constructions cited here are formulated in type IIB string theory, whereas the Standard Model constructions emphasized in this review are predominantly heterotic.
String dualities therefore offer another possible route, where suitable dual descriptions exist, to bring these ingredients into a common framework while tracking the effects of moduli stabilization and supersymmetry breaking on the spectrum and couplings.

A realistic compactification must account for moduli stabilization and permit controlled calculations of physical parameters as well as the light spectrum.
A Ricci-flat metric supplies an essential ingredient in many such calculations; the remaining compactification data and their dynamics must also be determined.
The ambition is to carry a specified compactification through to quantitative predictions that can be compared with experiment.
Progress should therefore be judged by how far we can follow the relation between the topology and geometry of the extra dimensions and the particle physics of the four dimensions we observe.
Given the continual advances in artificial intelligence and in its applications, it is a propitious moment for untamed ambition.

It is overwhelmingly likely that if there is one satisfactory construction of the real world in string theory, there will be many.
As to date, we do not have even one, we leave new avatars of the vacuum selection problem to the future.

\acknowledgments
We thank our collaborators on Calabi--Yau geometry and machine learning: Per Berglund, Kieran Bull, Giorgi Butbaia, Ben Campbell, Yang-Hui He, Elli Heyes, Edward Hirst, Tristan H\"ubsch, Andre Lukas, Viktor Mirjani\'c, Challenger Mishra, Justin Tan, Wati Taylor, and Andrew Turner.
We also thank Shahin Sheikh Jabbari for the invitation to write this review. 
VJ thanks the NSF Institute for Artificial Intelligence and Fundamental Interactions (IAIFI) and the Department of Physics at Northeastern University for hospitality during his sabbatical.
VJ is supported by the South African Research Chairs Initiative of the Department of Science, Technology, and Innovation (DSTI) and the National Research Foundation (NRF), grant 78554.

\appendix

\section{Machine learning architectures}\label{app:Appendix}

This appendix summarizes the tools used in the review.
Neural networks and transformers specify parameterized models; genetic algorithms and reinforcement learning specify ways of searching or learning.
They can be combined, for example by representing a reinforcement learning policy with a neural network.

\subsection{Neural networks}\label{app:neural-networks}

A neural network represents a parameterized map $f_\theta$ by composing simple operations.
In a fully connected feedforward network, the hidden layers take the form
\be\label{eq:app-network}
 h^{(0)}=x \,,\qquad
 h^{(\ell+1)}=\sigma_\ell\!\left(W_\ell h^{(\ell)}+b_\ell\right)\,,
 \qquad f_\theta(x)=W_Lh^{(L)}+b_L \,,
\ee
where the weights and biases comprise $\theta$, and the nonlinear activation functions $\sigma_\ell$ allow the network to represent nonlinear dependencies.
The output layer is adapted to the task, for example by converting scores to class probabilities.
In supervised learning, the parameters are fitted by minimizing a loss averaged over labeled examples; mean squared error is common for regression and cross-entropy for classification.
Backpropagation evaluates derivatives of the loss with respect to the parameters by the chain rule, enabling gradient based optimization~\cite{Rumelhart:1986gxv}.

In this review, inputs can encode a configuration matrix or polytope, with outputs predicting topological quantities, or can specify a point on a manifold, with outputs approximating a metric or K\"ahler potential.
In the latter setting, the physics informed losses of Section~\ref{sec:pinn} can supply the training signal without known target metrics.
Network design should respect relevant symmetries and provide the differentiability required by the geometric equations.
Performance must be assessed on independent examples or sample points: a small training loss alone does not establish generalization or control the global geometric error.

\subsection{Genetic algorithms}\label{app:ga}

A genetic algorithm searches by evolving a population of candidate solutions~\cite{Holland:1992Adaptation}.
Each candidate is encoded as a genome and assigned a fitness measuring how well it meets the desired objective.
Selection preferentially reproduces candidates with higher fitness.
Crossover combines parts of parent genomes, while mutation introduces changes that maintain exploration.
Repeated evaluation and selection concentrate effort on promising regions without requiring derivatives of the fitness function.
Population size, mutation rate, selection strength, and the representation itself influence the balance between exploration and refinement.

For Calabi--Yau searches, a genome can encode lattice vertices or combinatorial triangulation data.
The fitness can reward reflexivity, penalize violated constraints, or optimize a physical observable.
These choices underlie the construction of reflexive polytopes in~\cite{Berglund:2023ztk} and searches for triangulations with desired axion properties in~\cite{MacFadden:2024him}.
An encoding should either preserve the necessary geometric conditions under its genetic operations or permit efficient rejection or repair of invalid offspring.

The output is a collection of promising candidates whose geometric properties must be checked independently.
A finite run need not find the global optimum or every admissible solution.
Selection deliberately biases the sample toward the chosen fitness, so the resulting population does not directly estimate frequencies in the underlying landscape.

\subsection{Reinforcement learning}\label{app:rl}

Reinforcement learning formulates a search as a sequence of decisions.
At step $t$, an agent observes a state $s_t$, chooses an action $a_t$ according to a policy $\pi(a_t\mid s_t)$, and receives a reward $r_t$ as the environment changes.
For an episode ending after $T$ steps, a common objective is
\be\label{eq:app-return}
 J(\pi)=\mathbb E_\pi\!\left[\sum_{t=0}^{T-1}\gamma^t r_t\right]\,,
 \qquad 0\leq\gamma\leq1 \,.
\ee
The discount factor $\gamma$ controls the relative weight of later rewards.
Learning must assign credit to earlier decisions and balance exploration against exploiting actions already known to work.
Value based methods estimate future returns associated with states or actions, as in $Q$-learning~\cite{Watkins:1992bjs}; policy-based methods adjust the action probabilities directly.
Neural networks can approximate these functions when the state space is large.

For Calabi--Yau construction, states encode geometric data, actions modify the current candidate, and rewards express geometric or physical targets.
The triangulation searches of~\cite{Berglund:2024reu} illustrate this approach, including objectives involving fibrations and heterotic bundle constraints.
The state should retain the information needed to determine allowed moves and their consequences.
A high reward does not replace explicit verification of the desired conditions: necessary bundle conditions, for example, do not establish stability.
The learned policy also generally samples geometries non-uniformly.

\subsection{Transformers}\label{app:transformers}

Transformers process sequences through attention, which makes each token's representation depend on other tokens~\cite{Vaswani:2017lxt}.
Tokens may encode integers, lattice points, or simplices rather than words.
An embedding converts tokens to vectors; positional information records their order.
For one attention head, learned projections form query, key, and value matrices $Q,K,V$, and
\be\label{eq:app-attention}
 \operatorname{Attention}(Q,K,V)
 =\operatorname{softmax}\!\left(\frac{QK^{\mathsf T}}{\sqrt{d_k}}\right)V \,,
\ee
where $d_k$ is the query/key dimension and softmax acts along each row.
Thus, each output is a weighted combination of value vectors, with weights determined by query--key compatibility.
Multiple heads learn different such combinations; feedforward layers, residual connections, and normalization complete the basic block.

An encoder builds representations of an input sequence.
An autoregressive decoder generates an output one token at a time, conditioned on earlier tokens and, in an encoder--decoder model, on the encoded input.
A causal mask excludes future output tokens during training, which commonly minimizes next-token cross-entropy.
This separates parallel evaluation of training targets from sequential generation.

For example, \texttt{CYTransformer} learns to generate triangulations from polytope data~\cite{Yip:2025hon}, as discussed in Section~\ref{sec:newcy}.
Its output is a proposed combinatorial construction whose geometric conditions can be checked independently.
Attention can capture relations among widely separated tokens, but an arbitrary serialization does not automatically enforce the permutation or lattice symmetries of the underlying geometry.

\clearpage
\bibliographystyle{JHEP}
\bibliography{ref}

\end{document}